  \pdfoutput=1
  \documentclass[graphicx,amssymb,amsmath,enumerate,11pt]{article}
  \usepackage{fancyhdr}
  \usepackage{subfigure}
  \fancyheadoffset[]{0.1cm}
\usepackage[sort&compress,numbers]{natbib}
  \usepackage{epsfig}
  \usepackage{url}
  
\begin{document}      



\begin{center}
{\Large \bf  Sterile Neutrino Fits to Short Baseline Neutrino Oscillation Measurements}\\
{\it J.M. Conrad~\footnote{Massachusetts Institute of Technology, Cambridge, MA 02139, USA}, C.M. Ignarra~$^1$, G. Karagiorgi~\footnote{Columbia University, New York, NY 10027, USA}, M.H. Shaevitz~$^2$, J. Spitz~$^1$}
\end{center}

\vskip 0.5cm

\noindent{\bf Abstract:}\\
{\small This paper reviews short baseline oscillation experiments as
  interpreted within the context of one, two, and three sterile
  neutrino models associated with additional neutrino mass states in the $\sim 1$~eV range.
 Appearance and disappearance signals and limits are considered.
We show that fitting short baseline data sets to a (3+3) model, defined by three active and three sterile
neutrinos, results in an overall goodness of
fit of $67\%$, and a compatibility of $90\%$ among all data sets -- to be compared to the compatibility of 0.043\% and 13\% for a
(3+1) and a (3+2) model, respectively. While the (3+3) fit yields the highest quality overall, it still finds inconsistencies with the MiniBooNE appearance
data sets; in particular, the global fit fails to account for the observed MiniBooNE low-energy excess. Given the overall improvement, we recommend using the results of (3+2) and (3+3) fits,
rather than (3+1) fits, for future neutrino oscillation phenomenology.  These results
 motivate the pursuit of further short baseline experiments, such as those
 reviewed in this paper.}

\vskip 1cm

\section{Introduction}

Over the past 15~years, neutrino oscillations associated with small
splittings between the neutrino mass states have become well
established~\cite{SuperKsolar,SNO,Kamland,SuperKatmos, SoudanII, k2kosc,MinosCC1,MinosCC2,MinosCC3,MINOS4,MinosCC5,minostheta13,t2k,dc1stpub,dayabay,reno}.  Based on this, a phenomenological extension of the
Standard Model (SM) has been constructed involving three neutrino mass
states, over which the three known flavors of neutrinos ($\nu_e$,
$\nu_\mu$  and $\nu_\tau$) are distributed. This
is a minimal extension of the SM, requiring a lepton 
mixing matrix that is analogous to the quark sector, and introducing
neutrino mass. 

Despite its success, the model does not address
fundamental questions such as how neutrino masses should be
incorporated into a SM Lagrangian, or why the neutrino
sector has small masses and large mixing angles compared to the quark
sector.  As a result, while this structure makes
successful predictions, one would like to gain a deeper understanding of neutrino
phenomenology. This has led to searches for other unexpected properties of neutrinos that might lead to clues towards a more complete theory governing their behavior.

Recalling that the mass splitting is related to the frequency of oscillation, short baseline (SBL) experiments search for evidence of ``rapid" oscillations above the established solar ($\sim10^{-5}$~eV$^2$) and atmospheric ($\sim10^{-3}$~eV$^2$) mass splittings that
are incorporated into today's framework.  A key motivation is the
search for light sterile neutrinos---fermions that do not participate
in SM interactions but participate in mixing with the established SM neutrinos. Indications of oscillations between active and sterile neutrinos have been observed in the LSND~\cite{LSND}, MiniBooNE~\cite{MB} and reactor~\cite{reactor} experiments, though
many others have contributed additional probes of the effect, which are of comparable sensitivity and/or complementary to those above.

This paper examines these results within the context of models describing oscillations with sterile neutrinos. An oscillation formalism that introduces multiple sterile neutrinos is described in the next section. Following this, we review the SBL data sets used in the fits presented in this paper, which include
both positive signals and stringent limits.  We then detail  
the analysis approach, which we have developed in a series of past
papers~\cite{sorel, karagiorgi, viability}. 
The global fits are presented with one, two, and three
light sterile neutrinos.  While groups ~\cite{sorel,whitepaper,Maltoni:2007zf} have explored
fits with three sterile neutrinos in the past, the fits presented here
represent an important step forward. Specifically, we show that, for the first time, the 
(3+3) model resolves some disagreements between the data sets and substantially improves the overall goodness of fit as compared to the (3+2) model. Lastly, the future of SBL searches for sterile neutrinos is reviewed.

\section{Oscillations Involving Sterile Neutrinos}
\subsection{Light Sterile States}

Sterile neutrinos are additional states beyond the standard
electron, muon, and tau flavors, which do not interact via the
exchange of $W$ or $Z$ bosons~\cite{zwidth} 
and are thus ``sterile'' with respect to the weak interaction. These states are motivated by many Beyond Standard Model theories,
where they are often introduced as gauge singlets.   Traditionally, sterile neutrinos were introduced at very high mass scales within the context of grand unification and leptogenesis. For many years, sterile neutrinos with light masses were
regarded as less natural.  However, as recent data \cite{LSND,nubarminiosc,Gallium,reactor} has indicated the
potential existence of light sterile neutrinos,
the theoretical view has evolved to accommodate these light mass gauge
singlets.  At this point, it is generally accepted that the mass scale
for sterile neutrinos is not well predicted, and the existence of one or more sterile neutrinos accommodated by introducing extra neutrino mass states at the eV scale is possible.    An excellent review of the phenomenology of sterile neutrinos, as well as the data motivating light sterile models, is provided in Ref.~\cite{whitepaper}.

Within the expanded oscillation phenomenology, sterile neutrinos are handled as
additional non-interacting flavors, which are connected to additional
mass states via an extended mixing matrix with extra mixing angles and
possible CP violation phases. These additional mass states must be
mostly sterile, with only a small admixture of the active flavors, in
order to accommodate the limits on oscillations to sterile neutrinos
from the atmospheric and solar neutrino data. Experimental evidence
for these additional mass states would come from the disappearance
of an active flavor to a sterile neutrino state or
additional transitions from one active flavor to another through the
sterile neutrino state.

The number of light sterile neutrinos is not predicted by
theory. However, a natural tendency is to introduce three sterile
states.   Depending on how the states are distributed in mass scale, one, two, or all three states may be involved in 
SBL oscillations. These are referred to as (3+N) models where the 
``3'' refers to the three active flavors and the ``N" refers to the number of sterile neutrinos. 


Introducing sterile neutrinos can have implications in cosmological observations, especially measurements of the radiation density in the early universe. These are compounded if the extra neutrinos have significant mass ($>$1~eV) and do not decay. Currently, cosmological data
allow additional states.  For example, Ref.~\cite{atta} estimates the
effective number of neutrinos to be $N_{eff}=5.3\pm1.3$. 
Similar cosmological-based analyses favoring light sterile neutrinos can be found in Refs.~\cite{GG} and~\cite{Mena}. Upcoming Planck data~\cite{planck} is expected
to precisely measure $N_{eff}$.  This parameter, however, can be considered a model dependent one.  As an example, there are a variety
of classes of theories where the neutrinos do not thermalize in the early
universe~\cite{whitepaper}.  In these cases, the cosmological neutrino
abundance would substantially decrease, rendering cosmological
measurements of $N_{eff}$ invalid. Therefore,  while the community certainly looks forward to cosmological measurements of $N_{eff}$,
we think that SBL experiments are a largely better approach for probing sterile neutrinos and constraining their mixing properties.
We therefore proceed with a study of the SBL data, without further reference to the cosmological results.

\subsection{The Basic Oscillation Formalism}
\label{basic}
Before considering the phenomenology of light sterile neutrinos, it is
useful to introduce the idea of oscillations within a simpler model.   In this section, we first consider the two-neutrino formalism.  We then extend these ideas to form the well established three-active-flavor neutrino model.   Based on these concepts, we expand the discussion to include more states in the following section.

Neutrino oscillations require that 
1)  neutrinos have mass; 2)  the difference between the masses
is small; and 3) the mass eigenstates are rotations of the  
weak interaction eigenstates. These rotations are given in a simple two-neutrino model 
as:
\[
\begin{array}{l}
\nu _e=\cos \theta \;\nu _1+\sin \theta \;\nu _2 \\ 
\nu _\mu =-\sin \theta \;\nu _1+\cos \theta \;\nu _2~,
\end{array}
\]
where $\nu_i~(i=1,2)$ is the ``mass eigenstate", $\nu_\alpha~(\alpha=e,\mu)$ is the ``flavor eigenstate", and  $\theta$ is the ``mixing angle".
Under these conditions, a neutrino born in a pure flavor state through
a weak decay can
oscillate into another flavor as the state propagates in space,
due to the fact that the different mass eigenstate components
propagate with different frequencies.   The mass splitting between
the two states is $\Delta
m^2=\left| m_2^2-m_1^2\right|>0$.  
The oscillation probability for 
$\nu_\mu \rightarrow \nu_e$ oscillations is then given by:
\begin{equation}
{P}\left( \nu _\mu \rightarrow \nu _e\right) = \sin ^22\theta \;\sin
^2\left( \frac{1.27\;\Delta m^2\left( {\rm eV}^2\right) \,L\left({\rm km}%
\right) }{E \left({\rm GeV}\right) }\right)~,
  \label{prob}
\end{equation}
where $L$ is the distance from the source, and $E$ is the
neutrino energy.

From Eq.~\ref{prob}, one can see that 
the probability for observing  oscillations 
is large when $\Delta m^2 \sim E/L$.  In the discussions below, we
will focus on experiments with signals in the $\Delta m^2 \sim 1$~eV$^2$ range. These experiments are therefore designed with $E/L \sim
1$~GeV/km (or, alternatively, 1~MeV/m).   Typically, neutrino source energies range from a few MeV to a few GeV.  Thus, most of the experiments considered are located between a few
meters and a few kilometers from the source. This is not absolutely necessary--a very high energy experiment with a very long 
baseline is sensitive to oscillations in the   $\Delta m^2 \sim 1$~eV$^2$ range, as long as the ratio $E/L \sim 1$~GeV/km is maintained. In other words, ``short baseline experiments'' is something of a
misnomer -- what is meant is experiments with sensitivity to
$\Delta m^2 \sim 1$~eV$^2$ oscillations.  

In the case where $E/L\ll1~GeV/km $, such as in accelerator based experiments with long baselines 
(hundreds of kilometers),  one can see from Eq.~\ref{prob} that the 
oscillations will be rapid.    In the case of  
$\Delta m^2 \sim 1$~eV$^2$, sensitivity to the mass  splitting is
lost because the 
$\sin^2(1.27 \Delta m^2(L/E))$ term will average to 1/2 due to the finite
energy and position resolution of the experiment.   The
oscillation probability becomes $P=(\sin^2 2\theta)/2$ in this case. Thus, the information from
``long baseline experiments'' can be used to constrain the mixing angle,
but not the $\Delta m^2$.

The exercise of generalizing to a three-neutrino model is useful, 
since the inclusion of more states follows from this procedure. 
Within a three-neutrino model, the  
mixing matrix is written as:
\[
\left( 
\begin{array}{l}
\nu _e \\ 
\nu _\mu \\ 
\nu _\tau
\end{array}
\right) =\left( 
\begin{array}{lll}
U_{e1} & U_{e2} & U_{e3} \\ 
U_{\mu 1} & U_{\mu 2} & U_{\mu 3} \\ 
U_{\tau 1} & U_{\tau 2} & U_{\tau 3}
\end{array}
\right) \left( 
\begin{array}{l}
\nu _1 \\ 
\nu _2 \\ 
\nu _3
\end{array}
\right)~.
\]
The matrix elements are parametrized by three
mixing angles, analogous to the Euler angles.  As in the quark sector, the three-neutrino model can be extended to
include an imaginary term that introduces a CP-violating phase. This formalism is analogous to the quark sector, where strong 
and weak eigenstates are rotated and the resultant mixing is described 
conventionally by a unitary mixing matrix.  

The oscillation
probability for three-neutrino oscillations is typically written as:
\begin{eqnarray}
{ P}\left( \nu _\alpha \rightarrow \nu _\beta \right)=\delta_{\alpha \beta }- 4\sum\limits_{j>\,i}U_{\alpha i}U_{\beta i}^*U^*_{\alpha
j}U_{\beta j}\sin ^2\left( \frac{1.27\;\Delta m_{ij}^2 \,L }{E }%
\right)~,  \label{3-gen osc}
\end{eqnarray}
where $\Delta m_{i\,j}^2=m_j^2-m_i^2$, $\alpha$ and $\beta$ are
flavor-state indices $(e, \mu, \tau)$, and $i$ and $j$ are mass-state
indices ($1,2,3$ in the three-neutrino case, though Eq.~\ref{3-gen osc} holds for $n$-neutrino oscillations).    
Although in general there will be mixing among all three flavors
of neutrinos, if 
the mass scales are quite different ($m_3 \gg m_2  \gg m_1$), then the 
oscillation phenomena tend to decouple and the two neutrino 
mixing model is a good approximation in limited regions.

Three different $\Delta m^2$ parameters appear in Eq.~\ref{3-gen osc}; however, only two
are independent since the two small $\Delta m^2$ parameters must
sum to the largest.    If we consider the oscillation data measured at
$>5\sigma$~\cite{SuperKsolar,SNO,Kamland,SuperKatmos, SoudanII, k2kosc,MinosCC1,MinosCC2,MinosCC3,MINOS4,MinosCC5,minostheta13,t2k,dc1stpub,dayabay,reno}, then two $\Delta m^2$ ranges,  $7\times 10^{-5}$~eV$^2$  (solar) and
$3\times 10^{-3}$~eV$^2$ (atmospheric) are already defined.  These constrain the
third $\Delta m^2$, so that oscillation results at 1~eV$^2$, such as
those discussed in this paper, cannot be
accommodated within a three-neutrino model.

\subsection{\label{sec2}(3+$N$) Oscillation Formalism}

The sterile neutrino oscillation formalism followed in this paper assumes up to three additional neutrino mass eigenstates,
beyond the established three SM neutrino species.    We know, from solar and atmospheric oscillation observations, that three of the mass
states must be mostly active. Experimental hints point toward the existence of additional mass states that are mostly sterile, in the range of $\Delta m^2 = 0.01 - 100$~eV$^2$.     

Introducing extra mass states results in a large number of extra
parameters in the model.
Approximation is required to allow for efficient exploration of the
available parameters.    To this end, in our model we assume that the
three lowest states,  $\nu_1$, $\nu_2$, 
and $\nu_3$,  that are the mostly active states accounting for the solar and
atmospheric observations, have masses so small as to be effectively
degenerate with masses of zero.
This is commonly called the ``short baseline
approximation'' and it reduces the 
fit to two-, three-, and four-neutrino-mass oscillation models,
corresponding to (3+1), (3+2), and 
(3+3), respectively. 

The active (e, $\mu$, $\tau$) content of the $N$ additional mass eigenstates is 
assumed to be small; specifically, the $U_{\alpha i}$ elements of the extended (3+$N$)$\times$(3+$N$) mixing matrix for $i=4-6$ and $\alpha=e,\mu,\tau$, are restricted to values $|U_{\alpha i}|\le0.5$, while the following constraints are applied by way
of unitarity:
\begin{equation}
 \sum_{\alpha=e,\mu,\tau}|U_{\alpha i}|^2 \le 0.3~,
\end{equation}
for each $i=4-6$, and 
\begin{equation}
\sum_{i=4-6}|U_{\alpha i}|^2 \le 0.3~,
\end{equation}
for each $\alpha=e,\mu,\tau$. In our fits, since the SBL experiments considered have no $\nu_{\tau}$ sensitivity, we explicitly assume $|U_{\tau i}|=0$. The above restrictions therefore apply only for $\alpha=e,\mu$, and are consistent with solar and atmospheric neutrino experiments, which indicate that there can only be a small electron and muon flavor content in the fourth, fifth and sixth mass eigenstates~\cite{whitepaper}. 

In this formalism, the probabilities for $\nu_{\alpha}\rightarrow\nu_{\beta}$ oscillations can be deduced from the following equation:
\begin{eqnarray}
\label{eq:genoscprob}
P(\nu_{\alpha}\rightarrow\nu_{\beta})=\delta_{\alpha\beta}-\sum_{j<i}\large(4Re\{U_{\beta i}U^*_{\alpha i}U^*_{\beta j}U_{\alpha j}\}\sin^2(1.27\Delta m^2_{ij}L/E) \nonumber\\
-2Im\{U_{\beta i}U^*_{\alpha i}U^*_{\beta j}U_{\alpha j}\}\sin(2.53\Delta m^2_{ij}L/E)\large)~,
\end{eqnarray}
where $\Delta m^2_{ij}=m^2_i-m^2_j$ is in eV$^2$, $L$ is in m, and $E$
is in MeV. This formalism conserves CPT, but does not necessarily conserve CP.

To be explicit, for the (3+3) scenario, the mixing formalism is extended in
the following way:
\[
\left( 
\begin{array}{l}
\nu _e \\ 
\nu _\mu \\ 
\nu _\tau \\
\nu _{s_1} \\ 
\nu _{s_2} \\ 
\nu _{s_3} 
\end{array}
\right) =\left( 
\begin{array}{llllll}
U_{e1} & U_{e2} & U_{e3} & U_{e4} & U_{e5} & U_{e6} 
\\ 
U_{\mu 1} & U_{\mu 2} & U_{\mu 3} & U_{\mu 4} & U_{\mu 5} & U_{\mu 6} 

\\ 
U_{\tau 1} & U_{\tau 2} & U_{\tau 3} & U_{\tau 4} & U_{\tau 5} & U_{\tau 6}\\
U_{{s_1} 1} & U_{{s_1} 2} & U_{{s_1} 3} & U_{{s_1} 4} & U_{{s_1} 5} & U_{{s_1} 6}\\
U_{{s_2} 1} & U_{{s_2} 2} & U_{{s_2} 3} & U_{{s_2} 4} & U_{{s_2} 5} & U_{{s_3} 6}\\
U_{{s_3} 1} & U_{{s_3} 2} & U_{{s_3} 3} & U_{{s_3} 4} & U_{{s_3} 5} & U_{{s_3} 6}\\
\end{array}
\right) \left( 
\begin{array}{l}
\nu _1 \\ 
\nu _2 \\ 
\nu _3  \\
\nu _4 \\ 
\nu _5 \\ 
\nu _6
\end{array}
\right)
\]
The SBL approximation states that
$\nu_1\approx \nu_2 \approx \nu_3 \equiv0$.  With this assumption and for the case of the (3+3) scenario, the appearance ($\alpha\ne\beta$)
oscillation probability can be re-written as:
\begin{eqnarray}
P(\nu_{\alpha}\rightarrow\nu_{\beta})\simeq 
-4|U_{\alpha5}||U_{\beta5}||U_{\alpha4}||U_{\beta4}|\cos\phi_{54}\sin^2(1.27\Delta m^2_{54}L/E) \nonumber\\
-4|U_{\alpha6}||U_{\beta6}||U_{\alpha4}||U_{\beta4}|\cos\phi_{64}\sin^2(1.27\Delta m^2_{64}L/E) \nonumber\\
-4|U_{\alpha5}||U_{\beta5}||U_{\alpha6}||U_{\beta6}|\cos\phi_{65}\sin^2(1.27\Delta m^2_{65}L/E) \nonumber\\
+4(|U_{\alpha4}||U_{\beta4}|+|U_{\alpha5}||U_{\beta5}|\cos\phi_{54}+|U_{\alpha6}||U_{\beta6}|\cos\phi_{64})|U_{\alpha4}||U_{\beta4}|\sin^2(1.27\Delta m^2_{41}L/E) \nonumber\\
+4(|U_{\alpha4}||U_{\beta4}|\cos\phi_{54}+|U_{\alpha5}||U_{\beta5}|+|U_{\alpha6}||U_{\beta6}|\cos\phi_{65})|U_{\alpha5}||U_{\beta5}|\sin^2(1.27\Delta m^2_{51}L/E) \nonumber\\
+4(|U_{\alpha4}||U_{\beta4}|\cos\phi_{64}+|U_{\alpha5}||U_{\beta5}|\cos\phi_{65}+|U_{\alpha6}||U_{\beta6}|)|U_{\alpha6}||U_{\beta6}|\sin^2(1.27\Delta m^2_{61}L/E) \nonumber\\
+2|U_{\beta5}||U_{\alpha5}||U_{\beta4}||U_{\alpha4}|\sin\phi_{54}\sin(2.53\Delta m^2_{54}L/E) \nonumber\\
+2|U_{\beta6}||U_{\alpha6}||U_{\beta4}||U_{\alpha4}|\sin\phi_{64}\sin(2.53\Delta m^2_{64}L/E) \nonumber\\
+2|U_{\beta6}||U_{\alpha6}||U_{\beta5}||U_{\alpha5}|\sin\phi_{65}\sin(2.53\Delta m^2_{65}L/E) \nonumber\\
+2(|U_{\alpha5}||U_{\beta5}|\sin\phi_{54}+|U_{\alpha6}||U_{\beta6}|\sin\phi_{64})|U_{\alpha4}||U_{\beta4}|\sin(2.53\Delta m^2_{41}L/E) \nonumber\\
+2(-|U_{\alpha4}||U_{\beta4}|\sin\phi_{54}+|U_{\alpha6}||U_{\beta6}|\sin\phi_{65})|U_{\alpha5}||U_{\beta5}|\sin(2.53\Delta m^2_{51}L/E) \nonumber\\
+2(-|U_{\alpha4}||U_{\beta4}|\sin\phi_{64}-|U_{\alpha5}||U_{\beta5}|\sin\phi_{65})|U_{\alpha6}||U_{\beta6}|\sin(2.53\Delta
m^2_{61}L/E)~.  \label{3app}
\end{eqnarray}
CP violation appears in Eq.~\ref{3app} in the form of the three phases, defined by
\begin{eqnarray}
\label{cpvphase}
\phi_{54}=\mathrm{arg}(U_{e5}U_{\mu5}^*U_{e4}^*U_{\mu4})~, \\
\label{cpvphase2}
\phi_{64}=\mathrm{arg}(U_{e6}U_{\mu6}^*U_{e4}^*U_{\mu4})~,
\end{eqnarray}
and
\begin{eqnarray}
\label{cpvphase2}
\phi_{65}=\mathrm{arg}(U_{e6}U_{\mu6}^*U_{e5}^*U_{\mu5})~.
\end{eqnarray}
In each case, $\nu \rightarrow \bar \nu$ implies $\phi \rightarrow -\phi$.
In the case of disappearance  ($\alpha\equiv\beta$), the survival probability can be re-written as:
\begin{eqnarray}
P(\nu_{\alpha}\rightarrow\nu_{\alpha})\simeq
1-4|U_{\alpha4}|^2|U_{\alpha5}|^2\sin^2(1.27\Delta m^2_{54}L/E)\nonumber \\
-4|U_{\alpha4}|^2|U_{\alpha6}|^2\sin^2(1.27\Delta m^2_{64}L/E)-4|U_{\alpha5}|^2|U_{\alpha6}|^2\sin^2(1.27\Delta m^2_{65}L/E) \nonumber \\
-4(1-|U_{\alpha4}|^2-|U_{\alpha5}|^2-|U_{\alpha6}|^2)(|U_{\alpha4}|^2\sin^2(1.27\Delta m^2_{41}L/E) \nonumber \\
+|U_{\alpha5}|^2\sin^2(1.27\Delta
m^2_{51})+|U_{\alpha6}|^2\sin^2(1.27\Delta m^2_{61}L/E))~.  \label{disappeq}
\end{eqnarray}
This formula has no $\phi_{ij}$ dependencies because CP violation only
affects appearance.

We have discussed the formulas for (3+1) and (3+2) oscillations that
arise from Eq.~\ref{eq:genoscprob} in previous papers~\cite{sorel,karagiorgi,viability}. 
To reduce to a (3+2) model, the parameters $\Delta m^2_{61}$, $|U_{e6}|$,
$|U_{\mu6}|$, $\phi_{64}$ and $\phi_{65}$ are explicitly set to zero, in which case we have the following appearance and disappearance formulas for a (3+2) model:
\begin{eqnarray}
P(\nu_{\alpha}\rightarrow\nu_{\beta})\simeq 
-4|U_{\alpha5}||U_{\beta5}||U_{\alpha4}||U_{\beta4}|\cos\phi_{54}\sin^2(1.27\Delta m^2_{54}L/E) \nonumber\\
+4(|U_{\alpha4}||U_{\beta4}|+|U_{\alpha5}||U_{\beta5}|\cos\phi_{54})|U_{\alpha4}||U_{\beta4}|\sin^2(1.27\Delta m^2_{41}L/E) \nonumber\\
+4(|U_{\alpha4}||U_{\beta4}|\cos\phi_{54}+|U_{\alpha5}||U_{\beta5}|)|U_{\alpha5}||U_{\beta5}|\sin^2(1.27\Delta m^2_{51}L/E) \nonumber\\
+2|U_{\beta5}||U_{\alpha5}||U_{\beta4}||U_{\alpha4}|\sin\phi_{54}\sin(2.53\Delta m^2_{54}L/E) \nonumber\\
+2(|U_{\alpha5}||U_{\beta5}|\sin\phi_{54})|U_{\alpha4}||U_{\beta4}|\sin(2.53\Delta m^2_{41}L/E) \nonumber\\
+2(-|U_{\alpha4}||U_{\beta4}|\sin\phi_{54})|U_{\alpha5}||U_{\beta5}|\sin(2.53\Delta m^2_{51}L/E)~, \label{2app}
\end{eqnarray}
and
\begin{eqnarray}
P(\nu_{\alpha}\rightarrow\nu_{\alpha})\simeq
1-4|U_{\alpha4}|^2|U_{\alpha5}|^2\sin^2(1.27\Delta m^2_{54}L/E)\nonumber \\
-4(1-|U_{\alpha4}|^2-|U_{\alpha5}|^2)(|U_{\alpha4}|^2\sin^2(1.27\Delta m^2_{41}L/E) \nonumber \\
+|U_{\alpha5}|^2\sin^2(1.27\Delta m^2_{51}L/E))~.  \label{disappeq2}
\end{eqnarray}

 For a (3+1) model, $\Delta m^2_{61}$, $\Delta m^2_{51}$, $|U_{e6}|$, $|U_{\mu6}|$, $|U_{e5}|$,
 $|U_{\mu5}|$, $\phi_{64}$, $\phi_{65}$ and $\phi_{54}$ should be set to zero.  
This further simplifies the 
oscillation  probabilities, and one recovers 
the familiar two-neutrino appearance and disappearance probabilities. 
The appearance and disappearance formulas for a (3+1) model are then given by:
\begin{eqnarray}
P(\nu_{\alpha}\rightarrow\nu_{\beta})\simeq  4|U_{\alpha4}|^2 |U_{\beta4}|^2\sin^2(1.27\Delta m^2_{41}L/E)~, \label{1app}
\end{eqnarray}
and
\begin{eqnarray}
P(\nu_{\alpha}\rightarrow\nu_{\alpha})\simeq
1-4(1-|U_{\alpha4}|^2)|U_{\alpha4}|^2\sin^2(1.27\Delta m^2_{41}L/E)~. \label{disappeq2}
\end{eqnarray}

In principle,
the probability for neutrino oscillation is modified in the presence
of matter.   ``Matter effects'' arise because the electron neutrino flavor experiences
both Charged Current (CC) and Neutral Current (NC) elastic forward-scattering with electrons as it propagates through matter, while
the $\nu_\mu$ and $\nu_\tau$ experience only NC
forward-scattering. The sterile component
experiences no forward-scattering.   In practice, SM-inspired matter effects are very small
given the short baselines of the experiments, and so we
do not consider them further here.   Beyond-SM
matter effects are beyond the scope of this article, but are considered in Ref.~\cite{bsmmatter}.

\section{Experimental Data Sets} \label{sec:experiments}

This section provides an overview of the various types of past and existing
neutrino sources and detectors used in SBL experiments.
After introducing the experimental concepts, the specific experimental
data sets used in this analysis are discussed.  

The data fall
into two overall types: disappearance, where the active flavor is
assumed to have oscillated into a sterile neutrino and/or another flavor which is kinematically not allowed to interact or leaves no detectable signature; and appearance,
where the transition is between active flavors, but with mass
splittings corresponding to the mostly-sterile states.
Appearance and disappearance are natural divisions for testing the
compatibility of data sets,
as can be seen clearly from Eqs.~\ref{1app} and~\ref{disappeq2}. If 
$|U_{\alpha4}|^2$ and  $|U_{\beta4}|^2$ are shown to be small, then 
the effective mixing angle for appearance, $4|U_{\alpha4}|^2
|U_{\beta4}|^2$, cannot be large.   This constraint that the 
disappearance experiments place on appearance experiments 
extends to (3+2) and (3+3) models also.  

CPT conservation, which is assumed in the analysis, demands that neutrino and antineutrino disappearance probabilities are the same.
To test this, we divide the data into
antineutrino and neutrino sets and fit each set separately. If CP violation is already allowed in the oscillation formalism,
then any incompatibility found between respective neutrino and antineutrino fits could imply effective CPT violation,
as discussed in Ref.~\cite{viability}.

\begin{figure}[h!]
\begin{center}
\includegraphics[scale=.6]{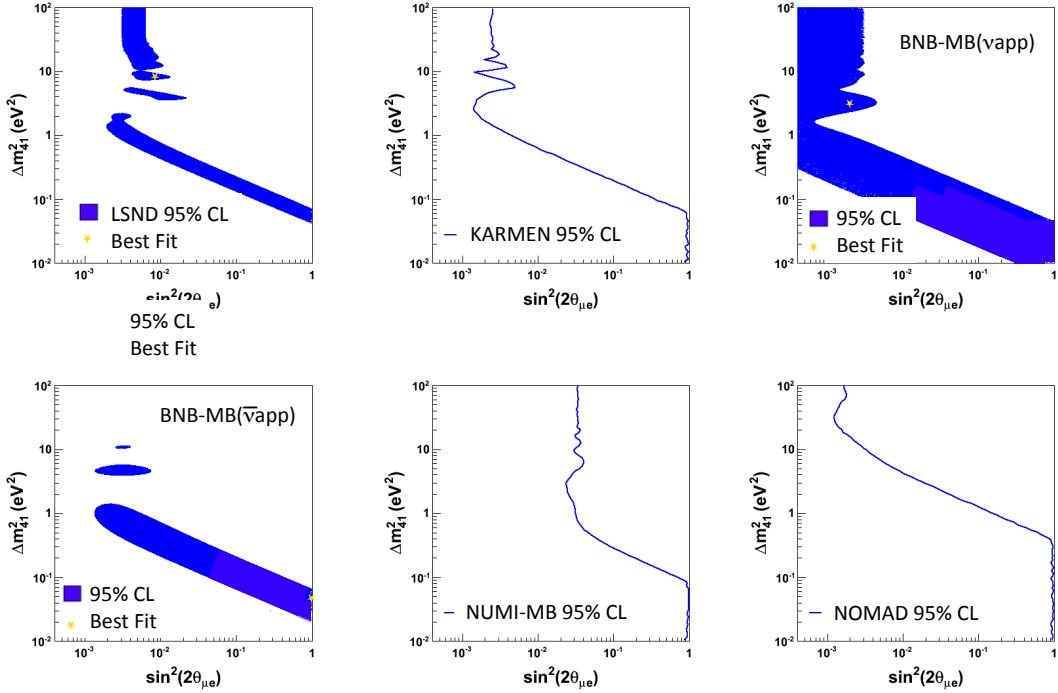}
\end{center}
\vspace{-2.5cm}
\caption{Summary of $\bar \nu_\mu \rightarrow \bar \nu_e$ and $
  \nu_\mu \rightarrow \nu_e$ results, shown at 95\% CL.  Top row:
  LSND, KARMEN, BNB-MB($\nu$app);  Bottom row: BNB-MB($\bar{\nu}$app),
NuMI-MB($\nu$app), NOMAD. See Sec.~\ref{fitexps} for details and references. 
\label{expts1}}
\end{figure}

\begin{figure}[h!]
\begin{center}
\includegraphics[scale=.6]{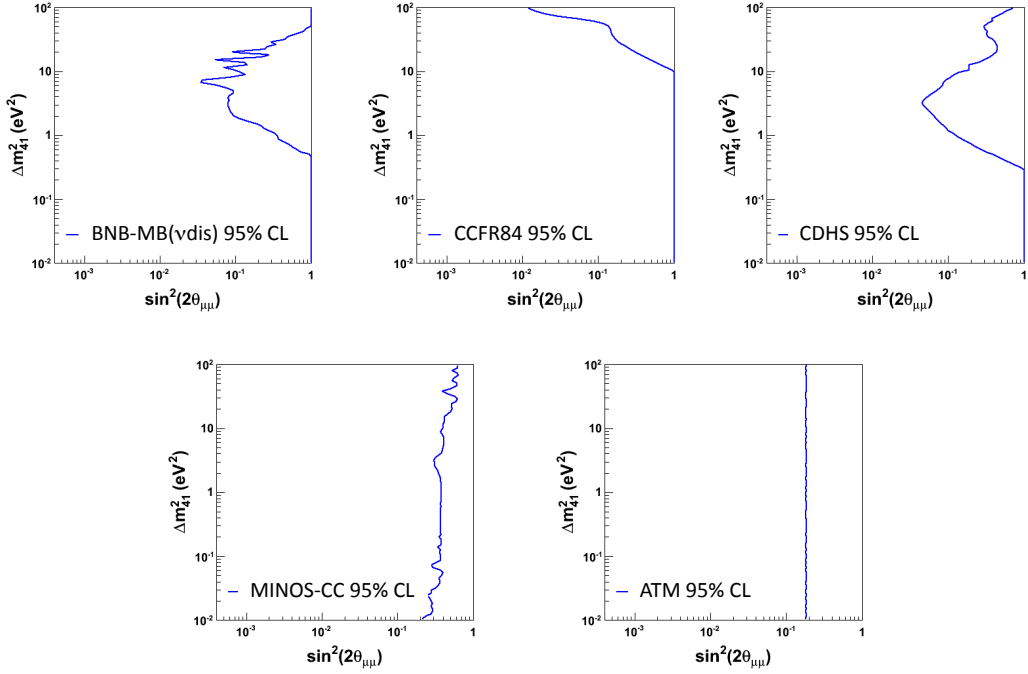}
\end{center}
\vspace{-2.5cm}
\caption{Summary of $\bar \nu_\mu \rightarrow \bar \nu_\mu$ and $
  \nu_\mu \rightarrow \nu_\mu$ results, shown at 95\% CL.  Top row:
BNB-MB($\nu$dis), CCFR84,  CDHS; Bottom row: MINOS-CC, ATM. See Sec.~\ref{fitexps} for details and references. 
\label{expts2}}
\end{figure}

\begin{figure}[h!]
\begin{center}
\includegraphics[scale=.6]{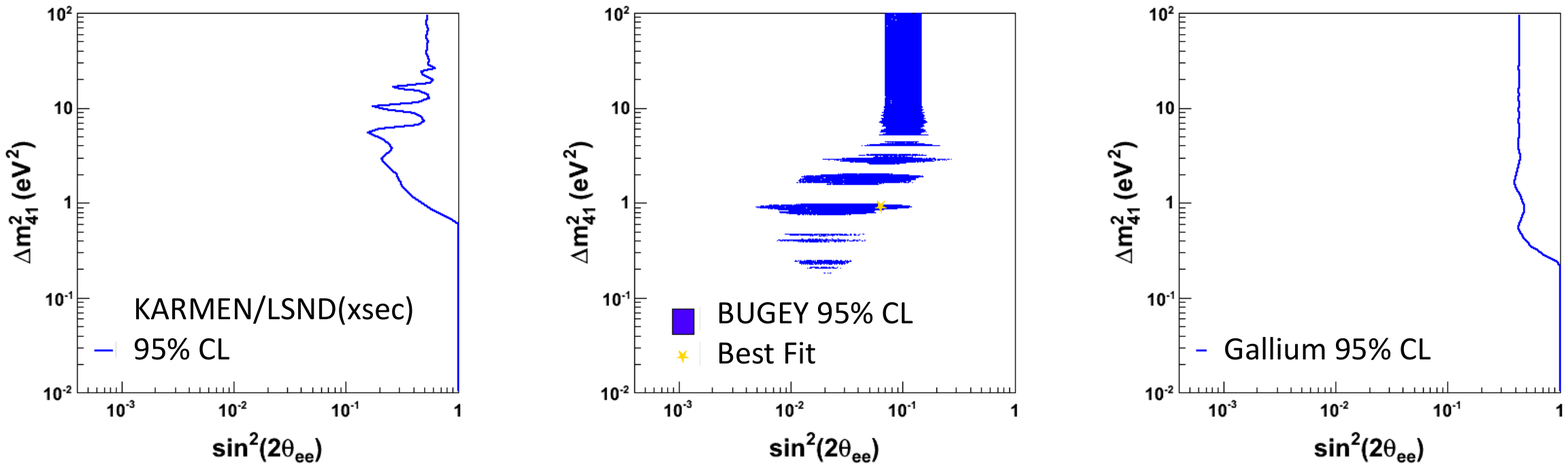}
\end{center}
\vspace{-7.cm}
\caption{Summary of $\bar \nu_e\rightarrow \bar \nu_e$ and $
  \nu_e \rightarrow \nu_e$ results, shown at 95\% CL.   From left:
  KARMEN/LSND(xsec),  Bugey, and Gallium. See Sec.~\ref{fitexps} for details and references. 
\label{expts3}}
\end{figure}

Figures~\ref{expts1}, \ref{expts2}, and \ref{expts3} provide summaries of the data sets, showing the
constraints they provide in a simple two-neutrino oscillation model,
which is functionally equivalent to the (3+1) scenario.  
Figure~\ref{expts1} shows the muon-to-electron flavor data sets in
neutrino and antineutrino mode at 95\% confidence level (CL).  Figures~\ref{expts2} and~\ref{expts3} show results for $\nu_{\mu}$ and $\bar \nu_{\mu}$, and $\nu_{e}$ and $\bar \nu_{e}$ disappearance, respectively.

\subsection{Sources and Detectors Used in Short Baseline Neutrino Experiments}

Before considering the data sets in detail, we provide an overview
of how SBL experiments are typically designed.   

\subsubsection{Sources of Neutrinos for Short Baseline Experiments}
\label{sub:sources}

The neutrino sources used in SBL experiments range
in energy from a few~MeV to hundreds of GeV and include man-made radioactive sources, reactors, and accelerator-produced beams. While the higher energy accelerator sources are mixtures of different neutrino flavors, the $<10$~MeV sources rely on beta decay and are thus pure electron neutrino flavor.

At the low energy end of the spectrum, the rate of electron neutrino interactions from
the beta decay of the $\sim$1~MCi sources $^{51}$Cr (half-life: 28
days) and $^{37}$Ar (half-life: 35 days) have been studied. These
sources were originally produced for the low energy ($\sim$1~MeV)
calibration of solar neutrino detectors~\cite{SAGE3,GALLEX3} but have proven themselves interesting as a probe of electron neutrino disappearance. 

Moving up in energy by a few~MeV, nuclear reactors are powerful sources of $\sim$2-8~MeV $\bar \nu_e$ through the
$\beta^+$-decaying elements produced primarily in the decay chains of 
$^{235}$U, $^{239}$Pu, $^{238}$U and $^{241}$Pu. While these four
isotopes are the progenitors of most of the reactor flux, modern reactor simulations include all fission sources~\cite{Christhesis}. Reactor simulations convolute predictions of fission rates over time with neutrino production per fission.
Recently, a reanalysis of the production cross section per
fission~\cite{whitepaper, Mueller, huber} has led to an increase in the predicted reactor flux. As their energy is too low for an appearance search (the neutrino energy is below the muon production kinematic threshold), reactor source neutrinos can only be used for $\bar \nu_e$ disappearance searches, where 
the neutrinos are detected using CC interactions with an outgoing $e^+$. 

The lowest neutrino energy (up to 53~MeV) accelerator sources used in existing SBL experiments are based on pion- and muon-decay-at-rest (DAR). The neutrino flux
comes from the stopped pion decay chain:
$\pi^{+}\rightarrow \mu^{+} \nu_{\mu} $ and $\mu^+ \rightarrow
e^{+}\bar{\nu}_{\mu}\nu_{e}$. Pions are produced in interactions of accelerator protons with a,  typically, graphite or water target. 
The contribution from the decay chain
$\pi^- \rightarrow \mu^- \bar \nu_\mu$ is suppressed by 
designing the target such that the $\pi^-$ mesons are captured with high probability.
The result is a source which has a well understood neutrino flavor
content and energy distribution, with a minimal ($< 10^{-3}$) $\bar \nu_e$ content~\cite{Burman:1996vx,Burman:1989dq}. 
This last point is important as $\bar \nu_\mu \rightarrow \bar \nu_e$ is 
the dominant channel used for oscillation searches by DAR sources.

In a conventional high energy (from $\sim$ 100~MeV to hundreds of GeV) accelerator-based neutrino beam, protons impinge on a target (beryllium and carbon are typical) to produce secondary mesons. The boosted mesons enter and subsequently decay inside of a long, often evacuated pipe. Neutrinos are primarily produced by $\pi^+$ and $\pi^-$ decay in flight (DIF). Pion sign selection, via a large magnet placed directly in the beamlines before the decay pipe, allows for nearly pure neutrino or antineutrino running, with only a few
percent ``wrong sign'' neutrino flux content in the case of neutrino running, and $\sim$15\%~\cite{AguilarArevalo:2008yp} in the case of antineutrino running.
These beams are generally produced by protons at 8~GeV and above. At these energies, in addition to pion production, kaon production contributes to the flux 
of both muon and electron neutrino flavor. 
There is often a substantial muon DIF content as well, contributing both $\nu_e$/$\bar{\nu}_e$ and $\bar{\nu}_\mu$/$\nu_{\mu}$ to the beam.
The result of the kaon and muon secondary content is that, while the neutrinos are predominantly muon-flavored, the beam will always have some intrinsic electron-flavor neutrino content, usually at the several percent level. Accelerator-based beams are predominantly used for $\nu_{\mu}\rightarrow\nu_e$ and $\bar{\nu}_{\mu}\rightarrow\bar{\nu}_e$ appearance searches, as well as $\nu_{\mu}$ and
$\bar{\nu}_{\mu}$ disappearance searches. An excellent review of methods in producing accelerator-based neutrino beams can be found in Ref.~\cite{kopp}. 

In contrast to lower-energy neutrino sources (DAR, reactor, and isotope sources), high energy accelerator-based neutrino sources are subject to significant energy dependent neutrino flux uncertainties, often at the level of 10-15\%, due to in-target meson production uncertainties. These uncertainties can affect the energy distribution, flavor content, and absolute normalization of a neutrino beam. Typically, meson production systematics are constrained with dedicated measurements by experiments such as HARP~\cite{harp} and MIPP~\cite{mipp}, which use replicated targets (geometry and material) and a wide range of proton beam energies to study meson production cross sections and kinematics directly. Alternatively, experiments can employ a two-detector design for comparing near-to-far event rate in energy to effectively reduce these systematics. However, due to the short baselines employed for studying sterile neutrino oscillations, a two-detector search is often impractical. {\it In-situ} measurements in single detector experiments can exploit flux (times cross section) correlations among different beam components and energies to reduce flux uncertainties, as has been done in the case of the MiniBooNE $\nu_e$ and $\bar{\nu}_e$ appearance searches described below.

\subsubsection{Short Baseline Neutrino Detectors}

Because low energy neutrino interaction cross sections are very small, the options for SBL detectors are typically limited to designs which can be constructed on a massive scale.
There
are several generic neutrino detection methods in use today: unsegmented scintillator
detectors,  unsegmented Cerenkov detectors, segmented
scintillator-and-iron calorimeters, and segmented
trackers. 

Neutrino oscillation experiments usually require sensitivity to charged current (CC) neutrino interactions, whereby one can definitively identify the flavor of the interacting neutrino by the presence of a charged lepton in the final state. However, in the case of sterile neutrino oscillation searches, neutral current (NC) interactions can also provide useful information, as they are directly sensitive to the sterile flavor content of the neutrino mass eigenstate, $|U_{si}|^2=1-|U_{ei}|^2-|U_{\mu i}|^2-|U_{\tau i}|^2$. 

Unsegmented scintillator detectors are typically used for few-MeV-scale 
SBL experiments, which require efficient electron neutrino identification and reconstruction. These detectors consist of large
tanks of oil-based (C$_n$H$_{2n}$) liquid scintillator surrounded by phototubes. The free protons in the oil provide a
target for the inverse beta decay interaction, $\bar \nu_e p \rightarrow e^+ n$.  The reaction threshold for this interaction is 1.8~MeV due to the mass difference between the proton and neutron and the mass of the positron.  The scintillation
light from the $e^+$, as well as light from the Compton scattering of
the 0.511~MeV annihilation photons provides an initial (``prompt'')
signal.  This is followed by $n$ capture on hydrogen and a
2.2~MeV flash of light, as the resulting $\gamma$ Compton-scatters in the scintillator. This coincidence sequence in time (positron followed by neutron capture) provides a clean, mostly background-free interaction signature.  Experiments often dope the liquid scintillator using an element with a high neutron capture cross section for improved event identification efficiency.

The CC interaction with the carbon in the oil (which produces either nitrogen or boron depending on whether
the scatterer is a neutrino or antineutrino) has a significantly
higher energy threshold than the free proton target scattering process. The CC
quasielastic interaction $\nu_e + C
\rightarrow e^- + N$ has an energy threshold of 13.4~MeV, which
arises from the carbon-nitrogen mass difference and the mass of the
electron.  In the case of both reactor and radioactive decay sources, the flux
cuts off below this energy threshold.  However, neutrinos from DAR sources
are at sufficiently high energy to produce these carbon scatters.

Unsegmented Cerenkov detectors  make use of a target which is a large volume of clear
medium (undoped oil or water are typical) surrounded by or
interspersed with phototubes.  Undoped oil has a larger refractive
index, leading to a larger Cerenkov opening angle.
Water is the only affordable medium once the detector size surpasses a
few kilotons.  In this paper, the only unsegmented Cerenkov detector
that is considered is the 450~ton oil-based MiniBooNE detector.  In such a detector, 
a track will project a ring with a sharp
inner and outer edge onto the phototubes.  Consider an electron
produced in a $\nu_e$ CC quasielastic interaction. As the
electron is low mass, it will multiple scatter and easily
bremsstrahlung, smearing the light projected on the tubes and
producing a ``fuzzy'' ring.  A muon produced by a CC quasielastic
$\nu_\mu$ interaction ($\nu_\mu n \rightarrow \mu^- p$) is heavier and will thus produce a sharper outer edge to the ring.  For the same visible energy, the track will also extend farther, filling the interior of the ring, and perhaps exit
the tank.   If the muon stops within the tank
and subsequently decays, the resulting electron provides an
added tag for particle identification.  In the case of the $\mu^-$,
18\% will capture in water, and thus have no electron tag,
while only 8\% will capture in the oil.

Scintillator and iron calorimeters provide an affordable detection technique for
$\sim$1~GeV and higher $\nu_\mu$ interactions.    At these energies,
multiple hadrons may be produced at the interaction vertex and will be
observed as hadronic showers.
In these devices, the iron provides the target, while
the scintillator provides information on energy deposition per unit
length.  This information allows separation between the hadronic shower, which
occurs in both NC and CC events, and the minimum ionizing track of an
outgoing muon, which occurs in CC events.  Transverse information can
be obtained if segmented scintillator strips are used, or if drift
chambers are interspersed.  The light from scintillator strips is
transported to tubes by wavelength-shifting fibers.
Information in the transverse plane improves separation of electromagnetic and
hadronic showers.  The iron can be magnetized to allow separation of
neutrino and antineutrino events based on the charge of the outgoing
lepton.

To address the problem of running at $\sim1$~GeV, where hadron track reconstruction is desirable, highly segmented
tracking designs have been developed.  The best resolution comes from
stacks of wire chambers,
where the material enclosing the gas provides the target.  However, a
more practical alternative has been stacks of thin, extruded
scintillator bars that are read out using wavelength shifting fibers.

\subsection{Data Used in The Sterile Neutrino Fits \label{fitexps}}

There are many SBL data sets that can be included in this analysis.
In this work, we have substantially expanded the number of data sets used beyond those in our past papers~\cite{sorel,karagiorgi,viability}. We identify and discuss new and updated data sets, as well as provide information on those used in past fits, below. The fit technique is described in Sec.~\ref{mthd}.

\subsubsection{Experimental Results from Decay at Rest Studies \label{DAR}}

In past sterile neutrino studies~\cite{sorel,karagiorgi,viability}, we have included the LSND and KARMEN appearance results described below.   Since that work, a new study that  constrains $\bar \nu_e$ 
disappearance from the relative LSND-to-KARMEN cross section measurements
was published~\cite{ConradShaevitz}. This new data set is included in this analysis.

~~\\
\noindent{\it LSND Appearance:}\\

LSND was a DAR experiment that ran in the
1990's, searching for $\bar \nu_\mu \rightarrow \bar \nu_e$.   The beam was produced using  800~MeV protons on target from the LAMPF accelerator at Los Alamos National
Laboratory, where a 1~mA
beam of protons impinged on a water target.  The center of the
8.75~m long, nearly cylindrical detector was located at 29.8~m from
the target, at an angle of 12$^\circ$ from the proton beam direction.
This was an unsegmented detector with a fiducial mass of 167~tons of
oil (CH$_2$), lightly doped with b-PBD scintillator. The intrinsic $\bar \nu_e$ content of the beam was $8\times10^{-4}$ of the $\bar \nu_\mu$ content. The
experiment observed a $\bar \nu_e$ excess of $87.9 \pm 22.4 \pm
6.0$ events above background, which was interpreted as oscillations
with a probability of $(0.264 \pm 0.067 \pm 0.045)\%$.  
Details are available in Ref.~\cite{LSND}.

This data set is referred to as {\bf LSND} in the analysis below and 
indicates a signal at 95\% CL shown in Fig.~\ref{expts1}.
This data covers energies between 20 and 53~MeV and contributes 5 energy bins to the global fit. Statistical errors
are taken into account by using a log-likelihood $\chi^2$ definition in the fit, while systematic errors on the background prediction are not included because these are small relative to the statistical error. Energy and baseline smearing are taken into account by averaging the oscillation probability over the energy bin width and over the neutrino flight path uncertainty.

~~\\
\noindent{\it KARMEN Appearance:}\\

KARMEN was another DAR experiment searching for $\bar \nu_\mu
\rightarrow \bar \nu_e$. KARMEN ran at the ISIS facility at Rutherford Laboratory, with
200 $\mu$A of protons impinging on a copper, tantalum, or uranium
target.    The neutrino detector was located at 
an angle of 100$^\circ$ with respect to the targeting protons to reduce
background from $\pi^-$ DIF. The resulting intrinsic $\bar \nu_e$ content was $6.4\times10^{-4}$ of the $\bar \nu_\mu$ content. 

The center of the approximately cubic, segmented scintillator detector was located at 17.7~m.   Thus, this detector was 60\% of the distance from the source
compared to LSND. The liquid scintillator target volume was 56 m$^3$ and consisted of 512 optically independent modules (17.4~cm $\times$ 17.8~cm $\times$ 353~cm) wrapped
in gadolinium-doped paper.  KARMEN saw no signal and set a limit on appearance.
More details are available in
Ref.~\cite{Karmen}.

This data set is referred to as {\bf KARMEN} in the analysis below and
indicates a limit at 95\% CL, as shown in Fig.~\ref{expts1}.
 This data sets contributes 9 energy bins, in the range 16 to 50~MeV.
As in the case of LSND, statistical errors
are taken into account by using a log-likelihood $\chi^2$ definition in the fit,
while systematic errors on the background prediction are not included.
Energy and baseline smearing are taken into account by averaging the $\sin^2(1.27\Delta m^2 L/E)$
and $\sin(2.53\Delta m^2 L/E)$ term contributions in the total signal prediction over energy bin widths.   The limit which is shown here is determined using a $\Delta
\chi^2$-based raster scan, as discussed in Sec \ref{mthd}

\clearpage
~~\\
\noindent{\it LSND and KARMEN Cross Section Measurements:}\\

Along with the oscillation searches, LSND and KARMEN measured  $\nu_e +^{12}{\rm C} \rightarrow ^{12}{\rm
  N}_{gs} + e^-$ scattering.  In this two-body interaction, with a
$Q$-value of 17.3~MeV, the neutrino energy can be reconstructed by
measuring the outgoing visible energy of the electron.  The $^{12}$N
ground state is identified by the subsequent $\beta$ decay,
$^{12}{\rm N}_{gs} \rightarrow ^{12}{\rm C} + e^+ + \nu_e$, which has
a $Q$-value of 16.3~MeV and a lifetime of 15.9~ms.  

The cross section is measured by both experiments under the
assumption that the $\nu_e$ flux has not oscillated, leading
to disappearance.  The
excellent agreement between the two results, as a function of energy,
allows a limit to be placed on $\nu_e$ oscillations.   The energy
dependence of the cross section, as well as the normalization, are well
predicted and both constraints are used in the analysis~\cite{ConradShaevitz}.

This data set is referred to as {\bf KARMEN/LSND(xsec)}  in the analysis
below,  and indicates a limit at 95\% CL as shown in
Fig.~\ref{expts3}. A total of
six (for KARMEN) plus five (for LSND) bins are used in the fit, which extend approximately from 28-50~MeV in the case of KARMEN and from 38-50~MeV in the case of LSND.   In
calculating the oscillation probability, the signal is averaged across the
lengths of the detectors. 
The experiments have
correlated systematics arising from the flux normalization due to a 
a shared underlying analysis for pion production in DAR experiments. 
This is addressed through application of pull-terms as described in
Ref.~\cite{ConradShaevitz}.

\subsubsection{The MiniBooNE Experimental Results \label{MBres}}

The MiniBooNE experiment provides multiple results from a single
detector.  This oil-based 450~t fiducial volume Cerenkov
detector was exposed to two conventional beams, the Booster Neutrino Beam (BNB) and the off-axis NuMI beam.   
The primary goal of MiniBooNE was to search for
$\nu_\mu \rightarrow \nu_e$ and $\bar \nu_\mu \rightarrow\bar \nu_e$ appearance, using the BNB, which
provides sensitivity to
$\Delta m^2\sim0.1-10$~eV$^2$ oscillations. The NuMI beam also provides some sensitivity to 
$\nu_\mu\rightarrow\nu_e$ appearance at a similar $\Delta m^2$.
In addition to the appearance searches,
MiniBooNE also looked for $\nu_{\mu}$ and $\bar{\nu}_{\mu}$ disappearance using the BNB. 


The MiniBooNE data sets included in our analysis have increased throughout the period that our group has been performing fits. Ref. \cite{karagiorgi} used a Monte Carlo prediction for neutrinos and antineutrinos to estimate MiniBooNE's sensitivity to sterile neutrinos. The full BNB neutrino and first published BNB antineutrino data sets from MiniBooNE form the experimental constraints in Ref.~\cite{viability}. Here, we have updated to include the full BNB  antineutrino data sets. A further update has been to employ a log-likelihood method for the BNB neutrino and antineutrino data sets from \cite{mbprlinprogress}, as this was recently adopted by the MiniBooNE collaboration~\cite{MBlog}. We also use the updated constraints on electron neutrino flux from kaons \cite{mbprlinprogress}. A partial data set from NuMI data taking was presented in Ref.~\cite{viability} and has not been updated, as the result was already systematics limited. In this analysis we also introduce the MiniBooNE disappearance search \cite{Kendall}.

In our fits to MiniBooNE appearance data, when drawing allowed regions and calculating compatibilities, which make use of $\Delta \chi^2$'s and not absolute  $\chi^2$'s, we use MiniBooNE's log-likelihood $\chi^2$ definition, summing over both $\nu_e$ and $\nu_{\mu}$ bins, as described in \cite{MBlog}. For consistency, the absolute MiniBooNE BNB $\nu_e$ and $\bar\nu_e$ appearance $\chi^2$ values quoted in our paper also correspond to the same definition, i.e. fitting to both $\nu_e$ and $\nu_{\mu}$ spectra; therefore, they differ from the ones published by MiniBooNE in \cite{mbprlinprogress}, which are obtained by fitting only to {\it a priori} constrained $\nu_e$ distributions. Note that the two definitions should yield consistent allowed regions and compatibility results.

~~\\
\noindent{\it The Booster Beam Appearance Search in Neutrino Running Mode:}\\

The BNB flux composition in neutrino mode consists of $>90\%$ $\nu_\mu$, 6\% $\bar{\nu}_{\mu}$, and 0.06\% $\nu_e$ and $\bar{\nu}_e$ combined  \cite{AguilarArevalo:2008yp}. In the MiniBooNE BNB search for $\nu_e$ appearance, the $\nu_e$ and $\bar{\nu}_e$ signal was
normalized to the $\nu_{\mu}$ and $\bar{\nu}_{\mu}$ CC quasielastic events observed in the detector, which
peaked at 700~MeV.

The global fits presented here use the full statistics of the 
MiniBooNE $\nu_\mu\rightarrow\nu_e$ data set, representing $6.46\times 10^{20}$ protons on target. 
In this data set, MiniBooNE has observed an excess of events at $200-1250$~MeV, corresponding to $161.9\pm49.0$ electron-like events \cite{mbprlinprogress}. The data set is  referred to as {\bf BNB-MB($\nu$app)} in
the analysis below.  

We include the BNB-MB($\nu$app) data set in our fits in the form of the full $\nu_e$ CC reconstructed energy distribution, in 11 bins in energy from 200 to 3000~MeV, fit simultaneously
with the full $\nu_{\mu}$ CC energy distribution, in 8 bins in energy up to 1900~MeV. We account for statistical and systematic uncertainties in each sample, as well as systematic correlations (from flux and cross section) among the $\nu_e$ signal and background and $\nu_{\mu}$ background
distributions. The systematic correlations are provided in the form of a full 19-bin$\times$19-bin fractional covariance matrix. By fitting the $\nu_e$ and $\nu_\mu$ spectra simultaneously, we are able to exploit the high-statistics $\nu_{\mu}$ CC sample as a constraint on background and signal event rates, by assuming no significant $\nu_{\mu}$ disappearance. For further information, see Ref.~\cite{MB}.

The data set results in a signal at 95\% CL, as shown in
Fig.~\ref{expts1}. This has changed slightly from our past analysis~\cite{viability} now that we are using updated constraints on intrinsic electron neutrinos from kaons and the log-likelihood method, but is in agreement with the equivalent analysis from the MiniBooNE Collaboration \cite{mbprlinprogress}.  


~~\\

\noindent{\it The Booster Beam Appearance Search in Antineutrino Running Mode:}\\

The BNB flux composition in antineutrino mode consists of $83\%$ $\bar{\nu}_\mu$, 0.6\% $\nu_e$ combined and $\bar{\nu}_e$, and a significantly larger wrong-sign composition than in neutrino mode, of 16\% $\nu_{\mu}$. As in the BNB $\nu_e$ appearance search, the electron flavor signal was normalized to the muon-flavor CC quasielastic events observed in the detector, which peaked at 500~MeV.

The global fits presented here use the full statistics of the
MiniBooNE $\bar \nu_\mu$
data set, representing $11.27\times10^{20}$ protons on target.
In this data set, MiniBooNE has observed an excess of events at $200-1250$~MeV, corresponding to $78.4\pm 30.8$ electron-like events.
The  data set is referred to as {\bf BNB-MB($\bar\nu$app)} in
the analysis below.  

As in neutrino mode, we fit the full $\bar{\nu}_e$ CC energy distribution, in 11 bins in energy from 200 to 3000~MeV, simultaneously
with the full $\bar{\nu}_{\mu}$ CC energy distribution, in 8 bins in energy up to 1900~MeV. The wrong-sign contamination in the beam ($\nu_{\mu}$) is assumed to not contribute to any oscillations; only $\bar{\nu}_{\mu}\rightarrow\bar{\nu}_e$ oscillations are assumed for this data set. We account for statistical and systematic uncertainties in each sample, as well as systematic correlations among the $\bar{\nu}_e$ and $\bar{\nu}_{\mu}$ distributions in the form of a full 19-bin$\times$19-bin fractional covariance matrix in each fit. For further information, see Ref.~\cite{MB}.

The data set results in a signal at 95\% CL, as shown in
Fig.~\ref{expts1}.

~~\\

\noindent{\it The NuMI Beam Appearance Search:}\\

The MiniBooNE detector is also exposed to the NuMI neutrino beam, arising from a 120~GeV proton beam
impinging on a carbon target. This beam is nominally used for the MINOS long baseline neutrino oscillation experiment.
NuMI events arrive out-of-time with the BNB-produced events.    
This 200~MeV to 3~GeV neutrino energy source is 
dominated by kaon decays near the NuMI target, which
is 110~mrad off-axis and located 745~m upstream of the MiniBooNE 
detector.   The beam consists of 81\% $\nu_\mu$, 13\% $\bar
\nu_\mu$, 5\% $\nu_e$ and 1\% $\bar \nu_e$.   
For more information on this data see Ref.~\cite{MBNuMI}.  

This data set is referred to as {\bf NuMI-MB($\nu$app)} in the
analysis below.   As seen in Fig.~\ref{expts1}, the data set provides
a limit at 95\% CL. In the fits presented 
here, this data is used to constrain electron flavor appearance in 
neutrino mode, with 10 bins used in the fit.  Statistical and systematic errors for this data set are 
added in quadrature.    

~~\\
\noindent{\it The Booster Beam Disappearance Search:}\\

The MiniBooNE experiment also searched for $\nu_{\mu}$ and $\bar{\nu}_{\mu}$ disappearance
using the Booster Neutrino Beam.   The neutrino (antineutrino) data set
corresponded to 5.6$\times 10^{20}$ ($3.4\times 10^{20}$) protons on
target, producing a beam 
covering the neutrino energy range up to 1.9~GeV.   The MiniBooNE $\nu_\mu$ disappearance result
provides restrictions on sterile neutrino oscillations which are comparable to 
those provided by the CDHS experiment, discussed below.
Therefore, we include that data set in these fits.   On the other hand, the 
$\bar\nu_\mu$ result was weaker due to the combination of fewer protons on
target and lower cross section.  The MINOS $\bar
\nu_\mu$ CC constraint, described below,  is substantially stronger, and
so we do not use the  MiniBooNE $\bar \nu_\mu$ data set.

The fit to the $\nu_\mu$ data set uses 16 bins ranging up to 1900~MeV in reconstructed
neutrino energy. A shape-only fit is performed, where the predicted spectrum given any set of
oscillation parameters is renormalized so that the total number of predicted events, after oscillations,
is equal to the total number of observed events. Then the normalized predicted spectrum is compared to the observed spectrum
in the form of a $\chi^2$ which accounts for statistical and shape-only systematic uncertainties and bin-to-bin correlations
in the form of a covariance matrix.

This data set is referred to as {\bf BNB-MB($\nu$dis)} in the analysis
below. Fig.~\ref{expts2} shows that this data sets a limit at 95\%
CL.   It should be noted that the published
MiniBooNE analysis used a Pearson $\chi^2$ method~\cite{Kendall}, and we are able to
reproduce those results.  However, to fold these results into our
analysis, we reverted to the  $\Delta
\chi^2$ definition used consistently among all data sets included in the fits (see Sec.~\ref{mthd}).

~~\\
\subsubsection{Results from Multi-GeV Conventional Short baseline
  $\nu_\mu$ Beams \label{Conv}}

The set of of multi-GeV conventional SBL $\nu_\mu$
experiments is the same as was used in previous fits. Our overview of these experiments is therefore very brief.

~~\\
\noindent{\it NOMAD Appearance Search:}\\

The NOMAD experiment~\cite{NOMAD1}, which ran at CERN using protons from the 450~GeV
SPS accelerator,  employed a
conventional neutrino beamline to create a wide band 2.5 to 40~GeV neutrino energy source.  These neutrinos were created with a
carbon-based, low-mass, tracking detector
located 600~m downstream of the target.   This detector had fine
spatial resolution and could search for muon-to-electron and
muon-to-tau oscillations.   No signal was observed in either mode.
In this analysis, we use the $\nu_\mu \rightarrow \nu_e$ constraint.

This data set is referred to as {\bf NOMAD} in the analysis below. 
This data set contributes 30 energy bins to the global fit.  The
statistical and systematic errors are added in quadrature.  This
experiment sets a limit at 95\% CL, as seen in Fig.~\ref{expts1}.

~~\\
\noindent{\it CCFR Disappearance Search:}\\

The CCFR data set was taken at Fermilab in 1984~\cite{CCFR84} with a
narrow band beamline, with meson energies set to 100, 140, 165, 200,
and 250~GeV, yielding $\nu_\mu$ and $\bar \nu_\mu$ beams that ranged
from 40 to 230~GeV in energy.  This was a two detector disappearance
search, with the near detector at 715~m and the far detector at 1116~m
from the center of the 352~m long decay pipe.  The calorimetric detectors were
constructed of segmented iron with scintillator and spark chambers,
and each had a downstream toroid to measure the muon momentum.

This data set is referred to as {\bf CCFR84} in the analysis below.
The data were published as the double ratios of the
observed-to-expected rates in a near-to-far ratio.  For each secondary
mean setting the data are divided into three energy bins.  The
systematic 
uncertainty is assumed to be energy independent and fully correlated
between the energy bins.   Due to the high beam energies and short
baselines,
this experiment sets a
limit at high $\Delta m^2$ in the muon flavor disappearance search
at 95\%~CL, as shown in Fig.~\ref{expts2}.

~~\\
\noindent{\it CDHS Disappearance Search:}\\

The CDHS experiment~\cite{CDHS} at CERN searched for $\nu_\mu$ disappearance with a two-detector design
of 
segmented calorimeters with iron and scintillator.  The
experiment used 19.2~GeV protons on a beryllium target to produce mesons
that
were subsequently focused into a 52~m decay channel.   The detectors were located
130~m and 885~m downstream of the target.

This data set is referred to as {\bf CDHS} in the analysis below. 
CDHS provides data and errors in 15 bins of muon energy,
provided in Table 1 of Ref.~\cite{CDHS}.  We associate  these bins 
to the neutrino energy distributions using the method 
described in
Ref.~\cite{sorel}:
the neutrino energy distribution for a given muon energy or range is
determined via the NUANCE~\cite{NUANCE}  neutrino cross section generator.
The experiment has a limit at 95\% CL and sets constraints that are 
comparable to the MiniBooNE $\nu_\mu$ disappearance limit described
above, but extending to slightly lower $\Delta m^2$.   See Fig.~\ref{expts2} for comparison.

\subsubsection{Reactor and Source Experiments \label{elecdis}}

The reactor experiment data set has been updated to reflect recent changes in
the predicted neutrino fluxes, as discussed below.
The source-based experimental data sets are both new to this paper,
having been published since our last set of fits~\cite{viability}.

~~\\
\noindent{\it Bugey Data Set:}\\

This analysis uses energy dependent data from the Bugey~3 
reactor experiment~\cite{Bugey}.   
The detector consisted of $^6$Li-doped liquid scintillator, with data taken at 15, 45 and 90~m from the 2.8~GW
reactor source.   The detectors are taken to be point-like in the analysis.

Recently, a reanalysis of reactor $\bar \nu_e$ flux predictions~\cite{Mueller, huber, whitepaper} has led to a reinterpretation of the
Bugey data. The data has transitioned from a limit on neutrino disappearance to an allowed region at 95\%~CL.  In this
analysis, we adjust the predicted Bugey flux spectra normalization according to the calculations from Ref.~\cite{whitepaper}.  

There are many other SBL reactor data sets in existence. However, we
have chosen to use only Bugey in these fits as the measurement has
the lowest combined errors.  Also, any global fit to multiple reactor data
sets must correctly account for the correlated systematics between
them, which is beyond the scope of our fits at present.

This data set is referred to as {\bf Bugey} in the analysis below.
As shown in Fig.~\ref{expts3}, this data set presents a signal at 95\% CL.  
The total number of bins in the analysis are 60, with the 15~m and 45~m baselines contributing 25 bins each
and the 90~m baseline contributing 10 bins, each extending from 1 to 6~MeV in positron energy. The fit follows the ``normalized energy spectra'' fit method
detailed in Ref.~\cite{Bugey}, and $\chi^2$ definition detailed within, which depends not only on the mass and mixing parameters we fit for, but also five large scale deformations of the positron spectrum due to systematic effects. Energy resolution and baseline smearing are taken into account. To fold in the flux normalization correction mentioned above, we update the theoretical prediction for the expected ratio by an overall normalization factor of 1.06237, 1.06197, and 1.0627 for the 15~m, 45~m, and 90~m baselines, respectively.

~~\\
\noindent{\it Gallium Calibration Data Set:}\\

Indications of $\nu_e$ disappearance have recently been published from calibration data taken
by the SAGE~\cite{SAGE3} and GALLEX~\cite{GALLEX3} experiments.  
These were solar neutrino experiments that 
used Mega-curie sources of $^{51}$Cr and $^{37}$Ar, which produce
$\nu_e$,  to calibrate the
detectors.   Each of the two experiments had two calibration periods.
The overall rates from these four measurements are
consistent, and show an overall deficit that has been reported to be 
consistent with electron flavor disappearance~\cite{Giunti1, Gallium}.
We use the four ratios of calibration data to expectation, 
as reported in Ref.~\cite{Gallium}, Table 2:  $1.00\pm0.10$,
$0.81\pm0.10$,
$095\pm 0.12$ and $0.79\pm0.10$.  These correspond to 
the two periods from GALLEX and the two periods from SAGE, respectively.
Our analysis of this data set, referred to as {\bf Gallium} below, follows that of Ref.~\cite{Gallium}; a 4-bin fit to the above measured calibration period rates is used. The predicted rates, after oscillations,
are obtained by averaging the oscillation probabilities taking into account
the detector geometry, the location of the source within the detector, and the 
neutrino energy distribution for each source (energy line and branching fraction).
The neutrino energies are approximately 430 and 750~keV for
$^{51}$Cr, and 812~keV for $^{37}$Ar. 
The data result in a limit
at 95\% CL, as shown in
Fig.~\ref{expts3}.

\subsubsection{Long Baseline Experimental Results Contributing to the
  Fits \label{lbl}}

While this study concentrates mainly on results from SBL
experiments, the data from experiments with baselines 
of hundreds of kilometers can be valuable.        At such long
baselines, the ability to identify the $\Delta m^2$ associated with any observed oscillation has disappeared due to the
rapid oscillations.  However, these experiments can place strong
constraints on the mixing parameters.  New to this paper is the
inclusion of the MINOS $\bar{\nu}_\mu$ CC constraint.    
We have included the atmospheric data set in our previous fits~\cite{sorel,karagiorgi,viability}.  

We note two long baseline results not included in this analysis.  First, we have 
dropped the Chooz data set that was included in previous
fits~\cite{sorel,karagiorgi,viability} due to the discovery that $\sin^2 2\theta_{13}$ is 
large~\cite{minostheta13,t2k,dc1stpub,dayabay,reno}, which
significantly complicates the use of this data for SBL
oscillation searches.  Second, the recent muon-flavor
disappearance results from IceCube~\cite{icecube} were published too late to be included in this iteration of fits.
However, the
MiniBooNE and MINOS muon-flavor disappearance results are more 
stringent than the IceCube  limits, and so we do not expect this to
affect the results.

~~\\
\noindent{\it MINOS $\bar{\nu}_{\mu}$ CC Disappearance Search:}\\

MINOS is a muon flavor disappearance experiment featuring two (near and far) iron-scintillator segmented calorimeter style detectors in the NuMI beamline (described above) at Fermilab. The near detector is located 1~km from the target while the far detector is located 730~m away. The wide band beam is peaked at about 4~GeV.   

MINOS ran in both neutrino and antineutrino mode.  We employ the
antineutrino data in our fits as it constrains the
allowed region for muon antineutrino disappearance when we divide the data sets into neutrino vs.~antineutrino fits.
The MINOS neutrino mode disappearance limit
is not as restrictive as the atmospheric result and so only the antineutrino data set is utilized. 

This result is referred to as {\bf MINOS-CC} in the analysis below.
The data present a limit at 95\% CL as discussed above and shown in Fig.~\ref{expts2}.
In our analysis of MINOS-CC, we fit both the antineutrino (right sign) data published by MINOS in antineutrino mode running~\cite{rightsignMINOS} and the antineutrino (wrong sign) data published by MINOS in neutrino mode running~\cite{wrongsignMINOS}. The right sign data are considered in 12 bins from 0 to 20~GeV, and the wrong sign in 13 bins from 0 to 50~GeV.  We account for possible oscillations in the near detector due to high $\Delta m^2$ values by using the ratio of the oscillation probabilities at the far and near detectors for each mass and mixing model.   As MINOS is sensitive to $\Delta m^2_{atm}$, we add an extra mass state to the oscillation probability using the best-fit atmospheric mass and mixing parameters from the MINOS experiment~\cite{MINOS4}. 
The data points and systematic errors are taken from~\cite{rightsignMINOS} and~\cite{wrongsignMINOS}.

~~\\
\noindent{\it Atmospheric Constraints on $\nu_\mu$ Disappearance Used in
Fits:} \\

Atmospheric neutrinos are produced when cosmic
rays interact with nuclei in the atmosphere to produce showers of mesons.  
The neutrino pathlength varies from a few to 12,800~km, while neutrino energies range from
sub- to few-GeV. Thus, this is a
long baseline source with sensitivity to primarily $\nu_{\mu}$ disappearance and effectively no sensitivity to $\Delta m^2_{ij}$. The former is a consequence of the atmospheric neutrino flux composition and the detector technology
used in atmospheric experiments. 
Thus, atmospheric neutrino measurements and long baseline accelerator-based $\nu_{\mu}$ disappearance experiments constrain the same parameters, and are treated in our fits in a similar way. 

As with our past fits, we include atmospheric constraints following
the prescription of Ref.~\cite{MaltoniValle}. We refer to this data set in our fits as {\bf ATM}.  This makes use of two 
data sets: (1) 1489 days of
Super-K muon-like and electron-like events with energies in the sub- to multi-GeV range, 
taking into account atmospheric flux predictions from~\cite{leptonicCPref33} and
treating systematic uncertainties according to~\cite{leptonicCPref34}; and (2) 
$\nu_{\mu}$ disappearance data from the long baseline, accelerator-based experiment K2K~\cite{leptonicCPref13_1,leptonicCPref13_2,leptonicCPref13_3}.
The atmospheric constraint is implemented in the form of a $\chi^2$ available to us as a function of the
parameter $d_{\mu}$, which depends on the muon flavor composition of $m_4$, $m_5$, and $m_6$ as follows:

\begin{equation}
d_{\mu}=\frac{1-\sqrt{1-4A}}{2}~,
\end{equation}
where
\begin{eqnarray}
A=(1-|U_{\mu4}|^2-|U_{\mu5}|^2-|U_{\mu6}|^2)(|U_{\mu4}|^2+|U_{\mu5}|^2+|U_{\mu6}|^2) \nonumber\\
+ |U_{\mu4}|^2|U_{\mu5}|^2+|U_{\mu4}|^2|U_{\mu6}|^2+|U_{\mu5}|^2|U_{\mu6}|^2~.
\end{eqnarray}
The atmospheric constraints set a limit at 95\% CL as shown in Fig.~\ref{expts2}.

\section{Analysis Description}

The analysis method follows the formalism described in Sec.~\ref{sec2}, and fits are performed to each of the (3+1), (3+2), and (3+3) hypotheses separately.

\subsection{\label{fitparameters}Fit Parameters}

The independent parameters considered in the (3+1) fit are $\Delta m^2_{41}$, representing the splitting between the (degenerate) first three mass eigenstates and the fourth mass eigenstate, and $|U_{e4}|$ and $|U_{\mu 4}|$, representing the electron and muon flavor content in the fourth mass eigenstate, which are assumed to be small. The (3+2) model introduces an additional, fifth mass eigenstate, where $\Delta m^2_{51} \ge \Delta m^2_{41}$, 
two additional mixing parameters, $|U_{e5}|$ and $|U_{\mu 5}|$,  as
well as the CP-violating phase $\phi_{54}$, defined by
Eq.~\ref{cpvphase}. The (3+3) model includes all the previous
parameters and yet another, sixth mass eigenstate, described by
$\Delta m^2_{61}$, where $\Delta m^2_{61} \ge \Delta m^2_{41},\ \Delta
m^2_{51}$, two additional mixing parameters, $|U_{e6}|$ and $|U_{\mu
  6}|$, as well as two more CP-violating phases, $\phi_{46}$ and
$\phi_{56}$. The above model parameters are allowed to freely vary
within the following ranges:  $\Delta m^2_{41}$, $\Delta m^2_{51}$ and
$\Delta m^2_{61}$ within 0.01-100~eV$^2$, $|U_{\alpha i}|$ within
0-0.5, and $\phi_{ij}$ within 0-2$\pi$, with the exception that for the $|U_{\alpha i}|$ there are additional
constraints imposed on the mixing parameters in order to conserve
unitarity of the full (3+$N$)$\times$(3+$N$) mixing matrix in each
scenario, as described in Sec.~\ref{sec2}.

\subsection{\label{mthd}Fitting Method}

The fitting method closely follows what has been done in Ref.~\cite{karagiorgi}. Given an oscillation model, (3+1), (3+2), or (3+3), the corresponding independent oscillation parameters are randomly generated within their allowed range, and then varied via a Markov Chain $\chi^2$ minimization procedure~\cite{markov}.  Each independent parameter $x$ is generated and varied according to
\begin{equation}
x = x_{old} + s(R-0.5)(x_{min} - x_{max})~,
\end{equation}
where $x_{old}$ is the value of parameter $x$ previously tested in the $\chi^2$ minimization chain,  $x_{min}$ and $x_{max}$ represent the boundaries on the parameter $x$ as described in Sec. \ref{fitparameters}, $R$ is a random number between 0 and 1, which is varied as one steps from $x_{old}$ to $x$, and $s$ is the ``stepsize'', a parameter of the Markov Chain.  By definition within the Markov Chain minimization method, the point is accepted based only on the point directly preceding it.  The acceptance of any new point $\vec{x}$ in the chain, where $\vec{x}$ is the new point in the oscillation parameter space, is determined by:
\begin{equation}
P = min(1,e^{-(\chi^2 -\chi^2_{old})/T})~,
\end{equation}
where $T$ is the Markov Chain parameter ``temperature''.  The step
size and temperature control how quickly the Markov Chain diffuses
toward the minimum $\chi^2$ value.   At every step in the chain, which
corresponds to a point in the oscillation parameter space, $\vec{x}$,
a $\chi^2_x$ is calculated by summing together the individual
$\chi^2_{x,d}$ contributed from each data set $d$ included in the fit,
where $d$ denotes any of the data sets described in Sec.~\ref{fitexps}.

In any given fit, we define possible signal indications at 90\% and 99\% CL by marginalizing over the full parameter space, and looking for closed contours formed about a global minimum, $\chi^2_{min}$, when projected onto any two-dimensional parameter space, assuming only two (2) degrees of freedom. We use the standard, two degree of freedom $\Delta\chi^2$ cuts of 4.61 for exploring allowed 90\% CL regions, 5.99 for exploring allowed 95\% CL regions (used only for Figs.~\ref{expts1}, \ref{expts2} and \ref{expts3}), and 9.21 for 99\% CL regions. If the null point ($U_{\alpha i}, U_{\beta i}=0$) is allowed at $>95\%$ CL, we instead proceed with drawing one-dimensional raster scan limits, obtained with the standard $\Delta\chi^2$ cuts of 2.70, 3.84 and 6.63 for 90\%, 95\% (used only for Figs.~\ref{expts1}, \ref{expts2} and \ref{expts3}), and 99\% CL, respectively.

\subsection{Parameter Goodness-of-Fit Test}

In any given fit, in addition to a standard $\chi^2$-probability,
which is quoted for the global $\chi^2_{min}$ and number of degrees of
freedom in the fit, we also report statistical compatibility
comparisons using the Parameter Goodness-of-fit test (PG test) from
Ref.~\cite{pgtest}.  This test reduces the bias imposed toward
data sets with a large number of bins in the standard $\chi^2$-
probability, in order to calculate the compatibility between data sets
simply on the basis of preferred parameters.  The compatibility, or PG
($\%$), can be calculated to quantify compatibility 
between any two or more data sets, or between combinations of data sets, according to
\begin{equation}
\chi^2_{PG}=\chi^2_{min,combined} - \sum_{i}\chi^2_{min,d}~,
\end{equation}  
where the $\chi^2_{min,combined}$ is the $\chi^2$-minimum of the combined fit of the data sets in consideration, and  $\chi^2_{min,d}$ is the $\chi^2$ of each data set or combination of data sets included in the combined fit when fit individually. The number of degrees of freedom ($ndf_{PG}$) for the PG test is given by
\begin{equation}
ndf_{PG}=\sum_{d} N_{p_{d}} - N_{p_{combined}}~.
\end{equation}
Here, $N_{p_{d}} $ represents the number of {\it independent} parameters involved in the fit of a particular data set and $ N_{p_{combined}}$ represents the number of {\it independent} parameters involved in the global fit.

\section{Results}

\begin{table}

\begin{center}

\begin{tabular}{|c|c|c|c|c|}
\hline
Tag & Section & Process  & $\nu$ vs.~$\bar \nu$ & App vs.~Dis \\ \hline
LSND & \ref{DAR} & $\bar \nu_\mu \rightarrow \bar \nu_e$ & $\bar \nu$ & App \\
KARMEN & \ref{DAR} & $\bar \nu_\mu \rightarrow \bar \nu_e$ & $\bar \nu$ & App \\
KARMEN/LSND(xsec) & \ref{DAR} & $\nu_e \rightarrow \nu_e$ & $\nu$ & Dis \\
BNB-MB($\nu$app) &  \ref{MBres} & $\nu_\mu \rightarrow \nu_e$  & $\nu$ &
App \\
BNB-MB($\bar\nu$app) & \ref{MBres} & $\bar \nu_\mu \rightarrow \bar
\nu_e$ & $\bar \nu$ & App \\
NuMI-MB($\nu$app) & \ref{MBres} & $\nu_\mu \rightarrow \nu_e$ & $\nu$
& App \\
BNB-MB($\nu$dis) & \ref{MBres} & $\nu_\mu \rightarrow \nu_\mu$ & $\nu$
& Dis \\
NOMAD & \ref{Conv} & $\nu_\mu \rightarrow \nu_e$  & $\nu$ & App \\
CCFR84 & \ref{Conv} & $\nu_\mu \rightarrow \nu_\mu$  & $\nu$ & Dis \\
CDHS & \ref{Conv} & $\nu_\mu \rightarrow \nu_\mu$ & $\nu$ & Dis \\
Bugey & \ref{elecdis} & $\bar \nu_e \rightarrow \bar \nu_e$ & $\bar \nu$ & Dis \\
Gallium & \ref{elecdis} & $\nu_e \rightarrow \nu_e$ & $\nu$ & Dis \\
MINOS-CC & \ref{lbl} & $\bar \nu_\mu \rightarrow \bar \nu_\mu$ & $\bar \nu$ & Dis \\
ATM & \ref{lbl} & $\nu_\mu \rightarrow \nu_\mu$ & $\nu$ & Dis \\
\hline
\end{tabular}
\caption{Data sets used in the fits and their
  corresponding use in the analysis.  Column 1 provides
  the tag for the data. Column 2
  references the description in Sec.~\ref{fitexps}.
 Column 3 lists the
  relevant oscillation process. 
   Column 4 lists which data sets
are included in the neutrino vs.~antineutrino analyses and column 5 lists
which data sets are included in the appearance vs.~disappearance study.}
\label{tab:tags}
\end{center}
\end{table}

This section presents the results of the analysis for the (3+1), (3+2), and (3+3) sterile neutrino models.  
For reference,
information about the data sets used in the analyses is provided
in Table~\ref{tab:tags}.  
Tables~\ref{tab:fitstats} and~\ref{tab:bfpoints} summarize the results
of the fits, which will be described in more detail below.   
Table~\ref{tab:fitstats} 
gives the fit results for the overall global fits and for various combinations of data sets.
When interpreting compatibilities, one should keep in mind that, along with 
a high compatibility among the individual data sets in a global fit, high compatibility values among 
groups of data sets is also important. Finally, Table~\ref{tab:bfpoints} provides the parameters for 
the best-fit points for each of the models.

\subsection{(3+1) Fit Results}

\begin{figure}[h!]
\begin{center}
\vspace{0.5cm}
\includegraphics[scale=.45]{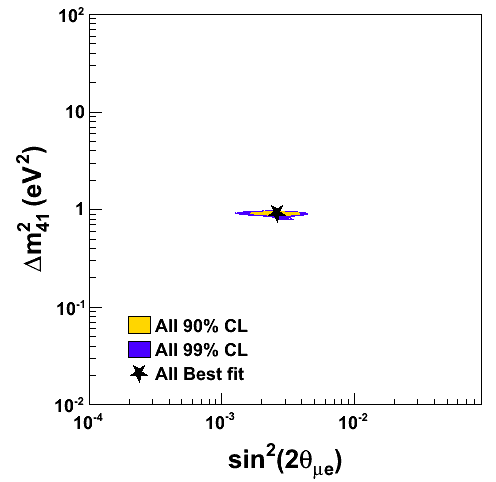}
\end{center}
\caption{The $\Delta m^2_{41}$ vs.~$\sin^2 2\theta_{\mu e}$ allowed space from fits
to all data---neutrino and  antineutrino---in a (3+1) model.}
\label{fig:3plus1_global}
\end{figure}

For a (3+1) model,  three parameters are determined:  
$\Delta m^2_{41},~|U_{e4}|$, and $|U_{\mu 4}|$.  A global (3+1) fit of all of the experiments
(Fig.~\ref{fig:3plus1_global}) yields a $\chi^2$-
probability of 55\% but a very low compatibility of 0.043\%, indicating a low compatibility among all individual data sets. 
Contrasting the good result from the $\chi^2$ test to the poor compatibility
illustrates how the $\chi^2$ test can be misleading.  As
discussed above, this
is due to some data sets dominating others due to the number of bins
in the fit, many of which may not have strong oscillation sensitivity.    
It is for this reason that most groups fitting for
sterile neutrinos now use the PG test as the figure of merit.

\begin{figure}
\begin{center}
\mbox{\subfigure{\includegraphics[scale=.35]{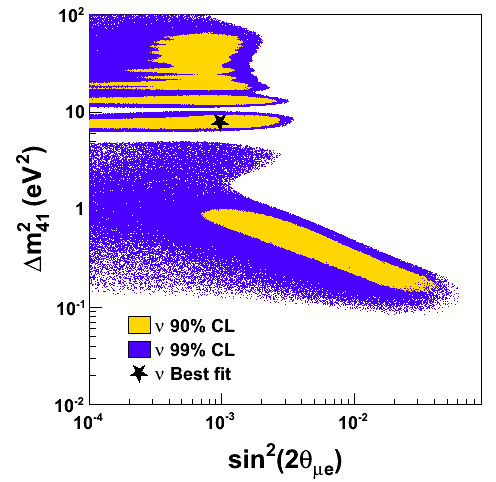}}\quad
\subfigure{\includegraphics[scale=.35]{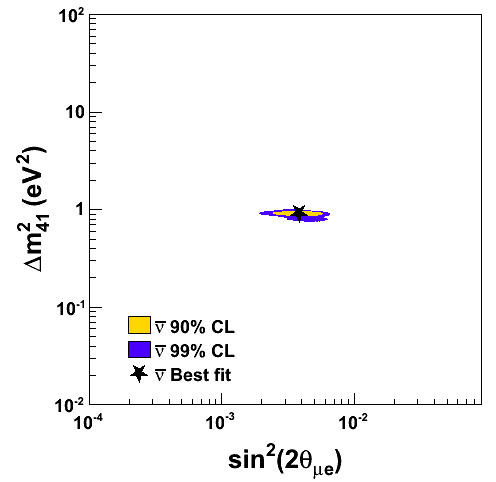} }}
\end{center}
\caption{The $\Delta m^2_{41}$ vs.~$\sin^2 2\theta_{\mu e}$ allowed space  from fits
to neutrino (left) and antineutrino (right) data in a (3+1) model.}\label{fig:3plus1_nunubar}
\end{figure}

In order to understand the source of the poor compatibility, the data sets are
subdivided, as shown in Table~\ref{tab:fitstats}, into separate neutrino and
antineutrino results.    Within each of these categories, the compatibility
values are 2.2\% and 11\% for neutrinos 
and antineutrinos respectively, which are reasonably good.   However,
the two data sets favor very different oscillation parameters,
as is seen in Table~\ref{tab:bfpoints} and Fig.~\ref{fig:3plus1_nunubar}.
This leads to a very low
compatibility of 0.14\% when the neutrino and antineutrino data are compared.  
The separation of the data sets into appearance and disappearance
also shows a strong incompatibility leading to an even lower compatibility
of 0.013\%.  These results imply that the (3+1) model is not sufficient
to describe all data sets simultaneously.

Looking at the best-fit values of Table~\ref{tab:bfpoints}, one also sees
that two different $\Delta m^2_{41}$ values are preferred for neutrino vs.~antineutrino
and for appearance vs.~disappearance.
This leads one to suspect that the data would prefer at least two mass splittings between the
mostly active and the mostly sterile states and, thus, encourages
the consideration of a (3+2) interpretation~\cite{sorel}.   Moreover, a (3+2)
model allows the introduction of a CP-violating phase, which can
address the differences between neutrino and antineutrino data sets~\cite{karagiorgi}.   Therefore, these results lead us to abandon (3+1), and move on to testing the (3+2) hypothesis.  It should be noted that the shortcomings of the (3+1) model have now been 
established by a number of independent analyses~\cite{whitepaper, JKopp,gli,Donini,viability,sorel}.

\begin{figure}
\begin{center}
\mbox{\subfigure{\includegraphics[scale=.35]{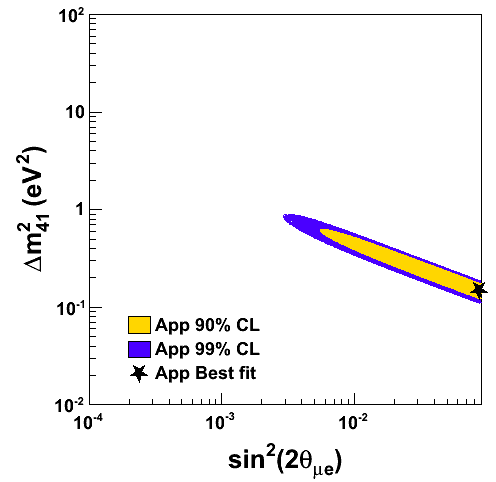}}\quad
\subfigure{\includegraphics[scale=.35]{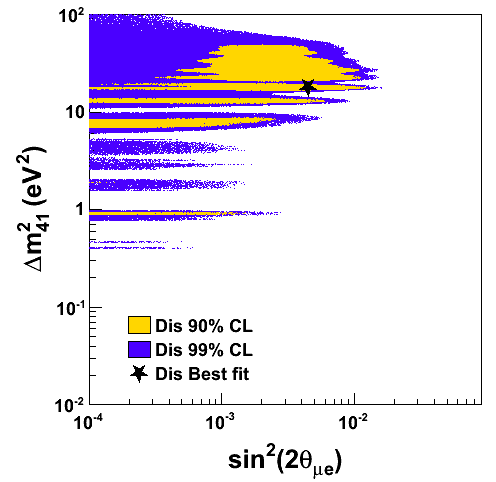} }}
\end{center}
\caption{The $\Delta m^2_{41}$ vs.~$\sin^2 2\theta_{\mu e}$ allowed space  from fits
to appearance (left) and disappearance (right) data in a (3+1) model.}\label{fig:3plus1_appdis}
\end{figure}

\begin{table}

\begin{center}
\begin{tabular}{|c|cccccl|}\hline
 & $\chi^2_{min}$ (dof) & $\chi^2_{null}$ (dof) &  $P_{best}$  & $P_{null}$  &
$\chi^2_{PG}$ (dof) & PG (\%)  \ \\  \hline\hline
{\bf3+1} & & & & & & \ \\  \hline
All & 233.9 (237) & 286.5 (240) & 55\% & 2.1\% & 54.0 (24) & 0.043\%  \ \\  \hline
App & 87.8 (87) & 147.3 (90) & 46\% & 0.013\% & 14.1 (9) & 12\%  \ \\  \hline
Dis & 128.2 (147) & 139.3 (150) & 87\% & 72\% & 22.1 (19) & 28\% \ \\  \hline
$\nu$  & 123.5 (120) & 133.4 (123) & 39\% & 25\% & 26.6 (14) & 2.2\% \ \\  \hline
$\overline{\nu}$ & 94.8 (114) & 153.1 (117) & 90\% & 1.4\% & 11.8 (7) & 11\%  \ \\  \hline
App vs.~Dis & - & - & -  & - &17.8 (2) & 0.013\%    \ \\  \hline
$\nu$ vs.~$\overline{\nu}$ & - & - & -  & - & 15.6 (3) & 0.14\%  \ \\  \hline \hline \hline  

{\bf3+2} & & & & & & \ \\  \hline
All  & 221.5 (233) & 286.5 (240) & 69\% & 2.1\% & 63.8 (52) & 13\%  \ \\  \hline
App & 75.0 (85) & 147.3 (90) & 77\% & 0.013\% & 16.3 (25) & 90\%  \ \\  \hline
Dis &122.6 (144) & 139.3 (150) & 90\% & 72\% & 23.6 (23) & 43\% \ \\  \hline
$\nu$ & 116.8 (116) & 133.4 (123) & 77\% & 25\% & 35.0 (29) & 21\% \ \\  \hline
$\overline{\nu}$ & 90.8 (110) & 153.1 (117) & 90\% & 1.4\% & 15.0 (16) & 53\%  \ \\  \hline 
App vs.~Dis & - & - & -  & - & 23.9 (4) & 0.0082\%   \ \\  \hline
$\nu$ vs.~$\overline{\nu}$ & - & - & -  & -  & 13.9 (7) & 5.3\% \ \\  \hline \hline \hline 

{\bf3+3} & & & & & & \ \\  \hline
All  & 218.2 (228) & 286.5 (240) & 67\% & 2.1\% & 68.9 (85) & 90\% \ \\  \hline
App & 70.8 (81) & 147.3 (90) & 78\% & 0.013\% & 17.6 (45) & 100\%   \ \\  \hline
Dis& 120.3 (141) & 139.3 (150) & 90\% & 72\% & 24.1 (34) & 90\% \ \\  \hline
$\nu$ &  116.7 (111) & 133.4 (123) & 34\% & 25\% & 39.5 (46) & 74\% \ \\  \hline
$\overline{\nu}$ & 90.6 (105) & 153 (117) & 84\% & 1.4\% & 18.5 (27) & 89\%  \ \\  \hline
App vs.~Dis & - & - & -  & - & 28.3 (6) & 0.0081\%    \ \\  \hline
$\nu$ vs.~$\overline{\nu}$ & - & - & -  & -  & 110.9 (12) & 53\% \ \\  \hline

\end{tabular}
\end{center}

\caption{The $\chi^2$ values, degrees of freedom (dof) and probabilities associated with the best-fit and null hypothesis in each scenario. Also shown are the results from the Parameter Goodness-of-fit tests.  $P_{best}$ refers to the $\chi^2$-probability at the best fit point and $P_{null}$ refers to the  $\chi^2$-probability at null. }
\label{tab:fitstats}
\end{table}

\begin{table}

\begin{center}
\begin{tabular}{|c|ccc|}\hline
{\bf 3+1} & $\Delta m^2_{41}$ & $|U_{\mu4}|$ & $|U_{e4}|$   \ \\  \hline\hline
All & 0.92 & 0.17 & 0.15   \ \\  \hline
App &0.15 & 0.39 & 0.39   \ \\  \hline
Dis & 18 & 0.18 & 0.18   \ \\  \hline
$\nu$ & 7.8 & 0.059 & 0.26  \ \\  \hline
$\overline{\nu}$& 0.92 & 0.23 & 0.13    \ \\  \hline 
\end{tabular}
\vspace*{5mm}

\begin{tabular}{|c|ccccccc|}\hline

{\bf 3+2} & $\Delta m^2_{41}$ & $\Delta m^2_{51}$ & $|U_{\mu4}|$  & $|U_{e4}|$ & $|U_{\mu5}|$ & $|U_{e5}|$ &
$\phi_{54}$  \ \\  \hline\hline
All & 0.92 & 17 & 0.13 & 0.15 & 0.16 & 0.069 & 1.8$\pi$\ \\  \hline
App & 0.31 & 1.0 & 0.31 & 0.31 & 0.17 & 0.17 & 1.1$\pi$ \ \\  \hline
Dis& 0.92 & 18 & 0.015 & 0.12 & 0.17 & 0.12 & N/A  \ \\  \hline
$\nu$ &7.6 & 17.6 & 0.05 & 0.27 & 0.18 & 0.052 & 1.8$\pi$ \ \\  \hline
$\overline{\nu}$ & 0.92 & 3.8 & 0.25 & 0.13 & 0.12 & 0.079 & 0.35$\pi$  \ \\  \hline
\end{tabular}

\vspace*{5mm}

\begin{tabular}{|c|cccccccccccc|}\hline

{\bf 3+3} & $\Delta m^2_{41}$ & $\Delta m^2_{51}$ & $\Delta m^2_{61}$ &$|U_{\mu4}|$  & $|U_{e4}|$ & $|U_{\mu5}|$ & $|U_{e5}|$ &$|U_{\mu6}|$ & $|U_{e6}|$&
$\phi_{54}$ & $\phi_{64}$ & $\phi_{65}$  \ \\  \hline\hline
All  &0.90 & 17 & 22 & 0.12 & 0.11 & 0.17 & 0.11 & 0.14 & 0.11 & 1.6$\pi$ & 0.28$\pi$ & 1.4$\pi$ \ \\  \hline
App &0.15 & 1.8 & 2.7 & 0.37 & 0.37 & 0.12 & 0.12 & 0.12 & 0.12 & 1.4$\pi$ & 0.32$\pi$ & 0.94$\pi$\ \\  \hline
Dis & 0.92 & 7.2 & 18 & 0.013 & 0.12 & 0.019 & 0.16 & 0.15 & 0.069 & N/A & N/A & N/A \ \\  \hline
$\nu$  & 13 & 17 & 26 & 0.076 & 0.24 & 0.16 & 0.067 & 0.10 & 0.017 & 1.1$\pi$ & 1.8$\pi$ & 0.037$\pi$ \ \\  \hline
$\overline{\nu}$ & 7.5 & 9.1 & 18 & 0.024 & 0.28 & 0.098 & 0.11 & 0.18 & 0.029 & 1.8$\pi$ & 2.0$\pi$ & 0.61$\pi$  \ \\  \hline

\end{tabular}

\end{center}

\caption{The oscillation parameter best-fit points in each scenario considered.  The values of  $\Delta m^2$ shown are in units of eV$^2$}
\label{tab:bfpoints}
\end{table}

\subsection{(3+2) Fit Results} \label{sec:3plus2}

\begin{figure}
\begin{center}
\vspace{0.5cm}
\includegraphics[scale=.45]{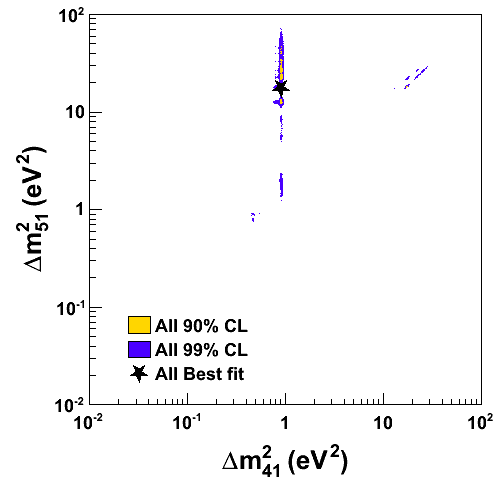}
\end{center}
\caption{The $\Delta m^2_{51}$ vs.~$\Delta m^2_{41}$ correlations from fits
to all data in a (3+2) model.}
\label{fig:3plus2_global}
\end{figure}

In a (3+2) model, there are seven parameters to determine:  
$\Delta m^2_{41}$, $\Delta m^2_{51}$, $|U_{e4}|$, $|U_{\mu 4}|$, $|U_{e5}|$, $|U_{\mu 5}|$, and $\phi_{54}$.
The best-fit values for these parameters from global fit to all data sets 
are given in Table~\ref{tab:bfpoints}.  The 90\% and 99\%
CL contours in marginalized ($\Delta m^2_{41},\Delta m^2_{51}$) space can be seen in Fig.~\ref{fig:3plus2_global}.

Adding a second mass eigenstate reduces the tension seen in the (3+1)
fits, bringing the overall compatibility to 13\% (see Table~\ref{tab:fitstats}), and reducing the $\chi^2$ of the global fit by 12.4 units, for four extra parameters introduced to the fit. For this compatibility test, the BNB-MB($\nu$app) data set has the worst $\chi^2$-probability. When considered by itself, the BNB-MB($\nu$app) data set gives a constrained (see Sec \ref{MBres}) $\chi^2$~(dof) of 19.2~(4) for the global best-fit parameters, which corresponds to a $\chi^2$-probability of 0.07\% .   This is one of the first indications that the MiniBooNE neutrino data
has some tension with the other data sets.

The need to introduce a CP-violating phase was established in previous studies of global fits~\cite{viability}.  This term affects only fits involving appearance data sets and results in a
difference in the oscillation probabilities for $\nu_\mu \rightarrow
\nu_e$ vs.~$\bar \nu_\mu \rightarrow \bar \nu_e$.   In particular, previous studies considered CP-violating fits in an attempt to reconcile the MiniBooNE neutrino appearance results with the MiniBooNE and LSND antineutrino appearance results.  

\begin{figure}
\begin{center}
\mbox{\subfigure{\includegraphics[scale=.35]{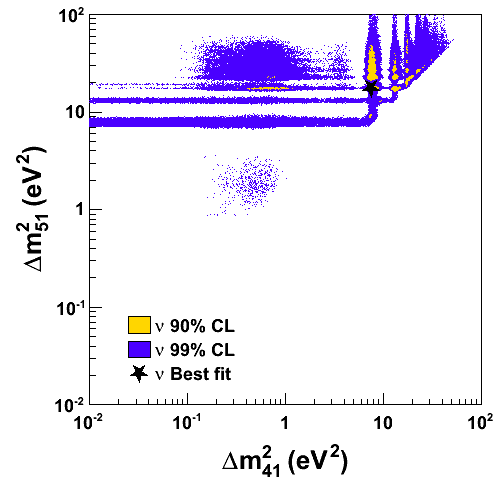}}\quad
\subfigure{\includegraphics[scale=.35]{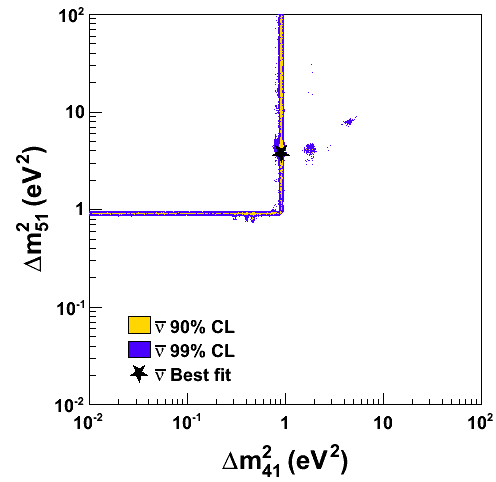} }}
\end{center}
\caption{The $\Delta m^2_{51}$ vs.~$\Delta m^2_{41}$ correlations from fits
to neutrino (left) and antineutrino (right) data in a (3+2) model.}\label{fig:3plus2_nunubar}
\end{figure}

Table~\ref{tab:fitstats} also gives the results for combinations of the data sets for cross
comparison.       We find that the separate neutrino and antineutrino data set fits
remain in good agreement and that
the compatibility between the neutrino and antineutrino data sets has risen to 
5.3\%---a significant improvement over the (3+1) result.    The
best-fit values and allowed regions are shown in Table~\ref{tab:bfpoints} and
in Fig.~\ref{fig:3plus2_nunubar} respectively. 

\begin{figure}
\begin{center}
\mbox{\subfigure{\includegraphics[scale=.35]{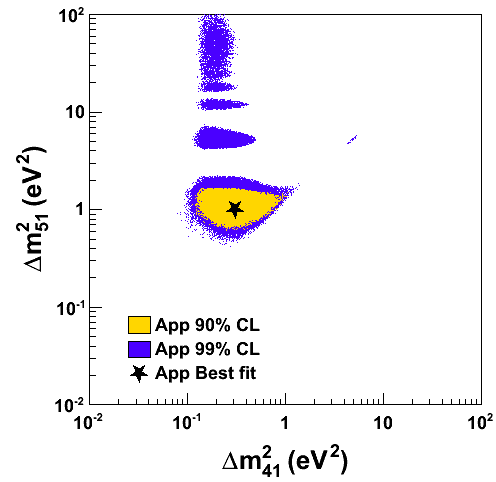}}\quad
\subfigure{\includegraphics[scale=.35]{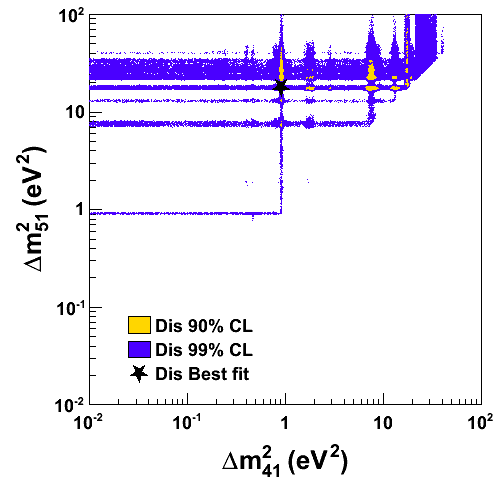} }}
\end{center}
\caption{The $\Delta m^2_{51}$ vs.~$\Delta m^2_{41}$ correlations from fits
to appearance (left) and disappearance (right) data in a (3+2) model.}\label{fig:3plus2_appdis}
\end{figure}

While the neutrino vs.~antineutrino discrepancy has been somewhat reduced, 
Table~\ref{tab:fitstats} points out a second important problem.
The appearance and disappearance data sets still have
very poor compatibility (0.0082\%), even in a (3+2) model.  The poor
compatibility can be partially traced to a discrepancy in the preferred mass
splittings for these two data sets.  As reported in Table~\ref{tab:bfpoints}, the appearance data sets prefer a low (0.31~eV$^2$) and a medium (1.0~eV$^2$) mass-squared splitting while the disappearance data sets
prefer a medium (0.92~eV$^2$) and a high (18~eV$^2$) splitting.
This is also illustrated in Fig.~\ref{fig:3plus2_appdis}. 
This suggests that three mass splittings may be required to reconcile appearance
and disappearance results, and motivates the consideration of a (3+3) model. 

\subsection{(3+3) Fit Results}  \label{sec:3plus2}

For a (3+3) model, there are 12 model parameters to be determined:  
$\Delta m^2_{41}, \Delta m^2_{51}, \Delta m^2_{61}$, $|U_{e4}|$, $|U_{\mu 4}|$, 
$|U_{e5}|$, $|U_{\mu 5}|$, $|U_{e6}|$, $|U_{\mu 6}|$, $\phi_{54}$, $\phi_{64}$, and $\phi_{65}$.
Adding a third mass eigenstate does not significantly change the global fit $\chi^2_{min}$; 
however, the tension between the individual data set fits is further reduced, raising the compatibility from 13\%, in (3+2), 
to 90\%.  The neutrino and antineutrino
compatibility rises by an order of magnitude from the (3+2) value to 53\%,
indicating that the (3+3) model can better accommodate the differences in these data sets.

It is interesting to note that the (3+3) fit prefers an ``inverted hierarchy'' among the three mostly
 sterile states, with $\Delta m^2_{54}=m^2_5-m^2_4=16$~eV$^2$ and $\Delta
m^2_{65}=m^2_6-m^2_5=5.0$~eV$^2$.  The overall splitting relative to the three
 mostly-active states is $\Delta m^2_{41} =m^2_4-m^2_{1}=0.90$~eV$^2$.  

\begin{figure}
\begin{center}
\mbox{\subfigure{\includegraphics[scale=.8]{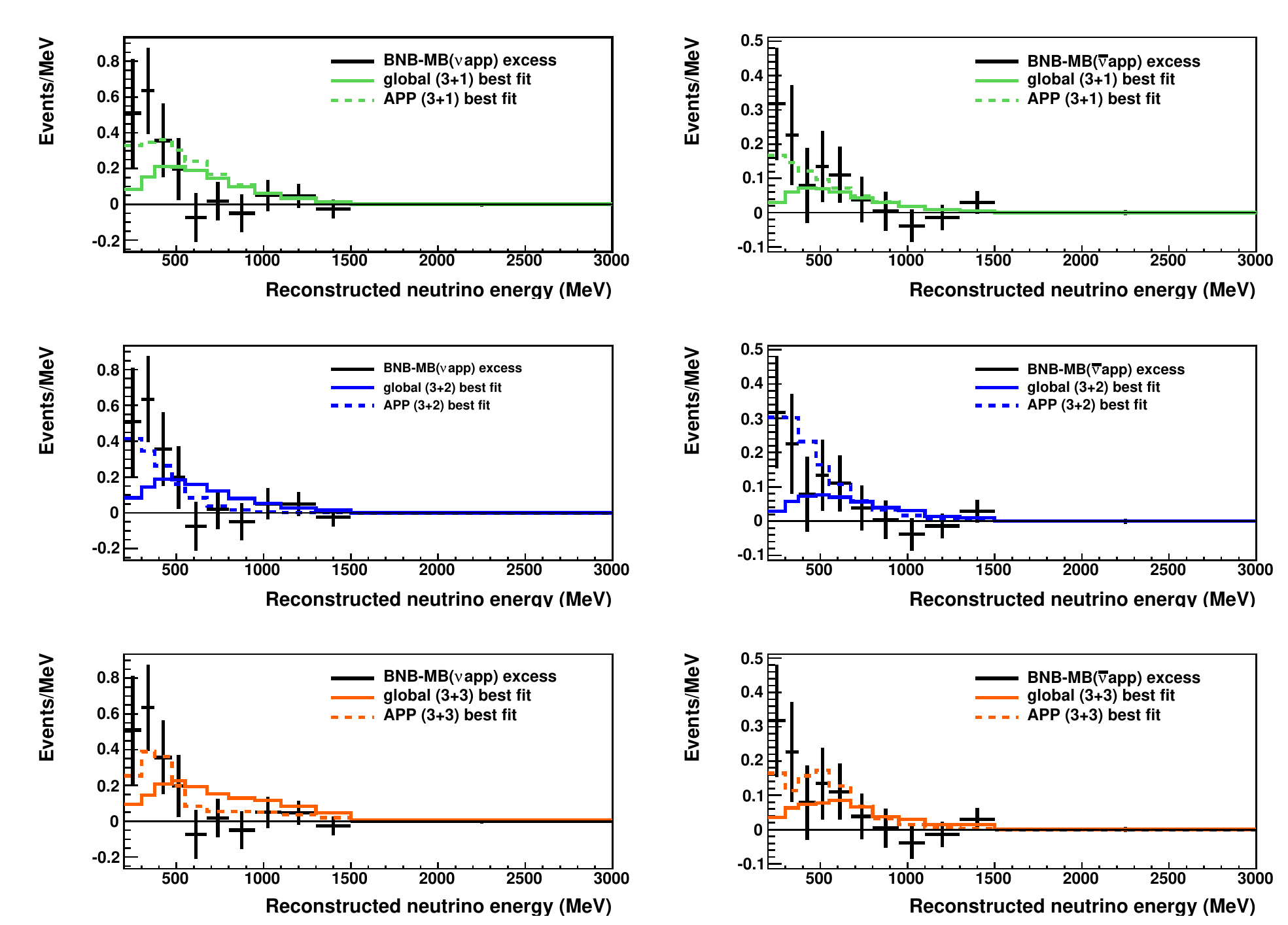}}}
\end{center}
\caption{\label{fig:MB_data_vs_models}A comparison of the BNB-MB($\nu$app) and BNB-MB($\bar{\nu}$app) excess data with 
the global best-fit oscillation signal predictions (solid colored lines) and with the appearance 
only best-fit predictions (dashed colored lines) for each of the models, (3+1), (3+2), and (3+3). The error bars 
on the excess correspond to data statistical and unconstrained background systematic errors, added in
quadrature.}
\end{figure}

The one puzzling discrepancy for the (3+3) fits is the tension 
still exhibited by the appearance vs.
disappearance data sets, where the compatibility remains low, at less than 
0.01\%.  We find that an important source of this incompatibility is
the BNB-MB($\nu$app) and BNB-MB($\bar \nu$app) data sets.  
The BNB-MB($\nu$app) data set has fairly 
small statistical and systematic uncertainties and therefore has a large impact on the
fits and compatibility calculations. This is shown in Fig. \ref{fig:MB_data_vs_models} where the MB data agrees well with the appearance-only fit but disagrees with the overall global fit.
Removing both the BNB-MB($\nu$app) 
and BNB-MB($\bar \nu$app) sets  
raises the compatibility to 3.5\%, corresponding to over two orders of magnitude improvement.
It has been known since the first MiniBooNE 
publication~\cite{MB} that the the BNB-MB($\nu$app) data was fairly consistent with no oscillations above 475~MeV; however, a significant low-energy excess was present below this energy.  The energy dependence
of the BNB-MB($\nu$app) excess does not fit very well with oscillation models
extracted from fits to global data sets, unless very low $\Delta m^2_{ij}$ with large mixing elements $|U_{ei}|$ and $|U_{\mu i}|$ are involved in the fit.
This may lead to the poor compatibility when included in appearance 
vs.~disappearance comparisons. Other possible explanations for this incompatibility include
downward fluctuations of the BNB-MB($\nu$app) data in the higher energy region or  
some other process contributing part of the low energy excess such as those suggested in Refs.~\cite{martini,axial}.

Statistical issues could be addressed with more MiniBooNE neutrino data that may become available over
the next several years.  In addition, the MicroBooNE experiment, which is expected to start running in 2014, will provide more information on the low-energy excess events and 
answer the question of whether the excess is associated with
outgoing electrons or photons~\cite{uboone}.  

\begin{figure}
\begin{center}
\mbox{\subfigure{\includegraphics[scale=.35]{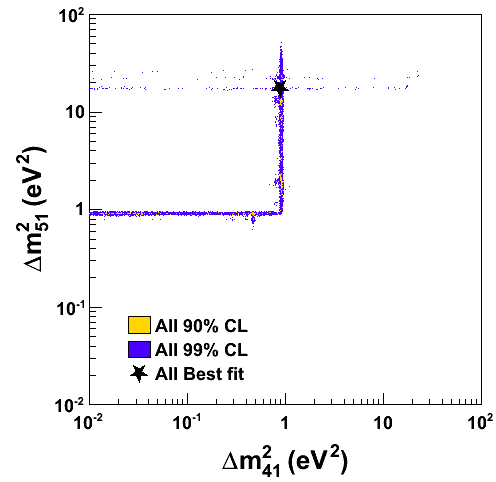}}\quad
\subfigure{\includegraphics[scale=.35]{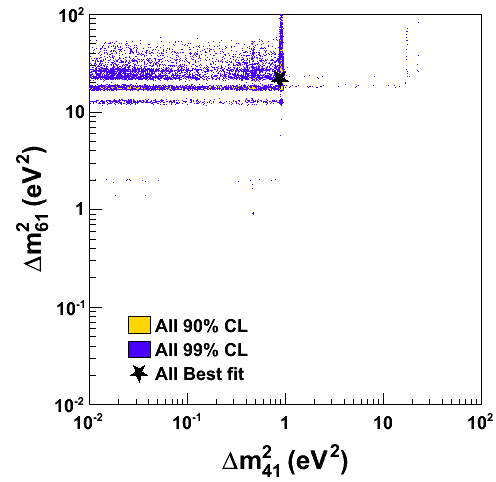} }}
\end{center}
\caption{The $\Delta m^2_{51}$ vs.~$\Delta m^2_{41}$ and $\Delta m^2_{61}$ vs.~$\Delta m^2_{41}$  correlations from fits
to all data in a (3+3) model.}\label{fig:3plus3_all}
\end{figure}

\begin{figure}
\begin{center}
\mbox{\subfigure{\includegraphics[scale=.35]{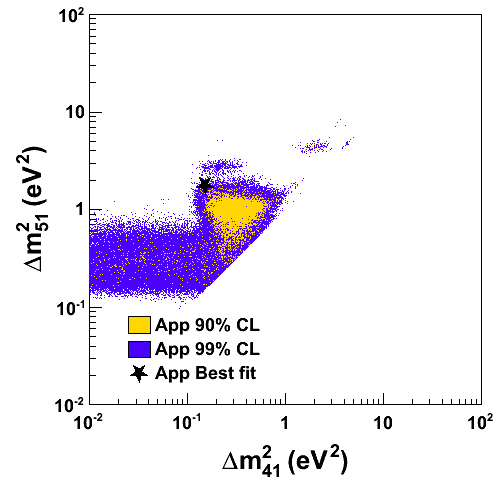}}\quad
\subfigure{\includegraphics[scale=.35]{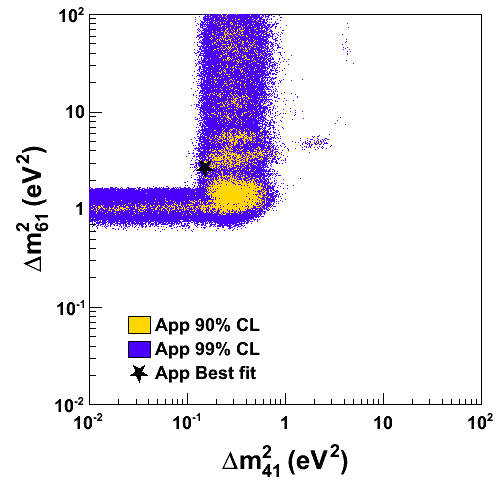} }}
\end{center}
\caption{The $\Delta m^2_{51}$ vs.~$\Delta m^2_{41}$ and $\Delta m^2_{61}$ vs.~$\Delta m^2_{41}$  correlations from fits
to appearance data in a (3+3) model.}\label{fig:3plus3_app}
\end{figure}

\begin{figure}
\begin{center}
\mbox{\subfigure{\includegraphics[scale=.35]{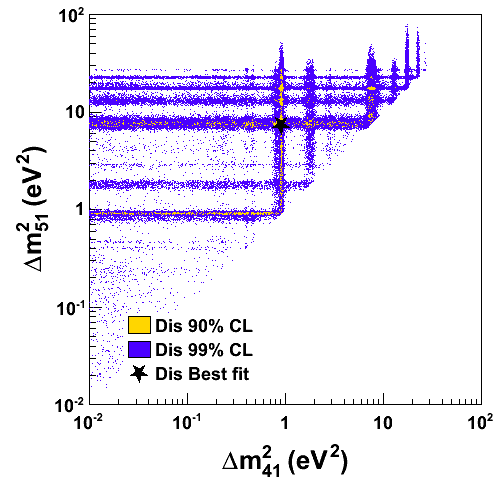}}\quad
\subfigure{\includegraphics[scale=.35]{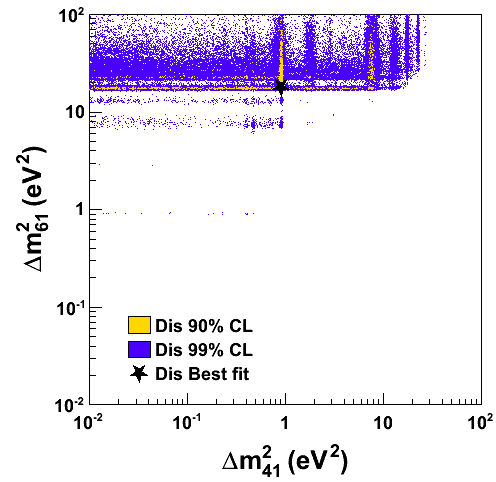} }}
\end{center}
\caption{The $\Delta m^2_{51}$ vs.~$\Delta m^2_{41}$ and $\Delta m^2_{61}$ vs.~$\Delta m^2_{41}$  correlations from fits
to disappearance data in a (3+3) model.}\label{fig:3plus3_dis}
\end{figure}

\subsection{Summary of Results}
The sterile neutrino fits to global data sets show that a (3+1) model is inadequate; multiple $\Delta m^2_{ij}$ values are needed along with CP-violating effects to explain the neutrino vs.~antineutrino differences.
A (3+2) model improves significantly on the (3+1) results but still shows some tension
for the neutrino vs.~antineutrino compatibility and cannot explain the appearance 
vs.~disappearance differences.  Among the three models considered, the (3+3) model gives the highest compatibility among the
various data set combinations but still has poor appearance vs.~disappearance
compatibility.  The BNB-MB($\nu$app) (and BNB-MB($\bar \nu$app)) data set is a prime contributor to this incompatibility and additional experimental information in this region should be available soon.  Figure~\ref{fig:MB_data_vs_models} gives a comparison of the 
BNB-MB($\nu$app) and BNB-MB($\bar\nu$app) data with the global best-fit predictions and with the appearance-only best-fit predictions for each of the three models, (3+1), (3+2), and (3+3).  One clearly sees that the appearance-only fit describes the data very well, especially for the (3+3) model. On the other hand, the global fit prediction is significantly below the data at low energy,  which contributes to the poor appearance vs.~disappearance compatibility.

In summary, out of the three sterile neutrino oscillation hypotheses considered in the analysis, we find that a (3+3) model provides the best description,
although the MiniBooNE appearance data continue to raise issues within the fits. As has been shown before, (3+1) scenarios provide a poor fit to the data, and should not be emphasized.
We therefore recommend continued investigations of (3+2) and (3+3) scenarios.

\clearpage
\section{The Future}

Establishing the existence of sterile neutrinos would have a major impact on particle physics. Motivated by this opportunity,
there are a number of existing and planned
experiments set to probe the parameter space
indicative of one or more sterile neutrinos. Such
experiments are necessary in order to confirm or
refute the observed anomalies in the $\Delta m^2 \sim
1~\mathrm{eV}^2$ region. The new experiments are
being designed to have improved sensitivity, with the
goal of 5$\sigma$ sensitivity
and the ability to observe oscillatory behavior in $L$
and/or $E$ within a single or between multiple detectors.  In these
experiments, the oscillation signal needs to be
clearly separated from any backgrounds.  

Sterile neutrino oscillation models are based on
oscillations associated with mixing between active
and sterile states and demand that there be both
appearance and disappearance.  It is
therefore imperative that the future program explore
both of these oscillation types.  Establishing
sterile neutrinos will require that the disappearance
and appearance rates are compatible with sterile
neutrino oscillation models. Future experiments will search for evidence of
sterile neutrino(s) using a variety of neutrino creation sources: (1)
pion/muon DIF
(e.g., Refs.~\cite{minosplus,boone,larlar,cernlar,uboone,nustorm}), (2)
pion or kaon DAR
(e.g., Refs.~\cite{coherent,sblcoherent,oscsns,superk,kdar}), (3)
unstable isotopes
(e.g., Refs.~\cite{lenssterile,ricochet,celand,dayabay,isodar}), (4)
atmospheric (see Ref.~\cite{whitepaper}), and (5) nuclear reactors
(e.g., Refs.~\cite{nucifer,neutrino4}). All of these experiments are under
development and the sensitivities are likely to change.   Therefore,
rather than displaying sensitivity curves for each future program, we instead focus on the conceptual ideas behind the experiments. Unless otherwise mentioned, the experiments below will provide ``significant'' sensitivity to a large portion of the favored sterile neutrino parameter space through searches for neutrino disappearance and/or appearance.


\subsection{The Importance of the $L/E$ Signature from Multiple Experiments}

Ultimately, in order to determine if there are zero, one, two, or three sterile
states contributing to oscillations in SBL experiments, it
will be necessary to demonstrate the expected $L/E$-dependent oscillation probabilities discussed in Sec.~\ref{sec2}.  Assuming that the SBL anomalies are confirmed, a consistent
$L/E$ dependence is the only signature which is distinct for
oscillations, and excludes other exotic explanations such as CPT violation~\cite{viability},  decays~\cite{Vanucci}, and Lorentz violation~\cite{Teppei}.
The ideal experiment would reconstruct the oscillation wave as a function of
$L/E$~\cite{Agarwalla}. The combined information from many
experiments, however, is more suitable for covering the widest possible range in $L/E$ as well as providing valuable flavor and neutrino vs.~antineutrino information.

\begin{figure}[h!]
\begin{center}
\includegraphics[scale=.7]{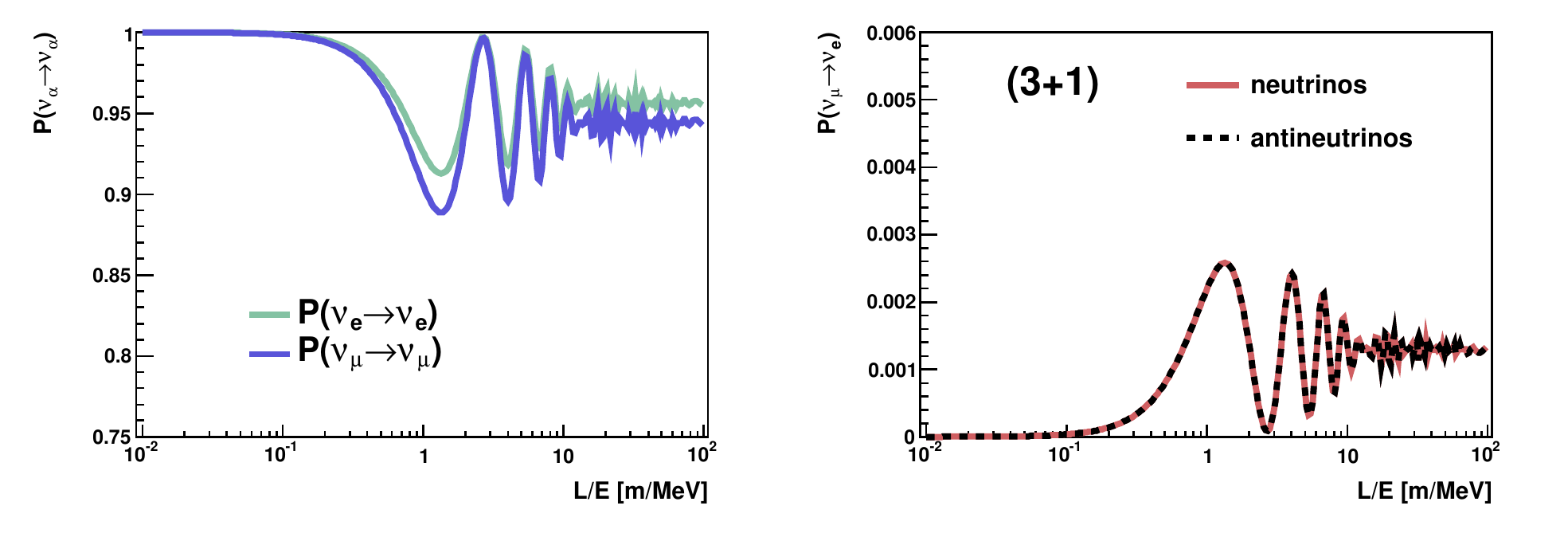}
\end{center}
\vspace{-1cm}
\caption{The (3+1) oscillation probabilities for the global best fit (``all"
data sets) values in Table 3 with 10\% resolution in \textit{L/E}.}
\label{LE31}
\end{figure}

\begin{figure}[h!]
\begin{center}
\includegraphics[scale=.7]{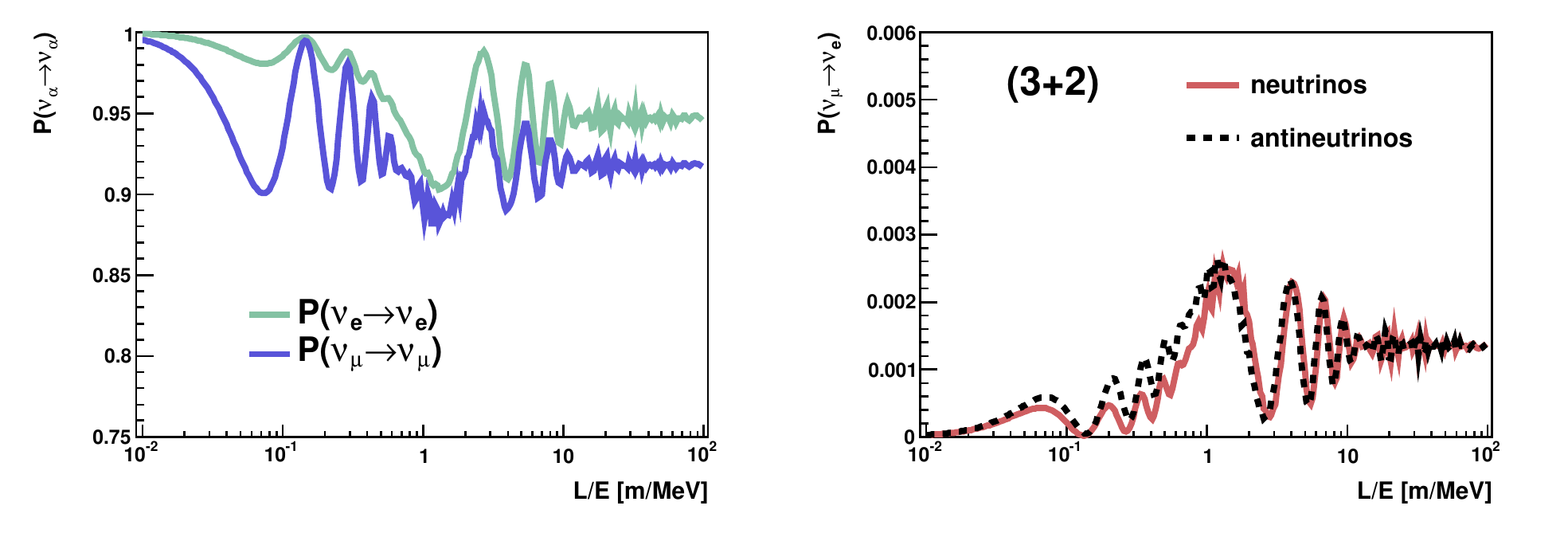}
\end{center}
\vspace{-1cm}
\caption{The (3+2) oscillation probabilities for the global best fit (``all"
data sets) values in Table 3 with 10\% resolution in \textit{L/E}.}
\label{LE32}
\end{figure}

\begin{figure}[h!]
\begin{center}
\includegraphics[scale=.7]{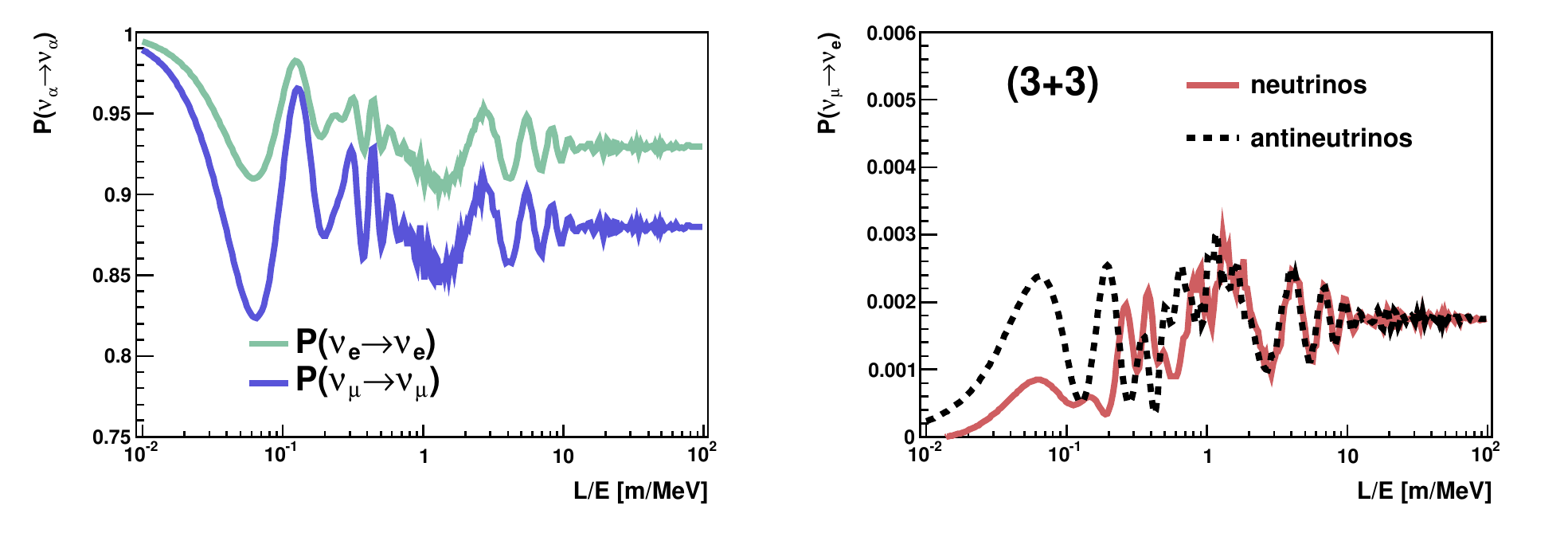}
\end{center}
\vspace{-1cm}
\caption{The (3+3) oscillation probabilities for the global best fit (``all"
data sets) values in Table 3 with 10\% resolution in \textit{L/E}.}
\label{LE33}
\end{figure}

The three models, (3+1), (3+2) and (3+3) have distinct signatures as a
function of $L/E$.
To illustrate this, we consider
the case of a hypothetical experiment with 10\% resolution in $L/E$,
assuming the best-fit values presented in Table~\ref{tab:bfpoints}.
In the case of (3+1), as shown in Fig.~\ref{LE31},
the disappearance (appearance) probabilities shown on the left (right),
have maxima and minima that evolve monotonically to $P=1/2 \sin^2                     
(2\theta)$, the long baseline limit discussed in
Sec.~\ref{basic}.    This can be contrasted with Figs.~\ref{LE32}
and~\ref{LE33}, where the structure of the oscillation wave, in the
approach to the long baseline limit, is more ``chaotic" due to the
interference between the various mass splitting terms.

In Figs.~\ref{LE31}, \ref{LE32}, and \ref{LE33}, the two curves on
the disappearance plots on the left refer to muon and electron flavor,
respectively. As the theory is CPT-conserving,
these disappearance curves should be identical for neutrinos and
antineutrinos.  The appearance
curves on the right also show the importance of neutrino and
antineutrino running, which can lead to very different $L/E$ dependencies
for the three models, and constrain CP-violating parameters.

In summary, it seems very unlikely that any {\it single} future experiment will be able to differentiate between the sterile neutrino models.  Multiple
experiments looking at different oscillation channels and covering a wide range of
$L/E$ regions are required.    Thus, the consideration of many independent and
relatively modest-size experiments, such as those listed in Table~\ref{summary_of_future}, is essential.

\begin{table}

\begin{center}

\begin{tabular}{|c|c|c|c|}
\hline
Source & App/Dis & Channel & Experiment \\
\hline
Reactor  &     Dis &   $\bar\nu_e\rightarrow \bar\nu_e$ & Nucifer, Stereo, SCRAMM, \\
& & &  NIST, Neutrino4, DANSS\\ \hline
Radioactive  &     Dis & $\nu_e\rightarrow\nu_e$    $\bar\nu_e\rightarrow \bar\nu_e$ & Baksan, LENS, Borexino, SNO+,  \\
& & &  Ricochet, CeLAND, Daya Bay\\ \hline
Accelerator-based &     Dis &   $\bar\nu_e\rightarrow \bar\nu_e$ &     IsoDAR \\
isotope& & &  \\ \hline
Pion / Kaon DAR & App \& Dis &
$\bar\nu_\mu\rightarrow \bar\nu_e$   $\nu_e \rightarrow \nu_e$   & OscSNS, DAE$\delta$ALUS,  \\
& & $\nu_\mu \rightarrow \nu_e$ &   KDAR\\ \hline
Accelerator (Pion DIF) & App \& Dis &  $\nu_\mu\rightarrow \nu_e$  $\bar\nu_\mu\rightarrow \bar\nu_e$    & MINOS+, MicroBooNE,  \\
& & $\nu_\mu\rightarrow\nu_\mu$    $\bar\nu_\mu\rightarrow \bar\nu_\mu$ & LAr1kton+MicroBooNE,  CERN SPS  \\ \hline
Low-energy $\nu$-Factory & App \& Dis & $\nu_e\rightarrow \nu_\mu$   $\bar\nu_e\rightarrow \bar\nu_\mu$ & $\nu$STORM \\

& &$\nu_\mu\rightarrow\nu_\mu$  $\bar\nu_\mu\rightarrow \bar\nu_\mu$ &  \\ 
\hline
\end{tabular} 
\caption{A summary of current and future sterile neutrino oscillation experiments.}
\label{summary_of_future}
\end{center}
\end{table}

\subsection{Future Experiments}
A summary of future sterile neutrino experiments is provided in Table~\ref{summary_of_future}.
\subsubsection{Pion Decay in Flight}
Muon neutrinos (antineutrinos) from positive (negative) pion DIF can be used to search for (anti)neutrino disappearance and electron (anti)neutrino appearance in the sterile neutrino region of interest. Given the usual neutrino energies for these experiments ($\mathcal{O}$(1~GeV)), the baseline for such an experiment can be considered ``short'' ($\mathcal{O}$(100-1000~m)).  

The BNB at Fermilab will provide pion-induced neutrinos to the MicroBooNE LArTPC-based detector starting in 2014~\cite{uboone}. MicroBooNE will probe the MiniBooNE low energy anomaly~\cite{miniboonelowe} with a $\sim$90 ton active volume about 100 meters closer to the neutrino source than MiniBooNE. Some coverage of the LSND allowed region in neutrino-mode is also expected, along with LArTPC development and needed precision neutrino-argon cross section measurements~\cite{anderson}. A design involving two LArTPC-based detectors in a near/far configuration, with MicroBooNE as the near detector, is also being considered for deployment in the BNB at Fermilab~\cite{larlar}. A similar two detector configuration in the CERN-SPS neutrino beam has recently been proposed~\cite{cernlar}. Two identical LArTPCs, in combination with magnetized spectrometers, would measure the mostly pion DIF induced muon neutrino composition of the beam as a function of distance (300~m, 1600~m) to probe electron neutrino appearance in the sterile neutrino parameter space.

Another BNB-based idea calls for a significant upgrade to the MiniBooNE experiment in which the current MiniBooNE detector becomes the 540~m baseline far detector in a two detector configuration and a MiniBooNE-like oil-based near detector is installed at a baseline of 200~m~\cite{boone}. Such a configuration could significantly reduce the now largely irreducible systematics associated with MiniBooNE-far-only, mainly coming from neutral pion background events and flux uncertainty. In conjunction with MicroBooNE, ``BooNE" could provide a sensitive study of LSND-like electron (anti)neutrino appearance, (anti)muon neutrino disappearance, and the MiniBooNE low energy anomaly.

A low energy 3-4~GeV/c muon storage ring could deliver a precisely known flux of electron neutrinos for a muon neutrino appearance search in the parameter space of interest for sterile neutrinos~\cite{nustorm}. The magnetized MINOS-like detectors, envisioned at 20-50~m (near) and $\sim$2000~m (far), would need to be magnetized in order to differentiate muon neutrino appearance from intrinsic muon antineutrinos created from the positive muon decay. Similar to most pion DIF beams, the muon storage ring could run in both neutrino-mode and antineutrino-mode. Such an experiment would also provide a technological demonstration of a muon storage ring with a ``simple" neutrino factory~\cite{neutfact}.

\subsubsection{Pion or Kaon Decay at Rest}
As discussed above, neutrinos from pion DAR and subsequent daughter muon DAR, with their well known spectrum, provide a source for an oscillation search. Notably, LSND employed muon antineutrinos from the pion daughter's muon DAR in establishing their 3.8$\sigma$ excess consistent with electron antineutrino appearance.

The 1~MW Spallation Neutron Source at Oak Ridge National Laboratory, a pion and muon DAR neutrino source, in combination with an LSND-style detector could directly probe the LSND excess with a factor of 100 lower steady-state background and higher beam power~\cite{oscsns}. A 1~MW source at a large liquid scintillator detector is also under consideration~\cite{Agarwalla}. Such an experiment could reconstruct appearance and disappearance oscillation waves across a $\sim$50~m length of detector.

If higher energy proton beams are targeted, then positive kaon DAR and the resulting monoenergetic (235.5~MeV) muon neutrino can also be used to search for sterile neutrinos through an electron neutrino appearance search with a LArTPC-based device~\cite{kdar}. An intense $>$3~GeV kinetic energy proton beam is required for such an experiment so as to produce an ample number of kaons per incoming proton.  

\subsubsection{Unstable Isotopes}
The disappearance of electron antineutrinos from radioactive isotopes is a direct probe of the reactor/gallium anomaly and an indirect probe of the LSND anomaly. As such neutrinos are in the ones-of-MeV range, the baseline for these experiments is generally on the order of tens of meters or so. Oscillation waves within a single detector can be observed if the neutrinos originate from a localized source, if the oscillation length is short enough, and if the detector has precise enough vertex resolution. 

The IsoDAR concept~\cite{isodar} calls for an intense 60~MeV proton source in combination with a kiloton-scale scintillation-based detector for sensitivity to the sterile neutrino. Such a source is being developed concurrently with the DAE$\delta$ALUS experiment, nominally a search for non-zero $\delta_{CP}$~\cite{daedalus}. Cyclotron-produced 60~MeV protons impinge on a beryllium-based target, mainly acting as a copious source of neutrons, which is surrounded by an isotopically pure shell of $^7$Li. $^8$Li, created via neutron capture on  $^7$Li inside the shell, decays to a 6.4~MeV mean energy electron antineutrino. Placing such an antineutrino source next to an existing detector such as KamLAND~\cite{Kamland} could quickly provide discovery-level sensitivity in the reactor anomaly allowed region. Furthermore, ISODAR has the ability to distinguish between one and multiple sterile neutrinos.

Another unstable isotope based idea involves the deployment of a radioactive source inside an existing kiloton-scale detector~\cite{celand} such as Borexino~\cite{borexino}, KamLAND~\cite{Kamland}, or SNO+~\cite{snoplus}. Electron antineutrinos from a small-extent, $\sim$2~PBq $^{144}$Ce or $^{106}$Ru beta source can be used to probe the sterile neutrino parameter space. For currently favored parameters associated with sterile neutrino(s), such antineutrinos are expected to disappear and reappear as a function of distance and energy inside of the detector, much like the IsoDAR concept described above.

\subsubsection{Nuclear Reactor}
A nuclear reactor can be used as a source for an electron antineutrino disappearance experiment with sensitivity to sterile neutrino(s). The Nucifer detector will likely be the first reactor-based detector to test the sterile neutrino hypothesis using antineutrino energy shape rather than just rate~\cite{nucifer}. The experiment will take data in 2012/2013. The idea is to place a 1~m$^3$-scale Gd-doped liquid scintillator device within a few tens of meters of a small-extent 70~MW research reactor in an attempt to observe antineutrino disappearance as a function of energy. The observation of an oscillation wave at high-$\Delta m^2$ would be unambiguous evidence for the existence of at least one sterile neutrino. Cosmic ray interactions and their products represent the largest source of background for this class of experiment.  

One of the challenges of a reactor-based search is the need for a relatively small reactor size, given the baseline required for maximal sensitivity to $\Delta m^2_{ij}\sim1$~eV$^2$; a large neutrino source size relative to the neutrino baseline smears $L$ and reduces $\Delta m^2_{ij}$ resolution. A sterile search at a GW-scale power reactor is possible, however. The SCRAAM experiment (see Ref.~\cite{whitepaper}) calls for a Gd-doped liquid scintillator detector at the San Onofre Nuclear Generation Station. 

\subsubsection{Neutral Current Based Experiments}
All of the future experiments previously discussed involve either disappearance or appearance of neutrinos and antineutrinos detected via the charged current. A NC-based disappearance experiment provides unique sensitivity to the sterile neutrino, however. In case a disappearance was observed in a NC experiment, one would know that the active flavor neutrino(s) in question had oscillated into the non-interacting sterile flavor. Specifically, such an experiment would provide a measure of $|U_{s4}|$, the sterile flavor composition of the fourth neutrino mass eigenstate, and definitively prove the existence of a sterile flavor neutrino, especially when considered in combination with CC based experiments. A full understanding of the mixing angles associated with sterile neutrino(s) will require a NC based experiment. The Ricochet concept~\cite{ricochet} calls for oscillometry measurements using NC coherent neutrino-nucleus scattering detected via low temperature bolometers~\cite{coherent,ricochet,sblcoherent}. Both reactor and isotope decay sources are being considered for these measurements, utilizing the as-yet-undetected coherent neutrino-nucleus scattering process.

\section{Conclusions}

This paper has presented results of SBL experiments
discussed within the context of oscillations involving sterile
neutrinos.  Fits to (3+1), (3+2), and (3+3)
models have been presented.     We have examined whether the 
(3+3) model addresses tensions observed with (3+1) and (3+2) fits.

Several issues arise when comparing data sets in (3+1) and (3+2) models.
In a (3+1) model, the compatibility of the neutrino vs.~antineutrino data sets 
is poor (0.14\%), and the compatibility among all data sets is 
only 0.043\%.   In a (3+2) model, there is a striking disagreement between
appearance and disappearance data sets, with a compatibility of 0.0082\%.


A (3+3) model fit achieves a compatibility of $90\%$ among all data sets.    The $\chi^2$-probability for the best fit is 67\%, compared to 2.1\% for the null (no oscillations) scenario.   Though this value is on the order of the $\chi^2$-probabilities found in (3+1) and (3+2) models, at 55\% and 69\%, respectively, the (3+3) fit resolves the incompatibility issues seen in (3+1) and (3+2) models, with the exception of the MiniBooNE appearance data sets. Therefore, we argue that (3+3) (and possibly (3+2)) fits should be the main focus of sterile neutrino phenomenological studies in the future. 

While the indications of sterile neutrino oscillations have historically been associated with only appearance-based SBL experiments, the recently realized suppression in observed $\bar\nu_e$ disappearance reactor experiments provides further motivation for these models. As we have shown one can consistently fit most results under the same, (3+3) hypothesis with improved compatibility. However, the need for additional information from both appearance and disappearance experiments provides strong motivation for pursuing the future experiments discussed in this paper.

\clearpage
\begin{center}
{ {\bf Acknowledgments}}
\end{center}
We thank William Louis and Zarko Pavlovic for valuable discussions. We thank the National Science Foundation for their support.
\bibliographystyle{apsrev4-1}
\bibliography{sbl}

\begin{thebibliography}{97}%
\makeatletter
\providecommand \@ifxundefined [1]{%
 \@ifx{#1\undefined}
}%
\providecommand \@ifnum [1]{%
 \ifnum #1\expandafter \@firstoftwo
 \else \expandafter \@secondoftwo
 \fi
}%
\providecommand \@ifx [1]{%
 \ifx #1\expandafter \@firstoftwo
 \else \expandafter \@secondoftwo
 \fi
}%
\providecommand \natexlab [1]{#1}%
\providecommand \enquote  [1]{``#1''}%
\providecommand \bibnamefont  [1]{#1}%
\providecommand \bibfnamefont [1]{#1}%
\providecommand \citenamefont [1]{#1}%
\providecommand \href@noop [0]{\@secondoftwo}%
\providecommand \href [0]{\begingroup \@sanitize@url \@href}%
\providecommand \@href[1]{\@@startlink{#1}\@@href}%
\providecommand \@@href[1]{\endgroup#1\@@endlink}%
\providecommand \@sanitize@url [0]{\catcode `\\12\catcode `\$12\catcode
  `\&12\catcode `\#12\catcode `\^12\catcode `\_12\catcode `\%12\relax}%
\providecommand \@@startlink[1]{}%
\providecommand \@@endlink[0]{}%
\providecommand \url  [0]{\begingroup\@sanitize@url \@url }%
\providecommand \@url [1]{\endgroup\@href {#1}{\urlprefix }}%
\providecommand \urlprefix  [0]{URL }%
\providecommand \Eprint [0]{\href }%
\providecommand \doibase [0]{http://dx.doi.org/}%
\providecommand \selectlanguage [0]{\@gobble}%
\providecommand \bibinfo  [0]{\@secondoftwo}%
\providecommand \bibfield  [0]{\@secondoftwo}%
\providecommand \translation [1]{[#1]}%
\providecommand \BibitemOpen [0]{}%
\providecommand \bibitemStop [0]{}%
\providecommand \bibitemNoStop [0]{.\EOS\space}%
\providecommand \EOS [0]{\spacefactor3000\relax}%
\providecommand \BibitemShut  [1]{\csname bibitem#1\endcsname}%
\let\auto@bib@innerbib\@empty
\bibitem [{\citenamefont {{K.~Abe {\it et al.} (Super-Kamiokande
  Collaboration)}}(2011)}]{SuperKsolar}%
  \BibitemOpen
  \bibfield  {author} {\bibinfo {author} {\bibnamefont {{K.~Abe {\it et al.}
  (Super-Kamiokande Collaboration)}}},\ }\href@noop {} {\bibfield  {journal}
  {\bibinfo  {journal} {Phys. Rev. D}\ }\textbf {\bibinfo {volume} {83}},\
  \bibinfo {pages} {052010} (\bibinfo {year} {2011})}\BibitemShut {NoStop}%
\bibitem [{\citenamefont {{B.~Aharmim {\it et al.} (SNO
  Collaboration)}}(2010)}]{SNO}%
  \BibitemOpen
  \bibfield  {author} {\bibinfo {author} {\bibnamefont {{B.~Aharmim {\it et
  al.} (SNO Collaboration)}}},\ }\href@noop {} {\bibfield  {journal} {\bibinfo
  {journal} {Phys. Rev. C}\ }\textbf {\bibinfo {volume} {81}},\ \bibinfo
  {pages} {055504} (\bibinfo {year} {2010})}\BibitemShut {NoStop}%
\bibitem [{\citenamefont {{S. Abe \textit{et al.} (KamLAND
  Collaboration)}}(2008)}]{Kamland}%
  \BibitemOpen
  \bibfield  {author} {\bibinfo {author} {\bibnamefont {{S. Abe \textit{et al.}
  (KamLAND Collaboration)}}},\ }\href@noop {} {\bibfield  {journal} {\bibinfo
  {journal} {Phys. Rev. Lett.}\ }\textbf {\bibinfo {volume} {100}},\ \bibinfo
  {pages} {221803} (\bibinfo {year} {2008})}\BibitemShut {NoStop}%
\bibitem [{\citenamefont {{R.~Wendell {\it et al.} (Kamiokande
  Collaboration)}}(2010)}]{SuperKatmos}%
  \BibitemOpen
  \bibfield  {author} {\bibinfo {author} {\bibnamefont {{R.~Wendell {\it et
  al.} (Kamiokande Collaboration)}}},\ }\href@noop {} {\bibfield  {journal}
  {\bibinfo  {journal} {Phys. Rev. D}\ }\textbf {\bibinfo {volume} {81}},\
  \bibinfo {pages} {092004} (\bibinfo {year} {2010})}\BibitemShut {NoStop}%
\bibitem [{\citenamefont {{W.W.M.~Allison {\it et al.} (Soudan-2
  Collaboration)}}(2005)}]{SoudanII}%
  \BibitemOpen
  \bibfield  {author} {\bibinfo {author} {\bibnamefont {{W.W.M.~Allison {\it et
  al.} (Soudan-2 Collaboration)}}},\ }\href@noop {} {\bibfield  {journal}
  {\bibinfo  {journal} {Phys. Rev. D}\ }\textbf {\bibinfo {volume} {72}},\
  \bibinfo {pages} {052005} (\bibinfo {year} {2005})}\BibitemShut {NoStop}%
\bibitem [{\citenamefont {{M.H.~Ahn \textit{et al.} (K2K
  Collaboration)}}(2006)}]{k2kosc}%
  \BibitemOpen
  \bibfield  {author} {\bibinfo {author} {\bibnamefont {{M.H.~Ahn \textit{et
  al.} (K2K Collaboration)}}},\ }\href@noop {} {\bibfield  {journal} {\bibinfo
  {journal} {Phys. Rev. D}\ }\textbf {\bibinfo {volume} {74}},\ \bibinfo
  {pages} {072003} (\bibinfo {year} {2006})}\BibitemShut {NoStop}%
\bibitem [{\citenamefont {{D.G. Michael {\it et al.} (MINOS
  Collaboration)}}(2006)}]{MinosCC1}%
  \BibitemOpen
  \bibfield  {author} {\bibinfo {author} {\bibnamefont {{D.G. Michael {\it et
  al.} (MINOS Collaboration)}}},\ }\href@noop {} {\bibfield  {journal}
  {\bibinfo  {journal} {Phys. Rev. Lett.}\ }\textbf {\bibinfo {volume} {97}},\
  \bibinfo {pages} {191801} (\bibinfo {year} {2006})}\BibitemShut {NoStop}%
\bibitem [{\citenamefont {{P. Adamson {\it et al.} (MINOS
  Collaboration)}}(2008{\natexlab{a}})}]{MinosCC2}%
  \BibitemOpen
  \bibfield  {author} {\bibinfo {author} {\bibnamefont {{P. Adamson {\it et
  al.} (MINOS Collaboration)}}},\ }\href@noop {} {\bibfield  {journal}
  {\bibinfo  {journal} {Phys. Rev. D}\ }\textbf {\bibinfo {volume} {77}},\
  \bibinfo {pages} {072002} (\bibinfo {year} {2008}{\natexlab{a}})}\BibitemShut
  {NoStop}%
\bibitem [{\citenamefont {{P. Adamson {\it et al.} (MINOS
  Collaboration)}}(2008{\natexlab{b}})}]{MinosCC3}%
  \BibitemOpen
  \bibfield  {author} {\bibinfo {author} {\bibnamefont {{P. Adamson {\it et
  al.} (MINOS Collaboration)}}},\ }\href@noop {} {\bibfield  {journal}
  {\bibinfo  {journal} {Phys. Rev. Lett.}\ }\textbf {\bibinfo {volume} {101}},\
  \bibinfo {pages} {131802} (\bibinfo {year} {2008}{\natexlab{b}})}\BibitemShut
  {NoStop}%
\bibitem [{\citenamefont {{P. Adamson {\it et al.} (MINOS
  Collaboration)}}(2011)}]{MINOS4}%
  \BibitemOpen
  \bibfield  {author} {\bibinfo {author} {\bibnamefont {{P. Adamson {\it et
  al.} (MINOS Collaboration)}}},\ }\href@noop {} {\bibfield  {journal}
  {\bibinfo  {journal} {Phys. Rev. Lett.}\ }\textbf {\bibinfo {volume} {106}},\
  \bibinfo {pages} {181801} (\bibinfo {year} {2011})}\BibitemShut {NoStop}%
\bibitem [{\citenamefont {{P. Adamson {\it et al.} (MINOS
  Collaboration)}}(2012)}]{MinosCC5}%
  \BibitemOpen
  \bibfield  {author} {\bibinfo {author} {\bibnamefont {{P. Adamson {\it et
  al.} (MINOS Collaboration)}}},\ }\href@noop {} {\bibfield  {journal}
  {\bibinfo  {journal} {Phys. Rev. Lett.}\ }\textbf {\bibinfo {volume} {107}},\
  \bibinfo {pages} {021801} (\bibinfo {year} {2012})}\BibitemShut {NoStop}%
\bibitem [{\citenamefont {{P. Adamson \textit{et al.} (MINOS
  Collaboration)}}(2011)}]{minostheta13}%
  \BibitemOpen
  \bibfield  {author} {\bibinfo {author} {\bibnamefont {{P. Adamson \textit{et
  al.} (MINOS Collaboration)}}},\ }\href@noop {} {\bibfield  {journal}
  {\bibinfo  {journal} {Phys. Rev. Lett.}\ }\textbf {\bibinfo {volume} {107}},\
  \bibinfo {pages} {181802} (\bibinfo {year} {2011})}\BibitemShut {NoStop}%
\bibitem [{\citenamefont {{K. Abe {\it et al.} (T2K
  Collaboration)}}(2011)}]{t2k}%
  \BibitemOpen
  \bibfield  {author} {\bibinfo {author} {\bibnamefont {{K. Abe {\it et al.}
  (T2K Collaboration)}}},\ }\href@noop {} {\bibfield  {journal} {\bibinfo
  {journal} {Phys. Rev. Lett.}\ }\textbf {\bibinfo {volume} {107}},\ \bibinfo
  {pages} {041801} (\bibinfo {year} {2011})}\BibitemShut {NoStop}%
\bibitem [{\citenamefont {{Y.~Abe {\it et al.} (Double Chooz
  Collaboration)}}(2012)}]{dc1stpub}%
  \BibitemOpen
  \bibfield  {author} {\bibinfo {author} {\bibnamefont {{Y.~Abe {\it et al.}
  (Double Chooz Collaboration)}}},\ }\href@noop {} {\bibfield  {journal}
  {\bibinfo  {journal} {Phys. Rev. Lett.}\ }\textbf {\bibinfo {volume} {108}},\
  \bibinfo {pages} {131801} (\bibinfo {year} {2012})}\BibitemShut {NoStop}%
\bibitem [{\citenamefont {{F.P. An {\it et al.} (Daya Bay
  Collaboration)}}(2012)}]{dayabay}%
  \BibitemOpen
  \bibfield  {author} {\bibinfo {author} {\bibnamefont {{F.P. An {\it et al.}
  (Daya Bay Collaboration)}}},\ }\href@noop {} {\bibfield  {journal} {\bibinfo
  {journal} {Phys. Rev. Lett.}\ }\textbf {\bibinfo {volume} {108}},\ \bibinfo
  {pages} {171803} (\bibinfo {year} {2012})}\BibitemShut {NoStop}%
\bibitem [{\citenamefont {Ahn}\ \emph {et~al.}(2012)\citenamefont {Ahn} \emph
  {et~al.}}]{reno}%
  \BibitemOpen
  \bibfield  {author} {\bibinfo {author} {\bibfnamefont {J.}~\bibnamefont
  {Ahn}} \emph {et~al.} (\bibinfo {collaboration} {RENO collaboration}),\
  }\href@noop {} {\bibfield  {journal} {\bibinfo  {journal} {Phys. Rev. Lett.}\
  }\textbf {\bibinfo {volume} {108}},\ \bibinfo {pages} {191802} (\bibinfo
  {year} {2012})}\BibitemShut {NoStop}%
\bibitem [{\citenamefont {Aguilar-Arevalo}\ \emph {et~al.}(2001)\citenamefont
  {Aguilar-Arevalo} \emph {et~al.}}]{LSND}%
  \BibitemOpen
  \bibfield  {author} {\bibinfo {author} {\bibfnamefont {A.}~\bibnamefont
  {Aguilar-Arevalo}} \emph {et~al.} (\bibinfo {collaboration} {LSND
  Collaboration}),\ }\href@noop {} {\bibfield  {journal} {\bibinfo  {journal}
  {Phys. Rev. D}\ }\textbf {\bibinfo {volume} {64}},\ \bibinfo {pages} {112007}
  (\bibinfo {year} {2001})}\BibitemShut {NoStop}%
\bibitem [{\citenamefont {Aguilar-Arevalo}\ \emph {et~al.}(2012)\citenamefont
  {Aguilar-Arevalo} \emph {et~al.}}]{MB}%
  \BibitemOpen
  \bibfield  {author} {\bibinfo {author} {\bibfnamefont {A.}~\bibnamefont
  {Aguilar-Arevalo}} \emph {et~al.} (\bibinfo {collaboration} {MiniBooNE
  Collaboration}),\ }\href@noop {} {\bibfield  {journal} {\bibinfo  {journal}
  {To be submitted to Phys. Rev. Lett.}\ } (\bibinfo {year}
  {2012})}\BibitemShut {NoStop}%
\bibitem [{\citenamefont {Mention}\ \emph {et~al.}(2011)\citenamefont
  {Mention}, \citenamefont {Fechner}, \citenamefont {Lasserre}, \citenamefont
  {Mueller}, \citenamefont {Lhuillier} \emph {et~al.}}]{reactor}%
  \BibitemOpen
  \bibfield  {author} {\bibinfo {author} {\bibfnamefont {G.}~\bibnamefont
  {Mention}}, \bibinfo {author} {\bibfnamefont {M.}~\bibnamefont {Fechner}},
  \bibinfo {author} {\bibfnamefont {T.}~\bibnamefont {Lasserre}}, \bibinfo
  {author} {\bibfnamefont {T.}~\bibnamefont {Mueller}}, \bibinfo {author}
  {\bibfnamefont {D.}~\bibnamefont {Lhuillier}},  \emph {et~al.},\ }\href@noop
  {} {\bibfield  {journal} {\bibinfo  {journal} {Phys. Rev. D}\ }\textbf
  {\bibinfo {volume} {83}},\ \bibinfo {pages} {073006} (\bibinfo {year}
  {2011})}\BibitemShut {NoStop}%
\bibitem [{\citenamefont {{M.~Sorel, J.M.~Conrad and
  M.~Shaevitz}}(2004)}]{sorel}%
  \BibitemOpen
  \bibfield  {author} {\bibinfo {author} {\bibnamefont {{M.~Sorel, J.M.~Conrad
  and M.~Shaevitz}}},\ }\href@noop {} {\bibfield  {journal} {\bibinfo
  {journal} {Phys. Rev. D}\ }\textbf {\bibinfo {volume} {70}},\ \bibinfo
  {pages} {073004} (\bibinfo {year} {2004})}\BibitemShut {NoStop}%
\bibitem [{\citenamefont {Karagiorgi}\ \emph {et~al.}(2007)\citenamefont
  {Karagiorgi}, \citenamefont {Aguilar-Arevalo}, \citenamefont {Conrad},
  \citenamefont {Shaevitz}, \citenamefont {Whisnant} \emph
  {et~al.}}]{karagiorgi}%
  \BibitemOpen
  \bibfield  {author} {\bibinfo {author} {\bibfnamefont {G.}~\bibnamefont
  {Karagiorgi}}, \bibinfo {author} {\bibfnamefont {A.}~\bibnamefont
  {Aguilar-Arevalo}}, \bibinfo {author} {\bibfnamefont {J.}~\bibnamefont
  {Conrad}}, \bibinfo {author} {\bibfnamefont {M.}~\bibnamefont {Shaevitz}},
  \bibinfo {author} {\bibfnamefont {K.}~\bibnamefont {Whisnant}},  \emph
  {et~al.},\ }\href@noop {} {\bibfield  {journal} {\bibinfo  {journal} {Phys.
  Rev. D}\ }\textbf {\bibinfo {volume} {5}},\ \bibinfo {pages} {013011}
  (\bibinfo {year} {2007})}\BibitemShut {NoStop}%
\bibitem [{\citenamefont {{G.~Karagiorgi, Z.~Djurcic, J.M.~Conrad,
  M.H.~Shaevitz and M.~Sorel}}(2009)}]{viability}%
  \BibitemOpen
  \bibfield  {author} {\bibinfo {author} {\bibnamefont {{G.~Karagiorgi,
  Z.~Djurcic, J.M.~Conrad, M.H.~Shaevitz and M.~Sorel}}},\ }\href@noop {}
  {\bibfield  {journal} {\bibinfo  {journal} {Phys. Rev. D}\ }\textbf {\bibinfo
  {volume} {80}},\ \bibinfo {pages} {073001} (\bibinfo {year}
  {2009})}\BibitemShut {NoStop}%
\bibitem [{\citenamefont {{K.N. Abazajian \textit{et al.}}}()}]{whitepaper}%
  \BibitemOpen
  \bibfield  {author} {\bibinfo {author} {\bibnamefont {{K.N. Abazajian
  \textit{et al.}}}},\ }\href@noop {} {\ }\Eprint
  {http://arxiv.org/abs/arXiv:1204.5379 [hep-ph] (2012)} {arXiv:1204.5379
  [hep-ph] (2012)} \BibitemShut {NoStop}%
\bibitem [{\citenamefont {Maltoni}\ and\ \citenamefont
  {Schwetz}(2007)}]{Maltoni:2007zf}%
  \BibitemOpen
  \bibfield  {author} {\bibinfo {author} {\bibfnamefont {M.}~\bibnamefont
  {Maltoni}}\ and\ \bibinfo {author} {\bibfnamefont {T.}~\bibnamefont
  {Schwetz}},\ }\href@noop {} {\bibfield  {journal} {\bibinfo  {journal} {Phys.
  Rev. D}\ }\textbf {\bibinfo {volume} {76}},\ \bibinfo {pages} {093005}
  (\bibinfo {year} {2007})}\BibitemShut {NoStop}%
\bibitem [{\citenamefont {{ALEPH Collaboration, DELPHI Collaboration, L3
  Collaboration, OPAL Collaboration, SLD Collaboration, LEP Electroweak Working
  Group, SLD Electroweak and Heavy Flavour Groups}}(2006)}]{zwidth}%
  \BibitemOpen
  \bibfield  {author} {\bibinfo {author} {\bibnamefont {{ALEPH Collaboration,
  DELPHI Collaboration, L3 Collaboration, OPAL Collaboration, SLD
  Collaboration, LEP Electroweak Working Group, SLD Electroweak and Heavy
  Flavour Groups}}},\ }\href@noop {} {\bibfield  {journal} {\bibinfo  {journal}
  {Phys. Rept.}\ }\textbf {\bibinfo {volume} {427}},\ \bibinfo {pages} {257}
  (\bibinfo {year} {2006})}\BibitemShut {NoStop}%
\bibitem [{\citenamefont {{A.A. Aguilar-Arevalo and others (MiniBooNE
  Collaboration)}}(2010)}]{nubarminiosc}%
  \BibitemOpen
  \bibfield  {author} {\bibinfo {author} {\bibnamefont {{A.A. Aguilar-Arevalo
  and others (MiniBooNE Collaboration)}}},\ }\href@noop {} {\bibfield
  {journal} {\bibinfo  {journal} {Phys. Rev. Lett.}\ }\textbf {\bibinfo
  {volume} {105}},\ \bibinfo {pages} {181801} (\bibinfo {year}
  {2010})}\BibitemShut {NoStop}%
\bibitem [{\citenamefont {Acero}\ \emph {et~al.}(2008)\citenamefont {Acero},
  \citenamefont {Giunti},\ and\ \citenamefont {Laveder}}]{Gallium}%
  \BibitemOpen
  \bibfield  {author} {\bibinfo {author} {\bibfnamefont {M.~A.}\ \bibnamefont
  {Acero}}, \bibinfo {author} {\bibfnamefont {C.}~\bibnamefont {Giunti}}, \
  and\ \bibinfo {author} {\bibfnamefont {M.}~\bibnamefont {Laveder}},\
  }\href@noop {} {\bibfield  {journal} {\bibinfo  {journal} {Phys. Rev. D}\
  }\textbf {\bibinfo {volume} {78}},\ \bibinfo {pages} {073009} (\bibinfo
  {year} {2008})}\BibitemShut {NoStop}%
\bibitem [{\citenamefont {Dunkley}\ \emph {et~al.}(2011)\citenamefont
  {Dunkley}, \citenamefont {Hlozek}, \citenamefont {Sievers}, \citenamefont
  {Acquaviva}, \citenamefont {Ade} \emph {et~al.}}]{atta}%
  \BibitemOpen
  \bibfield  {author} {\bibinfo {author} {\bibfnamefont {J.}~\bibnamefont
  {Dunkley}}, \bibinfo {author} {\bibfnamefont {R.}~\bibnamefont {Hlozek}},
  \bibinfo {author} {\bibfnamefont {J.}~\bibnamefont {Sievers}}, \bibinfo
  {author} {\bibfnamefont {V.}~\bibnamefont {Acquaviva}}, \bibinfo {author}
  {\bibfnamefont {P.}~\bibnamefont {Ade}},  \emph {et~al.},\ }\href@noop {}
  {\bibfield  {journal} {\bibinfo  {journal} {Astrophys. J.}\ }\textbf
  {\bibinfo {volume} {739}},\ \bibinfo {pages} {52} (\bibinfo {year}
  {2011})}\BibitemShut {NoStop}%
\bibitem [{\citenamefont {Gonzalez-Garcia}\ \emph {et~al.}(2010)\citenamefont
  {Gonzalez-Garcia}, \citenamefont {Maltoni},\ and\ \citenamefont
  {Salvado}}]{GG}%
  \BibitemOpen
  \bibfield  {author} {\bibinfo {author} {\bibfnamefont {M.}~\bibnamefont
  {Gonzalez-Garcia}}, \bibinfo {author} {\bibfnamefont {M.}~\bibnamefont
  {Maltoni}}, \ and\ \bibinfo {author} {\bibfnamefont {J.}~\bibnamefont
  {Salvado}},\ }\href@noop {} {\bibfield  {journal} {\bibinfo  {journal}
  {JHEP}\ }\textbf {\bibinfo {volume} {1008}},\ \bibinfo {pages} {117}
  (\bibinfo {year} {2010})}\BibitemShut {NoStop}%
\bibitem [{\citenamefont {Giusarma}\ \emph {et~al.}(2012)\citenamefont
  {Giusarma}, \citenamefont {Archidiacono}, \citenamefont {de~Putter},
  \citenamefont {Melchiorri},\ and\ \citenamefont {Mena}}]{Mena}%
  \BibitemOpen
  \bibfield  {author} {\bibinfo {author} {\bibfnamefont {E.}~\bibnamefont
  {Giusarma}}, \bibinfo {author} {\bibfnamefont {M.}~\bibnamefont
  {Archidiacono}}, \bibinfo {author} {\bibfnamefont {R.}~\bibnamefont
  {de~Putter}}, \bibinfo {author} {\bibfnamefont {A.}~\bibnamefont
  {Melchiorri}}, \ and\ \bibinfo {author} {\bibfnamefont {O.}~\bibnamefont
  {Mena}},\ }\href@noop {} {\bibfield  {journal} {\bibinfo  {journal} {Phys.
  Rev. D}\ }\textbf {\bibinfo {volume} {85}},\ \bibinfo {pages} {083522}
  (\bibinfo {year} {2012})}\BibitemShut {NoStop}%
\bibitem [{\citenamefont {Hamann}\ \emph {et~al.}(2011)\citenamefont {Hamann},
  \citenamefont {Hannestad}, \citenamefont {Raffelt},\ and\ \citenamefont
  {Wong}}]{planck}%
  \BibitemOpen
  \bibfield  {author} {\bibinfo {author} {\bibfnamefont {J.}~\bibnamefont
  {Hamann}}, \bibinfo {author} {\bibfnamefont {S.}~\bibnamefont {Hannestad}},
  \bibinfo {author} {\bibfnamefont {G.~G.}\ \bibnamefont {Raffelt}}, \ and\
  \bibinfo {author} {\bibfnamefont {Y.~Y.}\ \bibnamefont {Wong}},\ }\href@noop
  {} {\bibfield  {journal} {\bibinfo  {journal} {JCAP}\ }\textbf {\bibinfo
  {volume} {1109}},\ \bibinfo {pages} {034} (\bibinfo {year}
  {2011})}\BibitemShut {NoStop}%
\bibitem [{\citenamefont {Karagiorgi}\ \emph {et~al.}()\citenamefont
  {Karagiorgi}, \citenamefont {Shaevitz},\ and\ \citenamefont
  {Conrad}}]{bsmmatter}%
  \BibitemOpen
  \bibfield  {author} {\bibinfo {author} {\bibfnamefont {G.}~\bibnamefont
  {Karagiorgi}}, \bibinfo {author} {\bibfnamefont {M.~H.}\ \bibnamefont
  {Shaevitz}}, \ and\ \bibinfo {author} {\bibfnamefont {J.~M.}\ \bibnamefont
  {Conrad}},\ }\href@noop {} {\ }\Eprint {http://arxiv.org/abs/arXiv:1202.1024
  [hep-ph] (2012)} {arXiv:1202.1024 [hep-ph] (2012)} \BibitemShut {NoStop}%
\bibitem [{\citenamefont {Abdurashitov}\ \emph {et~al.}(2009)\citenamefont
  {Abdurashitov} \emph {et~al.}}]{SAGE3}%
  \BibitemOpen
  \bibfield  {author} {\bibinfo {author} {\bibfnamefont {J.}~\bibnamefont
  {Abdurashitov}} \emph {et~al.} (\bibinfo {collaboration} {SAGE
  Collaboration}),\ }\href@noop {} {\bibfield  {journal} {\bibinfo  {journal}
  {Phys. Rev. C}\ }\textbf {\bibinfo {volume} {80}},\ \bibinfo {pages} {015807}
  (\bibinfo {year} {2009})}\BibitemShut {NoStop}%
\bibitem [{\citenamefont {Kaether}\ \emph {et~al.}(2010)\citenamefont
  {Kaether}, \citenamefont {Hampel}, \citenamefont {Heusser}, \citenamefont
  {Kiko},\ and\ \citenamefont {Kirsten}}]{GALLEX3}%
  \BibitemOpen
  \bibfield  {author} {\bibinfo {author} {\bibfnamefont {F.}~\bibnamefont
  {Kaether}}, \bibinfo {author} {\bibfnamefont {W.}~\bibnamefont {Hampel}},
  \bibinfo {author} {\bibfnamefont {G.}~\bibnamefont {Heusser}}, \bibinfo
  {author} {\bibfnamefont {J.}~\bibnamefont {Kiko}}, \ and\ \bibinfo {author}
  {\bibfnamefont {T.}~\bibnamefont {Kirsten}},\ }\href@noop {} {\bibfield
  {journal} {\bibinfo  {journal} {Phys. Lett. B}\ }\textbf {\bibinfo {volume}
  {685}},\ \bibinfo {pages} {47} (\bibinfo {year} {2010})}\BibitemShut
  {NoStop}%
\bibitem [{\citenamefont {Jones}(2012)}]{Christhesis}%
  \BibitemOpen
  \bibfield  {author} {\bibinfo {author} {\bibfnamefont {C.}~\bibnamefont
  {Jones}},\ }\href@noop {} {\bibfield  {journal} {\bibinfo  {journal} {PhD
  Thesis, Massachusetts Institute of Technology}\ } (\bibinfo {year}
  {2012})}\BibitemShut {NoStop}%
\bibitem [{\citenamefont {Mueller}\ \emph {et~al.}(2011)\citenamefont
  {Mueller}, \citenamefont {Lhuillier}, \citenamefont {Fallot}, \citenamefont
  {Letourneau}, \citenamefont {Cormon} \emph {et~al.}}]{Mueller}%
  \BibitemOpen
  \bibfield  {author} {\bibinfo {author} {\bibfnamefont {T.}~\bibnamefont
  {Mueller}}, \bibinfo {author} {\bibfnamefont {D.}~\bibnamefont {Lhuillier}},
  \bibinfo {author} {\bibfnamefont {M.}~\bibnamefont {Fallot}}, \bibinfo
  {author} {\bibfnamefont {A.}~\bibnamefont {Letourneau}}, \bibinfo {author}
  {\bibfnamefont {S.}~\bibnamefont {Cormon}},  \emph {et~al.},\ }\href@noop {}
  {\bibfield  {journal} {\bibinfo  {journal} {Phys. Rev. C}\ }\textbf {\bibinfo
  {volume} {83}},\ \bibinfo {pages} {054615} (\bibinfo {year}
  {2011})}\BibitemShut {NoStop}%
\bibitem [{\citenamefont {Huber}(2011)}]{huber}%
  \BibitemOpen
  \bibfield  {author} {\bibinfo {author} {\bibfnamefont {P.}~\bibnamefont
  {Huber}},\ }\href@noop {} {\bibfield  {journal} {\bibinfo  {journal} {Phys.
  Rev. C}\ }\textbf {\bibinfo {volume} {84}},\ \bibinfo {pages} {024617}
  (\bibinfo {year} {2011})}\BibitemShut {NoStop}%
\bibitem [{\citenamefont {Burman}\ and\ \citenamefont {und
  Umwelt}(1996)}]{Burman:1996vx}%
  \BibitemOpen
  \bibfield  {author} {\bibinfo {author} {\bibfnamefont {R.}~\bibnamefont
  {Burman}}\ and\ \bibinfo {author} {\bibfnamefont {F.~K.~T.}\ \bibnamefont
  {und Umwelt}},\ }\href@noop {} {\emph {\bibinfo {title} {Neutrino Fluxes at
  KARMEN from Pion Decay in Flight}}}\ (\bibinfo  {publisher} {FZKA},\ \bibinfo
  {year} {1996})\BibitemShut {NoStop}%
\bibitem [{\citenamefont {Burman}\ \emph {et~al.}(1990)\citenamefont {Burman},
  \citenamefont {Potter},\ and\ \citenamefont {Smith}}]{Burman:1989dq}%
  \BibitemOpen
  \bibfield  {author} {\bibinfo {author} {\bibfnamefont {R.}~\bibnamefont
  {Burman}}, \bibinfo {author} {\bibfnamefont {M.}~\bibnamefont {Potter}}, \
  and\ \bibinfo {author} {\bibfnamefont {E.}~\bibnamefont {Smith}},\
  }\href@noop {} {\bibfield  {journal} {\bibinfo  {journal} {Nucl. Instrum.
  Meth. A}\ }\textbf {\bibinfo {volume} {291}},\ \bibinfo {pages} {621}
  (\bibinfo {year} {1990})}\BibitemShut {NoStop}%
\bibitem [{\citenamefont {{A.A. Aguilar-Arevalo \textit{et al.} (MiniBooNE
  Collaboration)}}(2009{\natexlab{a}})}]{AguilarArevalo:2008yp}%
  \BibitemOpen
  \bibfield  {author} {\bibinfo {author} {\bibnamefont {{A.A. Aguilar-Arevalo
  \textit{et al.} (MiniBooNE Collaboration)}}},\ }\href@noop {} {\bibfield
  {journal} {\bibinfo  {journal} {Phys. Rev. D}\ }\textbf {\bibinfo {volume}
  {79}},\ \bibinfo {pages} {072002} (\bibinfo {year}
  {2009}{\natexlab{a}})}\BibitemShut {NoStop}%
\bibitem [{\citenamefont {Kopp}(2007)}]{kopp}%
  \BibitemOpen
  \bibfield  {author} {\bibinfo {author} {\bibfnamefont {S.~E.}\ \bibnamefont
  {Kopp}},\ }\href@noop {} {\bibfield  {journal} {\bibinfo  {journal} {Phys.
  Rept.}\ }\textbf {\bibinfo {volume} {439}},\ \bibinfo {pages} {101} (\bibinfo
  {year} {2007})}\BibitemShut {NoStop}%
\bibitem [{\citenamefont {Catanesi}\ \emph {et~al.}(2007)\citenamefont
  {Catanesi} \emph {et~al.}}]{harp}%
  \BibitemOpen
  \bibfield  {author} {\bibinfo {author} {\bibfnamefont {M.}~\bibnamefont
  {Catanesi}} \emph {et~al.} (\bibinfo {collaboration} {HARP Collaboration}),\
  }\href@noop {} {\bibfield  {journal} {\bibinfo  {journal} {Nucl. Instrum.
  Meth. A}\ }\textbf {\bibinfo {volume} {571}},\ \bibinfo {pages} {527}
  (\bibinfo {year} {2007})}\BibitemShut {NoStop}%
\bibitem [{\citenamefont {Graf}\ \emph {et~al.}(2010)\citenamefont {Graf} \emph
  {et~al.}}]{mipp}%
  \BibitemOpen
  \bibfield  {author} {\bibinfo {author} {\bibfnamefont {N.}~\bibnamefont
  {Graf}} \emph {et~al.} (\bibinfo {collaboration} {MIPP Collaboration}),\
  }\href@noop {} {\bibfield  {journal} {\bibinfo  {journal} {Nucl. Instrum.
  Meth. A}\ }\textbf {\bibinfo {volume} {615}},\ \bibinfo {pages} {27}
  (\bibinfo {year} {2010})}\BibitemShut {NoStop}%
\bibitem [{\citenamefont {Conrad}\ and\ \citenamefont
  {Shaevitz}(2012)}]{ConradShaevitz}%
  \BibitemOpen
  \bibfield  {author} {\bibinfo {author} {\bibfnamefont {J.}~\bibnamefont
  {Conrad}}\ and\ \bibinfo {author} {\bibfnamefont {M.}~\bibnamefont
  {Shaevitz}},\ }\href@noop {} {\bibfield  {journal} {\bibinfo  {journal}
  {Phys. Rev. D}\ }\textbf {\bibinfo {volume} {85}},\ \bibinfo {pages} {013017}
  (\bibinfo {year} {2012})}\BibitemShut {NoStop}%
\bibitem [{\citenamefont {Armbruster}\ \emph {et~al.}(2002)\citenamefont
  {Armbruster} \emph {et~al.}}]{Karmen}%
  \BibitemOpen
  \bibfield  {author} {\bibinfo {author} {\bibfnamefont {B.}~\bibnamefont
  {Armbruster}} \emph {et~al.} (\bibinfo {collaboration} {KARMEN
  Collaboration}),\ }\href@noop {} {\bibfield  {journal} {\bibinfo  {journal}
  {Phys. Rev. D}\ }\textbf {\bibinfo {volume} {65}},\ \bibinfo {pages} {112001}
  (\bibinfo {year} {2002})}\BibitemShut {NoStop}%
\bibitem [{\citenamefont {{A.A. Aguilar-Arevalo \textit{et al.} (MiniBooNE
  Collaboration)}}()}]{mbprlinprogress}%
  \BibitemOpen
  \bibfield  {author} {\bibinfo {author} {\bibnamefont {{A.A. Aguilar-Arevalo
  \textit{et al.} (MiniBooNE Collaboration)}}},\ }\href@noop {} {\ }\Eprint
  {http://arxiv.org/abs/In Preparation} {In Preparation} \BibitemShut {NoStop}%
\bibitem [{\citenamefont {Aguilar-Arevalo}\ \emph {et~al.}(2008)\citenamefont
  {Aguilar-Arevalo} \emph {et~al.}}]{MBlog}%
  \BibitemOpen
  \bibfield  {author} {\bibinfo {author} {\bibfnamefont {A.}~\bibnamefont
  {Aguilar-Arevalo}} \emph {et~al.} (\bibinfo {collaboration} {MiniBooNE
  Collaboration}),\ }\href@noop {} {\bibfield  {journal} {\bibinfo  {journal}
  {Phys. Rev. D}\ }\textbf {\bibinfo {volume} {78}},\ \bibinfo {pages} {012007}
  (\bibinfo {year} {2008})}\BibitemShut {NoStop}%
\bibitem [{\citenamefont {Aguilar-Arevalo}\ \emph {et~al.}(2009)\citenamefont
  {Aguilar-Arevalo} \emph {et~al.}}]{Kendall}%
  \BibitemOpen
  \bibfield  {author} {\bibinfo {author} {\bibfnamefont {A.~A.}\ \bibnamefont
  {Aguilar-Arevalo}} \emph {et~al.} (\bibinfo {collaboration} {MiniBooNE
  Collaboration}),\ }\href@noop {} {\bibfield  {journal} {\bibinfo  {journal}
  {Phys. Rev. Lett.}\ }\textbf {\bibinfo {volume} {103}},\ \bibinfo {pages}
  {061802} (\bibinfo {year} {2009})}\BibitemShut {NoStop}%
\bibitem [{\citenamefont {Adamson}\ \emph {et~al.}(2009)\citenamefont {Adamson}
  \emph {et~al.}}]{MBNuMI}%
  \BibitemOpen
  \bibfield  {author} {\bibinfo {author} {\bibfnamefont {P.}~\bibnamefont
  {Adamson}} \emph {et~al.} (\bibinfo {collaboration} {MiniBooNE and MINOS
  Collaborations}),\ }\href@noop {} {\bibfield  {journal} {\bibinfo  {journal}
  {Phys. Rev. Lett.}\ }\textbf {\bibinfo {volume} {102}},\ \bibinfo {pages}
  {211801} (\bibinfo {year} {2009})}\BibitemShut {NoStop}%
\bibitem [{\citenamefont {Astier}\ \emph {et~al.}(2003)\citenamefont {Astier}
  \emph {et~al.}}]{NOMAD1}%
  \BibitemOpen
  \bibfield  {author} {\bibinfo {author} {\bibfnamefont {P.}~\bibnamefont
  {Astier}} \emph {et~al.} (\bibinfo {collaboration} {NOMAD Collaboration}),\
  }\href@noop {} {\bibfield  {journal} {\bibinfo  {journal} {Phys. Lett. B}\
  }\textbf {\bibinfo {volume} {570}},\ \bibinfo {pages} {19} (\bibinfo {year}
  {2003})}\BibitemShut {NoStop}%
\bibitem [{\citenamefont {Stockdale}\ \emph {et~al.}(1985)\citenamefont
  {Stockdale}, \citenamefont {Bodek}, \citenamefont {Borcherding},
  \citenamefont {Giokaris}, \citenamefont {Lang} \emph {et~al.}}]{CCFR84}%
  \BibitemOpen
  \bibfield  {author} {\bibinfo {author} {\bibfnamefont {I.}~\bibnamefont
  {Stockdale}}, \bibinfo {author} {\bibfnamefont {A.}~\bibnamefont {Bodek}},
  \bibinfo {author} {\bibfnamefont {F.}~\bibnamefont {Borcherding}}, \bibinfo
  {author} {\bibfnamefont {N.}~\bibnamefont {Giokaris}}, \bibinfo {author}
  {\bibfnamefont {K.}~\bibnamefont {Lang}},  \emph {et~al.},\ }\href@noop {}
  {\bibfield  {journal} {\bibinfo  {journal} {Z. Phys. C}\ }\textbf {\bibinfo
  {volume} {27}},\ \bibinfo {pages} {53} (\bibinfo {year} {1985})}\BibitemShut
  {NoStop}%
\bibitem [{\citenamefont {Dydak}\ \emph {et~al.}(1984)\citenamefont {Dydak},
  \citenamefont {Feldman}, \citenamefont {Guyot}, \citenamefont {Merlo},
  \citenamefont {Meyer} \emph {et~al.}}]{CDHS}%
  \BibitemOpen
  \bibfield  {author} {\bibinfo {author} {\bibfnamefont {F.}~\bibnamefont
  {Dydak}}, \bibinfo {author} {\bibfnamefont {G.}~\bibnamefont {Feldman}},
  \bibinfo {author} {\bibfnamefont {C.}~\bibnamefont {Guyot}}, \bibinfo
  {author} {\bibfnamefont {J.}~\bibnamefont {Merlo}}, \bibinfo {author}
  {\bibfnamefont {H.}~\bibnamefont {Meyer}},  \emph {et~al.},\ }\href@noop {}
  {\bibfield  {journal} {\bibinfo  {journal} {Phys. Lett. B}\ }\textbf
  {\bibinfo {volume} {134}},\ \bibinfo {pages} {281} (\bibinfo {year}
  {1984})}\BibitemShut {NoStop}%
\bibitem [{\citenamefont {Casper}(2002)}]{NUANCE}%
  \BibitemOpen
  \bibfield  {author} {\bibinfo {author} {\bibfnamefont {D.}~\bibnamefont
  {Casper}},\ }\href@noop {} {\bibfield  {journal} {\bibinfo  {journal} {Nucl.
  Phys. Proc. Suppl.}\ }\textbf {\bibinfo {volume} {112}},\ \bibinfo {pages}
  {161} (\bibinfo {year} {2002})}\BibitemShut {NoStop}%
\bibitem [{\citenamefont {Declais}\ \emph {et~al.}(1995)\citenamefont
  {Declais}, \citenamefont {Favier}, \citenamefont {Metref}, \citenamefont
  {Pessard}, \citenamefont {Achkar} \emph {et~al.}}]{Bugey}%
  \BibitemOpen
  \bibfield  {author} {\bibinfo {author} {\bibfnamefont {Y.}~\bibnamefont
  {Declais}}, \bibinfo {author} {\bibfnamefont {J.}~\bibnamefont {Favier}},
  \bibinfo {author} {\bibfnamefont {A.}~\bibnamefont {Metref}}, \bibinfo
  {author} {\bibfnamefont {H.}~\bibnamefont {Pessard}}, \bibinfo {author}
  {\bibfnamefont {B.}~\bibnamefont {Achkar}},  \emph {et~al.},\ }\href@noop {}
  {\bibfield  {journal} {\bibinfo  {journal} {Nucl. Phys. B}\ }\textbf
  {\bibinfo {volume} {434}},\ \bibinfo {pages} {503} (\bibinfo {year}
  {1995})}\BibitemShut {NoStop}%
\bibitem [{\citenamefont {Giunti}\ and\ \citenamefont
  {Laveder}(2011{\natexlab{a}})}]{Giunti1}%
  \BibitemOpen
  \bibfield  {author} {\bibinfo {author} {\bibfnamefont {C.}~\bibnamefont
  {Giunti}}\ and\ \bibinfo {author} {\bibfnamefont {M.}~\bibnamefont
  {Laveder}},\ }\href@noop {} {\bibfield  {journal} {\bibinfo  {journal} {Phys.
  Rev. C}\ }\textbf {\bibinfo {volume} {83}},\ \bibinfo {pages} {065504}
  (\bibinfo {year} {2011}{\natexlab{a}})}\BibitemShut {NoStop}%
\bibitem [{\citenamefont {Esmaili}\ \emph {et~al.}()\citenamefont {Esmaili},
  \citenamefont {Halzen},\ and\ \citenamefont {Peres}}]{icecube}%
  \BibitemOpen
  \bibfield  {author} {\bibinfo {author} {\bibfnamefont {A.}~\bibnamefont
  {Esmaili}}, \bibinfo {author} {\bibfnamefont {F.}~\bibnamefont {Halzen}}, \
  and\ \bibinfo {author} {\bibfnamefont {O.}~\bibnamefont {Peres}},\
  }\href@noop {} {\ }\Eprint {http://arxiv.org/abs/arXiv:1206.6903 [hep-ph]
  (2012)} {arXiv:1206.6903 [hep-ph] (2012)} \BibitemShut {NoStop}%
\bibitem [{\citenamefont {{P. Adamson \textit{et al.} (MINOS
  collaboration)}}()}]{rightsignMINOS}%
  \BibitemOpen
  \bibfield  {author} {\bibinfo {author} {\bibnamefont {{P. Adamson \textit{et
  al.} (MINOS collaboration)}}},\ }\href@noop {} {\ }\Eprint
  {http://arxiv.org/abs/arXiv:1202.2772 [hep-ex] (2012)} {arXiv:1202.2772
  [hep-ex] (2012)} \BibitemShut {NoStop}%
\bibitem [{\citenamefont {{P. Adamson \textit{et al.} (MINOS
  Collaboration)}}()}]{wrongsignMINOS}%
  \BibitemOpen
  \bibfield  {author} {\bibinfo {author} {\bibnamefont {{P. Adamson \textit{et
  al.} (MINOS Collaboration)}}},\ }\href@noop {} {\ }\Eprint
  {http://arxiv.org/abs/arXiv:1108.1509 [hep-ex] (2012)} {arXiv:1108.1509
  [hep-ex] (2012)} \BibitemShut {NoStop}%
\bibitem [{\citenamefont {Maltoni}\ \emph {et~al.}(2004)\citenamefont
  {Maltoni}, \citenamefont {Schwetz}, \citenamefont {Tortola},\ and\
  \citenamefont {Valle}}]{MaltoniValle}%
  \BibitemOpen
  \bibfield  {author} {\bibinfo {author} {\bibfnamefont {M.}~\bibnamefont
  {Maltoni}}, \bibinfo {author} {\bibfnamefont {T.}~\bibnamefont {Schwetz}},
  \bibinfo {author} {\bibfnamefont {M.}~\bibnamefont {Tortola}}, \ and\
  \bibinfo {author} {\bibfnamefont {J.}~\bibnamefont {Valle}},\ }\href@noop {}
  {\bibfield  {journal} {\bibinfo  {journal} {New J. Phys.}\ }\textbf {\bibinfo
  {volume} {6}},\ \bibinfo {pages} {122} (\bibinfo {year} {2004})}\BibitemShut
  {NoStop}%
\bibitem [{\citenamefont {M.~Honda}\ and\ \citenamefont
  {Midorikawa}(2004)}]{leptonicCPref33}%
  \BibitemOpen
  \bibfield  {author} {\bibinfo {author} {\bibfnamefont {K.~K.}\ \bibnamefont
  {M.~Honda}, \bibfnamefont {T.~Kajita}}\ and\ \bibinfo {author} {\bibfnamefont
  {S.}~\bibnamefont {Midorikawa}},\ }\href@noop {} {\bibfield  {journal}
  {\bibinfo  {journal} {Phys. Rev. D}\ }\textbf {\bibinfo {volume} {70}},\
  \bibinfo {pages} {043008} (\bibinfo {year} {2004})}\BibitemShut {NoStop}%
\bibitem [{\citenamefont {Gonzalez-Garcia}\ and\ \citenamefont
  {Maltoni}(2004)}]{leptonicCPref34}%
  \BibitemOpen
  \bibfield  {author} {\bibinfo {author} {\bibfnamefont {M.~C.}\ \bibnamefont
  {Gonzalez-Garcia}}\ and\ \bibinfo {author} {\bibfnamefont {M.}~\bibnamefont
  {Maltoni}},\ }\href@noop {} {\bibfield  {journal} {\bibinfo  {journal}
  {Phys.\ Rev. D}\ }\textbf {\bibinfo {volume} {70}},\ \bibinfo {pages}
  {033010} (\bibinfo {year} {2004})}\BibitemShut {NoStop}%
\bibitem [{\citenamefont {{S.H. Ahn {\it et al.} (K2K
  Collaboration)}}(2001)}]{leptonicCPref13_1}%
  \BibitemOpen
  \bibfield  {author} {\bibinfo {author} {\bibnamefont {{S.H. Ahn {\it et al.}
  (K2K Collaboration)}}},\ }\href@noop {} {\bibfield  {journal} {\bibinfo
  {journal} {Phys. Lett. B}\ }\textbf {\bibinfo {volume} {511}},\ \bibinfo
  {pages} {178} (\bibinfo {year} {2001})}\BibitemShut {NoStop}%
\bibitem [{\citenamefont {{M.H. Ahn {\it et al.} (K2K
  Collaboration)}}(2003)}]{leptonicCPref13_2}%
  \BibitemOpen
  \bibfield  {author} {\bibinfo {author} {\bibnamefont {{M.H. Ahn {\it et al.}
  (K2K Collaboration)}}},\ }\href@noop {} {\bibfield  {journal} {\bibinfo
  {journal} {Phys. Rev. Lett.}\ }\textbf {\bibinfo {volume} {041801}},\
  \bibinfo {pages} {90} (\bibinfo {year} {2003})}\BibitemShut {NoStop}%
\bibitem [{\citenamefont {{M.H. Ahn {\it et al.} (K2K
  Collaboration)}}(2006)}]{leptonicCPref13_3}%
  \BibitemOpen
  \bibfield  {author} {\bibinfo {author} {\bibnamefont {{M.H. Ahn {\it et al.}
  (K2K Collaboration)}}},\ }\href@noop {} {\bibfield  {journal} {\bibinfo
  {journal} {Phys. Rev. D}\ }\textbf {\bibinfo {volume} {072003}},\ \bibinfo
  {pages} {74} (\bibinfo {year} {2006})}\BibitemShut {NoStop}%
\bibitem [{\citenamefont {Br\^{a}emaud}(1999)}]{markov}%
  \BibitemOpen
  \bibfield  {author} {\bibinfo {author} {\bibfnamefont {P.}~\bibnamefont
  {Br\^{a}emaud}},\ }\href@noop {} {\emph {\bibinfo {title} {Markov chains:
  Gibbs fields, Monte Carlo simulation, and queues}}}\ (\bibinfo  {publisher}
  {Springer, New York},\ \bibinfo {year} {1999})\BibitemShut {NoStop}%
\bibitem [{\citenamefont {Maltoni}\ and\ \citenamefont
  {Schwetz}(2003)}]{pgtest}%
  \BibitemOpen
  \bibfield  {author} {\bibinfo {author} {\bibfnamefont {M.}~\bibnamefont
  {Maltoni}}\ and\ \bibinfo {author} {\bibfnamefont {T.}~\bibnamefont
  {Schwetz}},\ }\href@noop {} {\bibfield  {journal} {\bibinfo  {journal} {Phys.
  Rev. D}\ }\textbf {\bibinfo {volume} {68}},\ \bibinfo {pages} {033020}
  (\bibinfo {year} {2003})}\BibitemShut {NoStop}%
\bibitem [{\citenamefont {Kopp}\ \emph {et~al.}(2011)\citenamefont {Kopp},
  \citenamefont {Maltoni},\ and\ \citenamefont {Schwetz}}]{JKopp}%
  \BibitemOpen
  \bibfield  {author} {\bibinfo {author} {\bibfnamefont {J.}~\bibnamefont
  {Kopp}}, \bibinfo {author} {\bibfnamefont {M.}~\bibnamefont {Maltoni}}, \
  and\ \bibinfo {author} {\bibfnamefont {T.}~\bibnamefont {Schwetz}},\
  }\href@noop {} {\bibfield  {journal} {\bibinfo  {journal} {Phys. Rev. Lett.}\
  }\textbf {\bibinfo {volume} {107}},\ \bibinfo {pages} {091801} (\bibinfo
  {year} {2011})}\BibitemShut {NoStop}%
\bibitem [{\citenamefont {Giunti}\ and\ \citenamefont
  {Laveder}(2011{\natexlab{b}})}]{gli}%
  \BibitemOpen
  \bibfield  {author} {\bibinfo {author} {\bibfnamefont {C.}~\bibnamefont
  {Giunti}}\ and\ \bibinfo {author} {\bibfnamefont {M.}~\bibnamefont
  {Laveder}},\ }\href@noop {} {\bibfield  {journal} {\bibinfo  {journal} {Phys.
  Lett. B}\ }\textbf {\bibinfo {volume} {706}},\ \bibinfo {pages} {2007}
  (\bibinfo {year} {2011}{\natexlab{b}})}\BibitemShut {NoStop}%
\bibitem [{\citenamefont {Donini}\ \emph {et~al.}()\citenamefont {Donini},
  \citenamefont {Hernandez}, \citenamefont {Lopez-Pavon}, \citenamefont
  {Maltoni},\ and\ \citenamefont {Schwetz}}]{Donini}%
  \BibitemOpen
  \bibfield  {author} {\bibinfo {author} {\bibfnamefont {A.}~\bibnamefont
  {Donini}}, \bibinfo {author} {\bibfnamefont {P.}~\bibnamefont {Hernandez}},
  \bibinfo {author} {\bibfnamefont {J.}~\bibnamefont {Lopez-Pavon}}, \bibinfo
  {author} {\bibfnamefont {M.}~\bibnamefont {Maltoni}}, \ and\ \bibinfo
  {author} {\bibfnamefont {T.}~\bibnamefont {Schwetz}},\ }\href@noop {} {\
  }\Eprint {http://arxiv.org/abs/arXiv:1205.5230 [hep-ph] (2012)}
  {arXiv:1205.5230 [hep-ph] (2012)} \BibitemShut {NoStop}%
\bibitem [{\citenamefont {{M.~Matrini and M.~Ericson and
  G.~Chanfray}}()}]{martini}%
  \BibitemOpen
  \bibfield  {author} {\bibinfo {author} {\bibnamefont {{M.~Matrini and
  M.~Ericson and G.~Chanfray}}},\ }\href@noop {} {\ }\Eprint
  {http://arxiv.org/abs/arXiv:1202.4745v2 [hep-ph] (2012)} {arXiv:1202.4745v2
  [hep-ph] (2012)} \BibitemShut {NoStop}%
\bibitem [{\citenamefont {J.~A.~Harvey}\ and\ \citenamefont
  {Hill}(2007)}]{axial}%
  \BibitemOpen
  \bibfield  {author} {\bibinfo {author} {\bibfnamefont {C.~H.}\ \bibnamefont
  {J.~A.~Harvey}}\ and\ \bibinfo {author} {\bibfnamefont {R.}~\bibnamefont
  {Hill}},\ }\href@noop {} {\bibfield  {journal} {\bibinfo  {journal} {Phys.
  Rev. Lett}\ }\textbf {\bibinfo {volume} {99}},\ \bibinfo {pages} {261601}
  (\bibinfo {year} {2007})}\BibitemShut {NoStop}%
\bibitem [{\citenamefont {{H. Chen \textit{et al.} (MicroBooNE
  Collaboration)}}()}]{uboone}%
  \BibitemOpen
  \bibfield  {author} {\bibinfo {author} {\bibnamefont {{H. Chen \textit{et
  al.} (MicroBooNE Collaboration)}}},\ }\href@noop {} {\ }\Eprint
  {http://arxiv.org/abs/{``Proposal for a New Experiment Using the Booster and
  NuMI Neutrino Beamlines: MicroBooNE'' (2007)}} {{``Proposal for a New
  Experiment Using the Booster and NuMI Neutrino Beamlines: MicroBooNE''
  (2007)}} \BibitemShut {NoStop}%
\bibitem [{\citenamefont {{G. Tzanakos \textit{et al.} (MINOS+
  Collaboration)}}()}]{minosplus}%
  \BibitemOpen
  \bibfield  {author} {\bibinfo {author} {\bibnamefont {{G. Tzanakos \textit{et
  al.} (MINOS+ Collaboration)}}},\ }\href@noop {} {\ }\Eprint
  {http://arxiv.org/abs/{``MINOS+: a Proposal to FNAL to run MINOS with the
  medium energy NuMI beam, Tech. Rep.'' (2011)}} {{``MINOS+: a Proposal to FNAL
  to run MINOS with the medium energy NuMI beam, Tech. Rep.'' (2011)}}
  \BibitemShut {NoStop}%
\bibitem [{\citenamefont {{I. Stancu \textit{et al.}}}()}]{boone}%
  \BibitemOpen
  \bibfield  {author} {\bibinfo {author} {\bibnamefont {{I. Stancu \textit{et
  al.}}}},\ }\href@noop {} {\ }\Eprint {http://arxiv.org/abs/{``A Proposal to
  Build a MiniBooNE Near Detector: BooNE'' (2011)}} {{``A Proposal to Build a
  MiniBooNE Near Detector: BooNE'' (2011)}} \BibitemShut {NoStop}%
\bibitem [{\citenamefont {{G. Karagiorgi}}()}]{larlar}%
  \BibitemOpen
  \bibfield  {author} {\bibinfo {author} {\bibnamefont {{G. Karagiorgi}}},\
  }\href@noop {} {\ }\Eprint {http://arxiv.org/abs/{``Proceedings of TIPP
  2011'' (2011)}} {{``Proceedings of TIPP 2011'' (2011)}} \BibitemShut
  {NoStop}%
\bibitem [{\citenamefont {{C. Rubbia \textit{et al.}}}()}]{cernlar}%
  \BibitemOpen
  \bibfield  {author} {\bibinfo {author} {\bibnamefont {{C. Rubbia \textit{et
  al.}}}},\ }\href@noop {} {\ }\Eprint {http://arxiv.org/abs/{``A comprehensive
  search for anomalies from neutrino and anti-neutrino oscllations at large
  mass difference with two LArDTPC imaging detector at different distances from
  the CERN-PS, Tech. Rep.'' (2011)}} {{``A comprehensive search for anomalies
  from neutrino and anti-neutrino oscllations at large mass difference with two
  LArDTPC imaging detector at different distances from the CERN-PS, Tech.
  Rep.'' (2011)}} \BibitemShut {NoStop}%
\bibitem [{\citenamefont {{P. Kyberd \textit{et al.}}}()}]{nustorm}%
  \BibitemOpen
  \bibfield  {author} {\bibinfo {author} {\bibnamefont {{P. Kyberd \textit{et
  al.}}}},\ }\href@noop {} {\ }\Eprint {http://arxiv.org/abs/arXiv:1206.0294
  [hep-ex] (2012)} {arXiv:1206.0294 [hep-ex] (2012)} \BibitemShut {NoStop}%
\bibitem [{\citenamefont {{A.J. Anderson, J.M. Conrad, E. Figueroa-Feliciano,
  K. Scholberg, and J. Spitz}}(2011)}]{coherent}%
  \BibitemOpen
  \bibfield  {author} {\bibinfo {author} {\bibnamefont {{A.J. Anderson, J.M.
  Conrad, E. Figueroa-Feliciano, K. Scholberg, and J. Spitz}}},\ }\href@noop {}
  {\bibfield  {journal} {\bibinfo  {journal} {Phys. Rev. D}\ }\textbf {\bibinfo
  {volume} {84}},\ \bibinfo {pages} {113008} (\bibinfo {year}
  {2011})}\BibitemShut {NoStop}%
\bibitem [{\citenamefont {{A.J. Anderson, {\it et al.}}}()}]{sblcoherent}%
  \BibitemOpen
  \bibfield  {author} {\bibinfo {author} {\bibnamefont {{A.J. Anderson, {\it et
  al.}}}},\ }\href@noop {} {\ }\Eprint {http://arxiv.org/abs/arXiv:1201.3805
  [hep-ph] (2012)} {arXiv:1201.3805 [hep-ph] (2012)} \BibitemShut {NoStop}%
\bibitem [{\citenamefont {{I.~Stancu {\it et al.} (OscSNS
  Collaboration)}}()}]{oscsns}%
  \BibitemOpen
  \bibfield  {author} {\bibinfo {author} {\bibnamefont {{I.~Stancu {\it et al.}
  (OscSNS Collaboration)}}},\ }\href@noop {} {\ }\Eprint
  {http://arxiv.org/abs/{``The OscSNS White Paper'' (2008)}} {{``The OscSNS
  White Paper'' (2008)}} \BibitemShut {NoStop}%
\bibitem [{\citenamefont {{S.K. Agarwalla and P. Huber}}(2011)}]{superk}%
  \BibitemOpen
  \bibfield  {author} {\bibinfo {author} {\bibnamefont {{S.K. Agarwalla and P.
  Huber}}},\ }\href@noop {} {\bibfield  {journal} {\bibinfo  {journal} {Phys.
  Lett. B}\ }\textbf {\bibinfo {volume} {696}},\ \bibinfo {pages} {359}
  (\bibinfo {year} {2011})}\BibitemShut {NoStop}%
\bibitem [{\citenamefont {{J. Spitz}}(2012)}]{kdar}%
  \BibitemOpen
  \bibfield  {author} {\bibinfo {author} {\bibnamefont {{J. Spitz}}},\
  }\href@noop {} {\bibfield  {journal} {\bibinfo  {journal} {Phys. Rev. D}\
  }\textbf {\bibinfo {volume} {85}},\ \bibinfo {pages} {093020} (\bibinfo
  {year} {2012})}\BibitemShut {NoStop}%
\bibitem [{\citenamefont {{C. Grieb, J. Link, and R.S.
  Raghavan}}(2007)}]{lenssterile}%
  \BibitemOpen
  \bibfield  {author} {\bibinfo {author} {\bibnamefont {{C. Grieb, J. Link, and
  R.S. Raghavan}}},\ }\href@noop {} {\bibfield  {journal} {\bibinfo  {journal}
  {Phys. Rev. D}\ }\textbf {\bibinfo {volume} {75}},\ \bibinfo {pages} {093006}
  (\bibinfo {year} {2007})}\BibitemShut {NoStop}%
\bibitem [{\citenamefont {{J. Formaggio, E. Figueroa-Feliciano, A.J. Anderson
  }}(2012)}]{ricochet}%
  \BibitemOpen
  \bibfield  {author} {\bibinfo {author} {\bibnamefont {{J. Formaggio, E.
  Figueroa-Feliciano, A.J. Anderson }}},\ }\href@noop {} {\bibfield  {journal}
  {\bibinfo  {journal} {Phys. Rev. D}\ }\textbf {\bibinfo {volume} {85}},\
  \bibinfo {pages} {013009} (\bibinfo {year} {2012})}\BibitemShut {NoStop}%
\bibitem [{\citenamefont {{M. Cribier \textit{et al.}}}(2011)}]{celand}%
  \BibitemOpen
  \bibfield  {author} {\bibinfo {author} {\bibnamefont {{M. Cribier \textit{et
  al.}}}},\ }\href@noop {} {\bibfield  {journal} {\bibinfo  {journal} {Phys.
  Rev. Lett.}\ }\textbf {\bibinfo {volume} {107}},\ \bibinfo {pages} {201801}
  (\bibinfo {year} {2011})}\BibitemShut {NoStop}%
\bibitem [{\citenamefont {{A. Bungau \textit{et al.}}}()}]{isodar}%
  \BibitemOpen
  \bibfield  {author} {\bibinfo {author} {\bibnamefont {{A. Bungau \textit{et
  al.}}}},\ }\href@noop {} {\ }\Eprint {http://arxiv.org/abs/arXiv:1205.4419
  [hep-ex] (2012)} {arXiv:1205.4419 [hep-ex] (2012)} \BibitemShut {NoStop}%
\bibitem [{\citenamefont {{A. Porta \textit{et al.}}}(2010)}]{nucifer}%
  \BibitemOpen
  \bibfield  {author} {\bibinfo {author} {\bibnamefont {{A. Porta \textit{et
  al.}}}},\ }\href@noop {} {\bibfield  {journal} {\bibinfo  {journal} {IEEE
  Transactions on Nuclear Science}\ }\textbf {\bibinfo {volume} {57}},\
  \bibinfo {pages} {2732} (\bibinfo {year} {2010})}\BibitemShut {NoStop}%
\bibitem [{\citenamefont {Serebrov}()}]{neutrino4}%
  \BibitemOpen
  \bibfield  {author} {\bibinfo {author} {\bibfnamefont {A.}~\bibnamefont
  {Serebrov}},\ }\href@noop {} {\ }\Eprint
  {http://arxiv.org/abs/arXiv:1205.2955 [hep-ph] (2012)} {arXiv:1205.2955
  [hep-ph] (2012)} \BibitemShut {NoStop}%
\bibitem [{\citenamefont {Gninenko}(2012)}]{Vanucci}%
  \BibitemOpen
  \bibfield  {author} {\bibinfo {author} {\bibfnamefont {S.}~\bibnamefont
  {Gninenko}},\ }\href@noop {} {\bibfield  {journal} {\bibinfo  {journal}
  {Phys. Rev. D}\ }\textbf {\bibinfo {volume} {85}},\ \bibinfo {pages} {051702}
  (\bibinfo {year} {2012})}\BibitemShut {NoStop}%
\bibitem [{\citenamefont {{A. Aguilar-Arevalo \textit{et al.} (MiniBooNE
  Collaboration)}}()}]{Teppei}%
  \BibitemOpen
  \bibfield  {author} {\bibinfo {author} {\bibnamefont {{A. Aguilar-Arevalo
  \textit{et al.} (MiniBooNE Collaboration)}}},\ }\href@noop {} {\ }\Eprint
  {http://arxiv.org/abs/arXiv:1206.6915 [hep-ph] (2012)} {arXiv:1206.6915
  [hep-ph] (2012)} \BibitemShut {NoStop}%
\bibitem [{\citenamefont {{S.K. Agarwalla, J.M. Conrad, and M.H.
  Shaevitz}}(2011)}]{Agarwalla}%
  \BibitemOpen
  \bibfield  {author} {\bibinfo {author} {\bibnamefont {{S.K. Agarwalla, J.M.
  Conrad, and M.H. Shaevitz}}},\ }\href@noop {} {\bibfield  {journal} {\bibinfo
   {journal} {JHEP}\ }\textbf {\bibinfo {volume} {1112}},\ \bibinfo {pages}
  {085} (\bibinfo {year} {2011})}\BibitemShut {NoStop}%
\bibitem [{\citenamefont {{A.A. Aguilar-Arevalo \textit{et al.} (MiniBooNE
  Collaboration)}}(2009{\natexlab{b}})}]{miniboonelowe}%
  \BibitemOpen
  \bibfield  {author} {\bibinfo {author} {\bibnamefont {{A.A. Aguilar-Arevalo
  \textit{et al.} (MiniBooNE Collaboration)}}},\ }\href@noop {} {\bibfield
  {journal} {\bibinfo  {journal} {Phys. Rev. Lett.}\ }\textbf {\bibinfo
  {volume} {102}},\ \bibinfo {pages} {101802} (\bibinfo {year}
  {2009}{\natexlab{b}})}\BibitemShut {NoStop}%
\bibitem [{\citenamefont {{C. Anderson \textit{et al.} (ArgoNeuT
  Collaboration)}}(2012)}]{anderson}%
  \BibitemOpen
  \bibfield  {author} {\bibinfo {author} {\bibnamefont {{C. Anderson \textit{et
  al.} (ArgoNeuT Collaboration)}}},\ }\href@noop {} {\bibfield  {journal}
  {\bibinfo  {journal} {Phys. Rev. Lett.}\ }\textbf {\bibinfo {volume} {108}},\
  \bibinfo {pages} {161802} (\bibinfo {year} {2012})}\BibitemShut {NoStop}%
\bibitem [{\citenamefont {{S. Geer}}(1998)}]{neutfact}%
  \BibitemOpen
  \bibfield  {author} {\bibinfo {author} {\bibnamefont {{S. Geer}}},\
  }\href@noop {} {\bibfield  {journal} {\bibinfo  {journal} {Phys. Rev. D}\
  }\textbf {\bibinfo {volume} {57}},\ \bibinfo {pages} {6989} (\bibinfo {year}
  {1998})}\BibitemShut {NoStop}%
\bibitem [{\citenamefont {{J.M. Conrad and M.H. Shaevitz}}(2010)}]{daedalus}%
  \BibitemOpen
  \bibfield  {author} {\bibinfo {author} {\bibnamefont {{J.M. Conrad and M.H.
  Shaevitz}}},\ }\href@noop {} {\bibfield  {journal} {\bibinfo  {journal}
  {Phys. Rev. Lett.}\ }\textbf {\bibinfo {volume} {104}},\ \bibinfo {pages}
  {141802} (\bibinfo {year} {2010})}\BibitemShut {NoStop}%
\bibitem [{\citenamefont {{G. Alimonti \textit{et al.} (Borexino
  Collaboration)}}(2002)}]{borexino}%
  \BibitemOpen
  \bibfield  {author} {\bibinfo {author} {\bibnamefont {{G. Alimonti \textit{et
  al.} (Borexino Collaboration)}}},\ }\href@noop {} {\bibfield  {journal}
  {\bibinfo  {journal} {Astropart. Phys.}\ }\textbf {\bibinfo {volume} {16}},\
  \bibinfo {pages} {205} (\bibinfo {year} {2002})}\BibitemShut {NoStop}%
\bibitem [{\citenamefont {{M. Chen}}(2005)}]{snoplus}%
  \BibitemOpen
  \bibfield  {author} {\bibinfo {author} {\bibnamefont {{M. Chen}}},\
  }\href@noop {} {\bibfield  {journal} {\bibinfo  {journal} {Nucl. Phys. Proc.
  Suppl.}\ }\textbf {\bibinfo {volume} {145}},\ \bibinfo {pages} {65} (\bibinfo
  {year} {2005})}\BibitemShut {NoStop}%
\end{thebibliography}%


\end{document}